\documentclass[preprint2]{aastex6}
\usepackage{float}
\usepackage{graphicx}
\usepackage{amsmath}
\usepackage[utf8]{inputenc}
\usepackage{natbib}
\usepackage{color}
\usepackage{verbatim}
\usepackage{mathtools}
\usepackage{outlines}
\usepackage{ulem}
\usepackage{totcount}
\usepackage{textcomp}
\usepackage{afterpage}

\newtotcounter{citnum}
\def\oldbibitem{} \let\oldbibitem=\bibitem
\def\bibitem{\stepcounter{citnum}\oldbibitem}

\newcommand{\cII}{[C\,\textsc{ii}]}
\newcommand{\nII}{[N\,\textsc{ii}]}
\newcommand{\siII}{[Si\,\textsc{ii}]}
\newcommand{\feII}{[Fe\,\textsc{ii}]}
\newcommand{\oI}{[O\,\textsc{i}]}
\newcommand{\oIII}{[O\,\textsc{iii}]}
\newcommand{\oIV}{[O\,\textsc{iv}]}
\newcommand{\neII}{[Ne\,\textsc{ii}]}
\newcommand{\neIII}{[Ne\,\textsc{iii}]}
\newcommand{\neV}{[Ne\,\textsc{v}]}
\newcommand{\sIV}{[S\,\textsc{iv}]}
\newcommand{\sIII}{[S\,\textsc{iii}]}
\newcommand{\PAH}{\ensuremath{\textrm{\scriptsize PAH}}}
\newcommand{\TIR}{\ensuremath{\textrm{\scriptsize TIR}}}
\newcommand{\IR}{\ensuremath{\textrm{\scriptsize IR}}}
\newcommand{\um}{\,\textmu m}
\newcommand{\angstrom}{\ensuremath{\textrm{\AA}}}

\newcolumntype{s}{!{\extracolsep{8pt}}c!{\extracolsep{0pt}}}
\newcolumntype{t}{!{\extracolsep{-7pt}}c!{\extracolsep{0pt}}}
\newcolumntype{p}{!{\extracolsep{-3pt}}c!{\extracolsep{0pt}}}

\slugcomment{The Astrophysical Journal 855, 1 (2018) \hspace{0.5cm} Published 7 March 2018}

\shortauthors{Smercina \textit{et al.}}
\shorttitle{After The Fall}

\begin{document}

\title{After The Fall: The Dust and Gas in E+A Post-Starburst Galaxies}

\author{%
A. Smercina\altaffilmark{1,2},
J.D.T. Smith\altaffilmark{2,3},
D.A. Dale\altaffilmark{4},
K.D. French\altaffilmark{5,6,\#},
K.V. Croxall\altaffilmark{7},
S. Zhukovska\altaffilmark{8}, 
A. Togi\altaffilmark{9}, \\
E.F. Bell\altaffilmark{1}, 
A.F. Crocker\altaffilmark{10},
B.T. Draine\altaffilmark{11},
T.H. Jarrett\altaffilmark{12}, 
C. Tremonti\altaffilmark{13},
Yujin Yang\altaffilmark{14}, 
A.I. Zabludoff\altaffilmark{6}
}

\altaffiltext{1}{Department of Astronomy, University of Michigan, Ann Arbor, MI 48109, USA; \color{blue}{asmerci@umich.edu}}
\altaffiltext{2}{Ritter Astrophysical Research Center, University of Toledo, Toledo, OH 43606, USA}
\altaffiltext{3}{Max-Planck-Institut f\"{u}r Astronomie, K\"{o}nigstuhl 17, D-69117 Heidelberg, Germany}
\altaffiltext{4}{Department of Physics \& Astronomy, University of Wyoming, Laramie, WY 82071, USA}
\altaffiltext{5}{Observatories of the Carnegie Institute for Science, 813 Santa Barbara Street, Pasadena, CA 91101, USA}
\altaffiltext{6}{Department of Astronomy, University of Arizona, Steward Observatory, Tucson, AZ 85721, USA}
\altaffiltext{7}{Illumination Works LLC, 5550 Blazer Parkway, Suite 150, Dublin, OH 43017}
\altaffiltext{8}{Max Planck Institut f\"{u}r Astrophysik, Karl-Schwarzschild-Str. 1, 85748 Garching, Germany}
\altaffiltext{9}{Department of Physics and Astronomy, University of Texas at San Antonio, One UTSA Circle, San Antonio, TX 78249, USA}
\altaffiltext{10}{Department of Physics, Reed College, Portland, OR 97202, USA}
\altaffiltext{11}{Department of Astrophysical Sciences, Princeton University, Princeton, NJ 08544, USA}
\altaffiltext{12}{Department of Astronomy, University of Cape Town, Rondebosch 7701 South Africa}
\altaffiltext{13}{Department of Astronomy, University of Wisconsin--Madison, Madison, WI 53706, USA}
\altaffiltext{14}{Korea Astronomy and Space Science Institute, 776 Daedeokdae-ro, Yuseong-gu, Daejeon 34055, Korea}

\altaffiliation{$^{\#}$\,Hubble Fellow}

\begin{abstract}
The traditional picture of post-starburst galaxies as dust- and gas-poor merger remnants, rapidly transitioning to quiescence, has been recently challenged. Unexpected detections of a significant ISM in many post-starbursts raise important questions. Are they truly quiescent and, if so, what mechanisms inhibit further star formation? What processes dominate their ISM energetics? We present an infrared spectroscopic and photometric survey of 33 SDSS-selected E+A post-starbursts, aimed at resolving these questions. We find compact, warm dust reservoirs with high PAH abundances, and total gas and dust masses significantly higher than expected from stellar recycling alone. Both PAH/TIR and dust-to-burst stellar mass ratios are seen to decrease with post-burst age, indicative of the accumulating effects of dust destruction and an incipient transition to hot, early-type ISM properties. Their infrared spectral properties are unique, with dominant PAH emission, very weak nebular lines, unusually strong H$_{2}$\ rotational emission, and deep \cII\ deficits. There is substantial scatter among SFR indicators, and both PAH and TIR luminosities provide overestimates. Even as potential upper limits, all tracers show that the SFR has typically experienced a more than two order-of-magnitude decline since the starburst, and that the SFR is considerably lower than expected given both their stellar masses and molecular gas densities. These results paint a coherent picture of systems in which star formation was, indeed, rapidly truncated, but in which the ISM was \textit{not} completely expelled, and is instead supported against collapse by latent or continued injection of turbulent or mechanical heating. The resulting aging burst populations provide a ``high-soft'' radiation field which seemingly dominates the E+As' unusual ISM energetics. 
\end{abstract}
\keywords{galaxies: evolution -- galaxies: interactions -- galaxies: ISM -- galaxies: starburst}

\section{Introduction}
\label{sec:intro}
Once thought to be a simple evolutionary sequence, the pathways leading galaxies from the star-forming blue cloud to the quiescent red sequence have been revealed to be incredibly diverse \citep{barro2013,schawinski2014}. The cessation of star formation appears to happen on vastly different timescales, strongly dependent on a galaxy's growth history \citep{martin2007}. A class of unique objects called post-starbursts galaxies (PSBs) appear to be the remnants of the most violent of such ``quenching'' events. 

PSBs were originally characterized in galaxy clusters by both \cite{dressler1983} and \cite{couch&sharples1987} as galaxies with strong Balmer absorption features, but surprisingly weak nebular emission lines. Their optical spectra resemble a linear combination of a several hundred Myr-old (A-star dominated) stellar population and an old stellar population (early-type or K-star dominated), leading to their ``E(K)+A'' designation. E+As' unique spectral characteristics suggest that they experienced a burst of star formation, which was rapidly quenched (in $\lesssim$\ 100 Myr) several hundred Myr ago \citep{dressler1983,couch&sharples1987}. PSBs likely have various formation mechanisms --- such as ram pressure stripping from cluster infall \citep[see][]{blanton&moustakas2009} and intragroup interactions in both isolated groups and clusters \citep[e.g.,][]{zabludoff1996,zabludoff1998,alatalo2015b}. In the field, PSBs are nearly ubiquitously observed to host strong tidal features and kinematics indicative of violent relaxation due to major, late-stage mergers \citep{zabludoff1996,yang2004,yang2008}. Though locally rare, they are thought to be more common at high-redshift (\citealt{tran2004}; \citealt{zahid2016}; \citealt{kriek2016}) --- indeed, \cite{tran2004} estimate that $\sim$70\% of present-day ellipticals may have passed through a post-starburst phase at $z < 1$. This, as well as more recent results by \cite{wild2016}, suggests that PSBs may be a critical evolutionary component to the development of the present-day red sequence. 

In the past decade, much work has been done to constrain the physical mechanisms which could halt star formation on the short timescales required in PSB galaxies --- of order a single dynamical period. Based on hydrodynamic simulations, \cite{hopkins2006} presented a unified, merger-driven evolutionary sequence for galaxies, beginning with mergers of gas-rich disks, progressing to late-stage mergers with heavily dust-obscured, central starbursts, coupled with SMBH fueling and subsequent quasar-mode feedback, and culminating in the development of quiescent spheroids resembling present-day ellipticals. Observationally, late-stage gas-rich mergers appear to manifest as (ultra) luminous infrared galaxies --- (U)LIRGs \citep{kormendy&sanders1992,kartaltepe2010,draper2012,carpineti2015}. 

This basic framework has been modified with the addition of E+As \citep[e.g.,][]{hopkins2008,snyder2011}. Hydrodynamic simulations have focused on the brief time-steps directly following the late-stage merger phase, suggesting that E+As represent an intermediate stage in the evolution of gas-rich major mergers, prior to the development of the spheroidal remnant. These results predict that the E+A signature remains centrally-concentrated throughout --- a direct link to their compact progenitor starbursts. \cite{snyder2011} found that the E+A phase generally lasted $\le 300$~Myr and only in rare cases persisted on Gyr timescales. 

Elegant as this evolutionary picture is, observational evidence to support it has remained elusive, and the detailed evolution of the interstellar medium (ISM) in E+As remains almost completely unknown. \cite{alatalo2016a,alatalo2016b} have recently characterized a sample of galaxies possessing strong Balmer decrements, but stronger emission lines than classically selected E+As (see \S\,\ref{sec:sample-selection}). These objects appear to be in very early post-starburst phases, likely probing both the latest stages of the burst and earliest stages of star formation decline. By selection, their sample possesses strong classical shocks, inferred from optical emission line ratios indicative of shock excitation. The vast majority are likely still forming stars at a relatively high rate, based on their nebular lines, and they show evidence of AGN-driven outflows in many cases, consistent with the \cite{hopkins2006} evolutionary scenario. They detect significant reservoirs of molecular gas in all cases \citep{alatalo2016b}. 

E+As appear to be heterogeneous in their H\,I content --- some possess large reservoirs, while others harbor little atomic gas \citep{chang2001,buyle2006,zwaan2013}. Recent work by \cite{french2015} (FYZ15, hereafter), studying the molecular content of the sample presented here, has shown that, like the \cite{alatalo2016b} sample, E+As also host significant molecular reservoirs, implying that their molecular fuel is not completely expelled in starburst or AGN-driven outflows. Additionally, \cite{roseboom2009} and \cite{rowlands2015} found dust and molecular gas in two small samples of E+As with a variety of ages. Combined with the FYZ15 discovery of molecular reservoirs, these results suggests that E+As may retain at least a portion of their ISM past the period of star formation cessation. 

The evolution of post-starburst systems is inextricably tied to the fate of their gas and dust, which, due to inhomogeneous selection criteria and results, is at best inconclusive.  Thus, the evolutionary pathway from merger-induced starbursts to passive, gas-poor ellipticals, though theoretically compelling, remains unclear. In order to investigate directly this transition, a coherent description of the ISM properties of post-burst galaxies is required:
\begin{itemize}
\item Do they maintain a significant ISM mass and, if so, from where does it originate and what processes dominate its energetics? 
\item At what level does star formation continue, is it consistent with the density of molecular material, and what mechanisms set the level? 
\item How is dust distributed in these systems, and how does it respond to aging starlight post-burst?
\item Do their ISM properties support the view of E+As as rapidly transitioning systems?
\end{itemize}
Here we attempt to illuminate these questions with a detailed analysis of the broad-ranging infrared properties found in the first results of a survey of 33 E+A PSBs. The combined photometric and spectroscopic view from \textit{Spitzer}, WISE, and \textit{Herschel} constrains many of the physical conditions of the ISM in these unique objects. We present the sample in \S\,\ref{sec:sample-selection}. 

\section{Sample Selection}
\label{sec:sample-selection}
The parent sample of this survey was drawn from the Sloan Digital Sky Survey (SDSS) Data Release 5 (DR5). Sources were required to possess H$\alpha$\ equivalent widths (EQWs) $<3\,\mbox{\AA}$\ (vs. typically $\gg$10\,$\textrm{\AA}$\ in normal star-forming galaxies) and Lick H$\delta$~absorption indices $> 4\,\mbox{\AA}$, defining a parent sample of 1122 galaxies. Sources were fitted using \cite{bruzual&charlot2003} stellar population synthesis models, to infer burst properties and post-burst ages. The majority of sources have existing GALEX NUV and FUV wavelengths coverage, the addition of which significantly alleviates age-reddening degeneracies. French et al. (2017, submitted) details the UV-optical spectrophotometric fitting methodology which provides, among other physical parameters, reliable post-starburst ages over the range 100-1500 Myr, with typical uncertainties of $\sim$20\%. 

Of the SDSS parent sample, five galaxies had serendipitous overlap in existing \emph{Spitzer} continuum survey fields. Though these five sources were too faint to observe with IRS, the remarkable similarity in the shape of their 0.5--4\um\ spectral energy distributions (SEDs) permitted an extrapolation of the optical/NIR (2MASS) photometry, via template fitting, to the full E+A sub-sample. Sources with extrapolated 8\um\ and 24\um\ flux densities $>$4 mJy were considered for \textit{Spitzer} followup, resulting in a sub-sample of 26 galaxies. An additional equivalent width cut of EW([O\,\textsc{iii}], 5007$\mbox{\AA}$)~$< 1\mbox{\AA}$, was imposed to eliminate sources with strong AGN activity --- leaving 15 galaxies selected for \textit{Spitzer} photometric and spectroscopic follow-up (Program 40757, PI J.D. Smith).

The rest of the present sample (unobserved with \textit{Spitzer}) were drawn from Wide-Field Infrared Survey Explorer (WISE) detections, in which most of the SDSS E+A sample were detected. We selected 18 additional galaxies from WISE, by imposing a flux cut at W4/22\um\ of 5 mJy, for a total sample of 33 galaxies. These 33 sources were all targeted for photometric and spectroscopic follow-up with \textit{Herschel} (Program OT2\_jsmith01\_2, PI J.D. Smith). These same sources were also followed up with ground-CO imaging by FYZ15.  

As required due to sensitivity constraints, the WISE infrared-based selection undoubtedly resulted in selection of E+As which were typically more infrared-bright. Indeed, the distribution of optical-to-infrared luminosity ratios is slightly skewed toward lower values for the WISE-selected sample (though with a broad distribution), compared to the original sample of 15. However, the sample as a whole displays optical-to-infrared ratios consistent with normal star-forming galaxies (see \S\,\ref{sec:halpha-nuv}).

A list of relevant sample parameters and luminosities is given in Table \ref{tab:sample}. The SDSS spectra for the sample can be found in Figure \ref{fig:sdss_spec} of the \textsc{appendix}.

\section{Observations, Reduction, and Modeling}
\label{sec:analys-reduc}

\subsection{Photometry}
All IRAC and MIPS photometric data were reduced as described in \cite{dale2005}. All PACS and SPIRE photometric data were reduced as described in \cite{dale2012}. The available WISE photometry was calibrated and reduced following the method presented in \cite{jarrett2013}. \textit{Spitzer} photometry from IRAC and MIPS (3-160\micron) is available for 15/33 objects. WISE photometry is available for the full sample in all passbands. \textit{Herschel} PACS photometry is also available for the full sample, but SPIRE data were only obtained for 15 sources. A more detailed description on the reduction for each dataset can be found below. The photometry is presented in its entirety in Tables \ref{tab:spitzer-phot}, \ref{tab:herschel-phot}, and \ref{tab:wise-phot} in the \textsc{appendix}.

\subsubsection{Spitzer and Herschel}
\label{sec:sp-her_reduction}
The photometric apertures were chosen by eye to encompass essentially all of the emission at every wavelength; the aperture corrections for the \textit{Spitzer}/IRAC photometry follow those described in \cite{dale2009}. The sky was subtracted in the aperture photometry process. This subtraction was accomplished via a set of sky apertures that collectively circumscribe each galaxy, projected on the sky close enough to the galaxy to measure the ``local'' sky but far enough away to avoid containing any galaxy emission. 

Uncertainties in the integrated photometry, $\epsilon_{\rm total}$, were formulated by quadrature sum of the calibration uncertainty, $\epsilon_{\rm cal}$, and the measurement uncertainty, $\epsilon_{\rm sky}$, based on the measured sky fluctuations, and the areas covered by the galaxy and the sum of the sky apertures, i.e., 

\floattable
\begin{deluxetable*}{rccccccccccccc}
\rotate
\tablecaption{Sample Parameters\label{tab:sample}}
\tablecolumns{14}
\tabletypesize{\scriptsize}
\tablehead{%
\colhead{Galaxy} &
\colhead{Alt. ID} &
\colhead{} &
\colhead{} &
\colhead{} &
\colhead{} &
\colhead{D$_{\mathrm{L}}$} &
\colhead{$\mathrm{\log(M_{*})}$} &
\colhead{$\mathrm{R_{90}}$} &
\colhead{$\mathrm{\log(L_{H\alpha})}$} &
\colhead{$\mathrm{SFR_{H\alpha}}$} &
\colhead{$\mathrm{\log(L_{NUV})}$} &
\colhead{$\mathrm{\log(L_{B})}$} &
\colhead{$\mathrm{\log(L^{\prime}_{CO})}$} \\
%%%
\colhead{(SDSS)} &
\colhead{(FYZ15)} &
\colhead{Source} &
\colhead{R.A.} &
\colhead{Decl.} &
\colhead{z} &
\colhead{(Mpc)} &
\colhead{(M$_{\odot}$)} &
\colhead{(arcsec)} &
\colhead{(L$_{\odot}$)} &
\colhead{(M$_{\odot}$\,yr$^{-1}$)} &
\colhead{(L$_{\odot}$)} &
\colhead{(L$_{\odot}$)} &
\colhead{(K km s$^{-1}$ pc$^{2}$)} \\
%%%
\colhead{(1)} & 
\colhead{(2)} & 
\colhead{(3)} & 
\colhead{(4)} & 
\colhead{(5)} & 
\colhead{(6)} & 
\colhead{(7)} & 
\colhead{(8)} & 
\colhead{(9)} & 
\colhead{(10)} & 
\colhead{(11)} & 
\colhead{(12)} &
\colhead{(13)} &
\colhead{(14)}
}
\startdata
0336\_469\_51999&EAH17&H&12:44:51.69&$-$01:45:35.6&0.048&222.01&10.05&4.67&5.93&0.08&7.72&9.44&$<$8.41\\
0379\_579\_51789&EAS15&S&22:55:06.80&+00:58:39.9&0.053&247.20&10.83&9.70&6.48&0.68&7.79&10.22&8.48\\
0413\_238\_51929&EAS02&S&03:16:54.91&$-$00:02:31.1&0.023&104.86&10.08&11.83&5.80&0.10&6.84&9.50&8.11\\
0480\_580\_51989&EAH08&H&09:48:18.68&+02:30:04.2&0.060&280.94&10.41&4.48&5.86&0.18&7.87&9.78&8.55\\
0570\_537\_52266&EAS05&S&09:44:26.96&+04:29:56.8&0.047&215.33&10.57&8.42&6.28&0.58&7.58&9.97&8.48\\
0598\_170\_52316&EAH14&H&11:53:06.45&+64:17:56.5&0.062&289.99&10.04&3.15&6.38&0.21&8.22&9.74&$<$8.65\\
0623\_207\_52051&EAS12&S&16:13:30.19&+51:03:35.6&0.034&153.66&10.01&18.20&5.78&0.04&7.81&9.78&7.94\\
0637\_584\_52174&EAS14&S&21:05:08.67&$-$05:23:59.4&0.083&390.19&11.31&9.68&6.86&1.78&8.34&10.48&9.10\\
0656\_404\_52148&EAS01&S&00:44:59.24&$-$08:53:22.9&0.020&88.58&10.24&13.08&5.36&0.07&7.03&9.86&$<$7.79\\
0755\_042\_52235&EAH06&H&07:45:49.50&+31:22:42.2&0.044&202.74&10.53&5.52&6.47&1.08&\ldots&9.96&$<$8.40\\
0756\_424\_52577&EAS03&S&07:51:14.31&+34:25:05.5&0.063&292.66&10.86&7.46&6.48&1.18&\ldots&10.01&9.16\\
0815\_586\_52374&EAS13&S&16:27:02.56&+43:28:33.9&0.046&213.17&10.95&11.41&6.18&0.62&8.03&10.25&$<$8.52\\
0870\_208\_52325&EAH16&H&09:26:57.69&+42:31:36.6&0.111&536.37&10.74&4.52&6.95&0.81&8.37&10.45&$<$9.19\\
0951\_128\_52398&EAS07&S&11:19:07.62&+58:03:14.3&0.033&148.45&10.54&7.23&5.90&0.12&7.50&10.13&$<$8.03\\
0962\_212\_52620&EAS06&S&10:37:57.36&+46:14:40.3&0.023&103.02&10.14&7.10&5.52&0.08&7.25&9.24&8.63\\
0986\_468\_52443&EAH04&H&21:14:00.54&+00:32:06.4&0.027&122.39&10.18&8.98&6.15&0.27&7.12&9.84&7.96\\
1001\_048\_52670&EAH15&H&10:52:20.45&+05:49:41.6&0.041&188.94&10.40&4.72&6.02&0.34&7.56&9.96&$<$8.44\\
1003\_087\_52641&EAH11&H&11:05:40.71&+05:59:54.3&0.054&251.09&10.61&4.67&6.43&0.32&8.00&10.01&$<$8.59\\
1039\_042\_52707&EAS10&S&13:05:25.83&+53:35:30.3&0.038&174.40&10.53&6.44&5.52&0.05&7.76&10.20&$<$8.19\\
1170\_189\_52756&EAS11&S&16:10:20.49&+41:51:17.6&0.040&181.41&10.74&10.55&6.13&0.51&7.80&10.37&$<$8.24\\
1279\_362\_52736&EAS09&S&12:46:26.84&+50:47:31.4&0.027&122.73&10.56&10.68&5.90&0.17&7.34&9.94&8.52\\
1352\_610\_52819&EAH09&H&15:08:55.09&+37:33:29.8&0.029&132.32&10.21&5.34&6.10&0.28&7.39&9.65&7.89\\
1604\_161\_53078&EAH07&H&11:11:17.96&+11:33:15.8&0.038&174.24&10.65&12.50&6.59&2.17&8.08&10.38&$<$8.01\\
1616\_071\_53169&EAS08&S&12:39:36.05&+12:26:20.0&0.041&187.12&10.67&13.55&5.99&0.34&7.59&10.14&$<$8.00\\
1853\_070\_53566&EAH18&H&16:21:00.81&+21:10:06.1&0.031&140.93&10.64*&\ldots&\ldots&\ldots&\ldots&9.96&\ldots\\
1927\_584\_53321&EAS04&S&08:27:01.40&+21:42:24.4&0.015&68.73&9.99&8.84&5.59&0.63&6.94&9.61&$<$7.14\\
2001\_473\_53493&EAH05&H&12:17:02.43&+39:04:37.3&0.065&305.06&10.00&3.44&6.04&0.23&8.11&9.58&8.96\\
2276\_444\_53712&EAH01&H&08:34:33.71&+17:20:46.3&0.048&220.73&10.45&9.35&6.01&0.38&7.58&9.97&9.11\\
2360\_167\_53728&EAH02&H&09:26:19.29&+18:40:41.0&0.054&250.84&9.96&5.26&5.65&0.09&\ldots&9.05&8.93\\
2365\_624\_53739&EAH13&H&10:22:00.79&+22:09:47.4&0.113&544.68&11.00&4.61&7.04&1.53&8.65&10.65&9.29\\
2376\_454\_53770&EAH10&H&10:33:42.71&+21:07:40.8&0.105&505.46&10.24&3.44&5.87&0.04&7.88&10.04&9.26\\
2750\_018\_54242&EAH12&H&14:55:05.45&+13:16:51.6&0.083&390.55&10.55&2.83&6.51&0.29&7.81&9.88&$<$8.82\\
2777\_258\_54554&EAH03&H&14:48:16.05&+17:33:05.9&0.045&206.72&10.34&7.06&5.54&0.12&7.76&9.49&9.20\\
\enddata
\tablecomments{(1) SDSS Plate\_Fiber\_MJD notation. 
(2) FYZ15 EA designation.
(3) Primary telescope(s) used for targeting; S = \textit{Spitzer} + \textit{Herschel}, H = \textit{Herschel}-only. All objects have been observed by \textit{Herschel}.
(4)-(5) Right Ascension and Declination. 
(6) SDSS spectroscopic redshift. 
(7) Redshift-derived luminosity distance, assuming \cite{planck-cosmo} cosmology. 
(8) Stellar mass presented in \cite{french2015}, derived from the MPA-JHU emission line analysis of the SDSS DR7 data products \citep{kauffmann2003a}. *Stellar mass calculated using the GAMA G12/G15 WISE date products \citep{cluver2014}. 
(9) Radius encompassing 90\% of the SDSS Petrosian flux in $r$-band. 
(10) H$\alpha$\ luminosity derived from the MPA-JHU emission line analysis \citep{aihara2011} and corrected for foreground extinction using the updated extinction maps of \cite{schlegel1998}. 
(11) Aperture-corrected, H$\alpha$-based SFR from \cite{french2015}, corrected for in-situ extinction using the H$\alpha$/H$\beta$\ Balmer decrement.
(12) NUV luminosity from the GALEX 2274$\mbox{\AA}$~filter. 
(13) $B$-band luminosity, converted from SDSS photometry (see \S\,\ref{sec:halpha-nuv}). 
(14) CO(1--0) integrated line luminosities reported in \cite{french2015}.}
\end{deluxetable*}

\clearpage

\begin{equation}
\epsilon_{\rm total} = \sqrt{\epsilon_{\rm cal}^2 + \epsilon_{\rm sky}^2}
\end{equation}
with 
\begin{equation}
\epsilon_{\rm sky} ~ = \sigma_{\rm sky} \Omega_{\rm pix} ~ \sqrt{N_{\rm pix}+{N_{\rm pix}}^2/N_{\rm sky}}, 
\end{equation}

\noindent where $\sigma_{\rm sky}$ is the standard deviation of the sky surface brightness fluctuations, $\Omega_{\rm pix}$ is the solid angle subtended per pixel, and $N_{\rm pix}$ and $N_{\rm sky}$ are the number of pixels in the galaxy and (the sum of) the sky apertures, respectively. For non-detections, 5$\sigma$ upper limits are derived from the sky background, assuming a galaxy spans all $N_{\rm pix}$ pixels in the aperture, i.e., 
\begin{equation}
f_\nu(5\sigma~{\rm upper~limit}) ~ = ~ 5 ~ \epsilon_{\rm sky}.
\end{equation}

\subsection{IRS Spectra}
\label{sec:irs-reduction}
\textit{Spitzer}/IRS spectral maps were obtained in small 3$\times$1 half-slit width stepped Short-Low (SL, 5.2--14.5\um) maps and single width Long-Low (LL, 14.5--38.5\um) maps. All IRS data were reduced with the IRS spectral cube analysis tool, CUBISM \citep{cubism_smith2007}. The four spectral segments were stitched together using data from the four SL and LL sub-modules: SL2 (5.25-7.6\um), SL1 (7.5-14.5\um), LL2 (14.5-20.75\um), and LL1 (20.5-38.5\um). In the overlapping regions between sub-modules (i.e. SL2-SL1 and LL2-LL1), scaling factors were obtained by first trimming the excess pixels near the overlapping regions, then stitching them together using low-order polynomial fits from either side. The combined SL and LL spectra were initially offset, which we attributed to varying slit losses. Indeed, the color-corrected synthetic photometry computed from the SL and LL spectral segments differed significantly from the observed IRAC 8\um\ and MIPS 24\um\ values. The ratio between synthetic photometry and true photometry were, on average, $\sim$55\% for stitched SL/IRAC and $\sim$45\% for stitched LL/MIPS.  SL and LL spectra were separately scaled to match the observed IRAC/MIPS fluxes, which eliminated the spectral offset without further scaling between spectral segments. The respective scaling factors can be found in Table \ref{tab:spitzer-phot} in the \textsc{Appendix}. 

Once scaled, the SL and LL segments were blended, again using low-order polynomial fitting.  After the respective SL/8\um\ and LL/24\um\ scaling, final spectra are then directly tied to the global galaxy photometry, significantly easing comparisons with other global results. The spectral uncertainties from the IRS pipeline are adopted, with additional uncertainty propagated from the spectral decomposition. The decomposition method is described in greater detail in \S\,\ref{sec:specdecomp}.

The spectra were initially decomposed using PAHFIT \citep[][SDD07 hereafter; see \S\,\ref{sec:specdecomp}]{smith2007}, with only the pipeline uncertainties. The resulting fits were then subtracted from a featureless section of the continuum and the resultant residuals were added in quadrature with the spectral uncertainties before running the final PAHFIT decomposition.

\subsection{PACS Spectra}
\label{sec:pacs}
Spectra of the \cII\,158\um\ and \oI\,63\um\ lines were obtained for each source. PACS spectral observations were obtained in the Un-Chopped mapping mode and reduced using the Herschel Interactive Processing Environment (HIPE) version 12.0.2765 \citep{hipe}. The reductions applied the standard spectral response functions, a custom flat field correction, and flagged instrument artifacts and bad pixels \citep[see][]{pacs,kingfish}. The residual dark signal not removed by chopping was determined from each individual observation, and subtracted during processing. \textit{Herschel}'s baseline exhibits significant baseline drifts and distinctive instrumental transients are common occurrences. These instabilities result in a variable non-astrophysical continuum, which is dominated by emission from \textit{Herschel} itself. 

Transient signals are strongly correlated with motions of the PACS grating and of {\it Herschel}. Using fits of the \citet{draine2007} dust model to spectral energy distributions of galaxies in the KINGFISH sample we estimate the expected astrophysical continuum is less that 2\% of the spectral line flux detected at the line positions. Thus, the continuum adjacent to the expected location of the lines can be approximated as constant and is used to correct for transients. 

The averages of the clean off-observations obtained were subtracted from observations to correct for the thermal background contributed by {\it Herschel}. Subsequently, all spectra within a given spatial element were combined. Final spectral cubes with 2\farcs06 spatial pixels were created by combining individual pointings using the Drizzle algorithm implemented in HIPE. In-flight flux calibrations\footnote{Calibration Version 65} were applied to the data. These calibrations resulted in absolute flux uncertainties of 15\% with relative flux uncertainties between each {\it Herschel} pointing of $\sim$10\%. 

\subsection{SED Modeling}
\label{sec:sed}
The spectral energy distributions are fit with the models of \cite{draineli2007} (DL07) as updated by \cite{aniano2012}. Following \cite{dale2001}, \cite{draineli2007} model interstellar dust heating with a $\delta$-function in interstellar radiation field (ISRF) intensity $U$, coupled with a power-law distribution $U_{\rm min} < U < U_{\rm max}$,
\begin{equation}
\begin{split}
dM{}&_{\rm dust}/dU = M_{\rm dust} \times \\
& \left[ (1-\gamma) \delta(U-U_{\rm min}) + \gamma \frac{\alpha-1}{U_{\rm min}^{1-\alpha} - U_{\rm max}^{1-\alpha}} U^{-\alpha} \right],
\end{split}
\end{equation}
where $U$\ is normalized to the local Galactic ISRF, $dM_{\rm dust}$\ is the differential dust mass heated by a range of starlight intensities $[U,U+dU]$, $M_{\rm dust}$\ is the total dust mass, and $(1-\gamma)$\ is the fraction of the dust heated by the diffuse ISRF defined by $U=U_{\rm min}$. The minimum and maximum ISRF intensities span $0.01 < U_{\rm min} < 30$\ and $3 < \log U_{\rm max} < 8$. 

%\afterpage{\clearpage}

A sum of three different spectral energy distributions is fit to each galaxy: a blackbody of temperature $T_*=5000$\,K --- appropriate for the stellar contribution to the IR at $\lambda > 3$\um\ \citep[see][]{smith2007,draine2007} --- along with two related dust components. Following \cite{draine2007}, the sum can be expressed as
\begin{equation}
\begin{split}
f{}&_{\nu}^{\rm model} = \Omega_* B_\nu(T_*) + \frac{M_{\rm dust}}{4 \pi D^2} \times \\
& \left[ (1-\gamma) p_\nu^{(0)}(q_{\rm PAH},U_{\rm min}) + \gamma p_\nu(q_{\rm PAH},U_{\rm min},U_{\rm max},\alpha) \vphantom{p_\nu^{(0)}} \right],
\end{split}
\end{equation}
where $\Omega_*$ is the solid angle subtended by stellar photospheres, $D$ is the distance to the galaxy, and $\gamma$ and $(1-\gamma)$ are the fractions of the dust mass heated by the ``power-law'' and ``delta-function'' starlight distributions, respectively. $p_\nu^{(0)}(q_{\rm PAH},U_{\rm min})$ and $p_\nu(q_{\rm PAH},U_{\rm min},U_{\rm max},\alpha)$ are, respectively, the emitted power per unit frequency per unit dust mass for dust heated by a single starlight intensity $U_{\rm min}$, and the same for dust heated by a power-law distribution of starlight intensities $dM/dU \propto U^{-\alpha}$ extending from $U_{\rm min}$ to $U_{\rm max}$. Finally, the fractional contribution to total dust mass from PAHs, denoted as $q_{\rm PAH}$, varies in the model suite between 0\% and 12\% with a grid spacing of 0.1\% in $q_{\rm PAH}$. 

We adopt the choice of \cite{draine2007} to fix $U_{\rm max}=10^6$ and $\alpha=2$ to minimize the number of free parameters. We use a minimum value of 0.01 for $U_{\rm min}$. The remaining free parameters $\Omega_*$, $M_{\rm dust}$, $q_{\rm PAH}$, $U_{\rm min}$, and $\gamma$ are found via $\chi^2$ minimization:
\begin{equation}
\chi^2 = \sum_b \frac{(f_{\nu,b}^{\rm obs} - f_{\nu,b}^{\rm model})^2}{(\sigma_{b}^{\rm obs})^2 + (\sigma_{b}^{\rm model})^2}, 
\end{equation}
where $f_{\nu,b}^{\rm model}$ is the model flux density obtained after convolving the model with the $b^\textrm{th}$ filter bandpass, $\sigma^{\rm obs}_{b}$ is the uncertainty in the observed flux density, and $\sigma^{\rm model}_{b}$ is set to $0.1 f^{\rm model}_{\nu,b}$ to allow for the uncertainty intrinsic to the model.

To derive the TIR luminosities quoted in this paper, we subtract the stellar component of the fit, to isolate the dust emission and numerically integrate the result from 3--1100\um. The starlight-subtracted full SED fitting method agrees well with the three-band (24\um, 70\um, 160\um) prescription of \cite{dalehelou2002}, to within $< 15\%$. Failing to remove the starlight component can result in substantial discrepancies between the two estimates. The average starlight fraction of the \textit{full} TIR luminosity (stellar blackbody not subtracted) is 40\%, with a sample dispersion of $\pm$23\% --- typical for nearby star-forming galaxies, but considerably lower than for most early-type galaxies, which are stellar-dominated.

Though the radiation field present in E+As likely differs from star-forming galaxies (see \S\,\ref{sec:discuss} for an in-depth discussion), quantities such as dust mass and TIR luminosity, output by the DL07 models, are relatively unaffected by these differences. \cite{draine2014} found that in M31's bulge, dust mass was unaffected beyond a $\sim$50\% level when considering a MW-type radiation field vs. a more appropriate, old star-dominated template. As the radiation field in E+As will likely be intermediate between these two extremes, changes to the assumed radiation field are not a dominant source of uncertainty. Rather, the emissivity of large grains is the dominant uncertainty in the dust mass. It should be noted that recent results \citep[e.g., the PHAT survey][]{dalcanton2012} indicate that the DL07 dust opacities may be higher than extinction-derived measures by a factor of $\sim$2, leading to a consistent overestimation of dust mass. Despite this, we use here the original models, which have been consistently adopted for \textit{all other} extragalactic samples, such as SINGS/KINGFISH (see \S\,\ref{sec:comparison_samples}). 

\subsection{Comparison Samples}
\label{sec:comparison_samples}
This work makes frequent comparisons between the E+A sample and other infrared-focused samples of galaxies, in an effort to provide useful context. In this section, we briefly describe the three most frequently referenced samples and point the reader to relevant references for further reading. 

\begin{figure*}[t]
\centering
\leavevmode
\includegraphics[width={0.95\linewidth}]{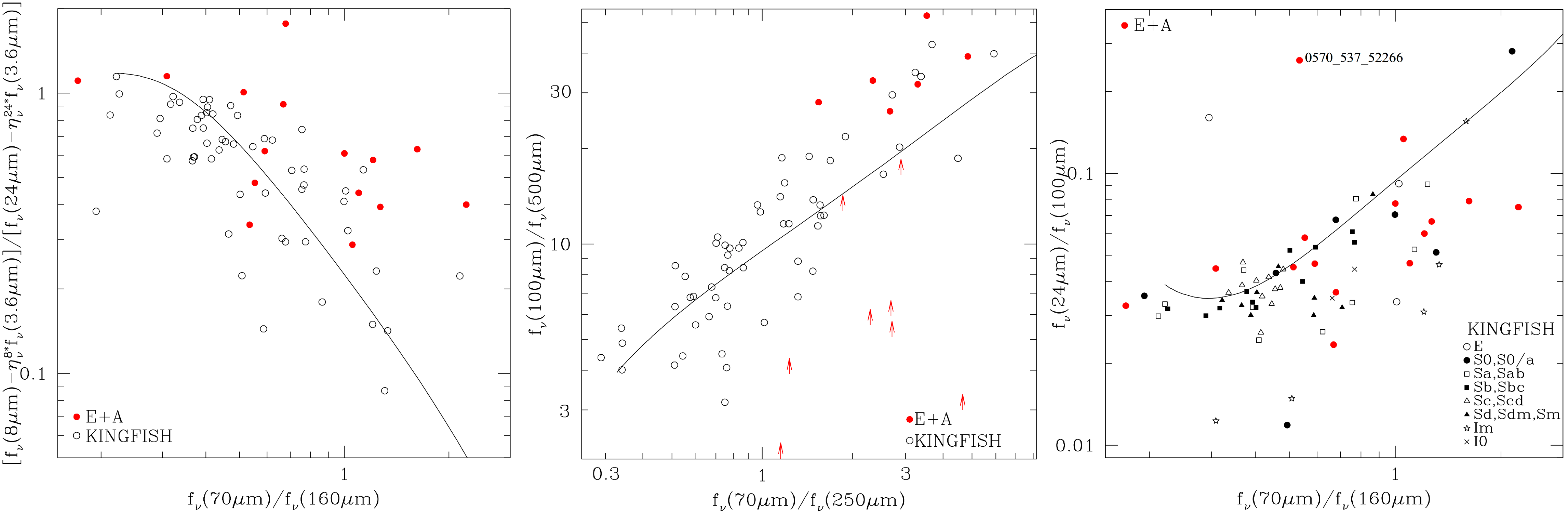}
\caption{Far-infrared/sub-millimeter color--color diagrams. In the leftmost panel, the stellar contribution to the 8\um\ and 24\um\ photometry has been removed. The solid line indicates the sequence of model spectral energy distributions for normal star-forming galaxies from \cite{dale2014}. Upward arrows denote sources with 5$\sigma$~upper limits on the 500\um\ photometry.}
\label{fig:fir_col-col}
\end{figure*}

The \textit{Spitzer} Infrared Nearby Galaxies Survey (SINGS) surveyed 75 nearby, primarily star-forming galaxies (see \citealt{kennicutt2003} for survey introduction). The survey targeted galaxies with a wide range of stellar masses, including examples of both the most massive early-type spiral galaxies (e.g., M81) and lower-mass, low-metallicity dwarf galaxies (e.g., Holmberg II). The sample also spans the full range of nuclear classifications, including Seyfert, LINER, and H\,II-dominated. See SDD07 for an in-depth analysis of the low-resolution mid-infrared spectral properties of the SINGS sample, and \cite{draine2007} for analysis of the infrared SEDs and derivation of dust masses and PAH abundances. 

The KINGFISH (Key Insights on Nearby Galaxies: A Far-Infrared Survey with Herschel) survey is directly descendant from the SINGS survey (see \citealt{kennicutt2011} for survey introduction), targeting a 61-galaxy subset of the original sample. \textit{Herschel} allowed, for the first time, mapping of the cold dust and far-infrared cooling lines in these galaxies. See \cite{dale2012,dale2017} for a description of the far-IR/sub-mm photometry for the sample, and \cite{smith2017} for an analysis of the \cII\ cooling-line deficit in KINGFISH sources. 

The Great Observatories All-sky LIRG Survey (GOALS) is a flux-limited survey of over 200 luminous and ultra-luminous infrared galaxies --- those with L$_{\mathrm{IR}} \geqslant 10^{11}$\,L$_{\odot}$\ (LIRG) and 10$^{12}$\,L$_{\odot}$\ (ULIRG), respectively --- at z $<$\ 0.09, identified by the infrared astronomical satellite (IRAS). GOALS is a multiwavelength survey, combining data from \textit{Chandra}, GALEX, HST, and \textit{Spitzer} (see \citealt{armus2009} for survey introduction). Like SINGS/KINGFISH, GOALS targeted galaxies spanning the full range of optically-classified nuclear types (Seyferts, LINERs, and H\,II). Focusing on the infrared properties, \cite{inami2013} provides details of the IRS spectra and fine-structure emission-line fluxes for the sample, while \cite{stierwalt2014} details the physics of the dust and gas, including properties of PAH and H$_{2}$\ emission, and dust composition. 

No single, dedicated survey has produced a similarly complete analysis of the infrared properties of early-type galaxies (ETGs). Of the SINGS sample, there are seven galaxies of either S0 or elliptical type. Additionally, several \textit{Spitzer} archival studies have yielded relatively large samples of ETGs, with various photometric and spectroscopic coverage. \cite{temi2009} composed an archival atlas of 225 nearby ETGs (E, E-S0, and S0; within 70 Mpc) which were covered in at least one of the three MIPS passbands: 24\um, 70\um, and 160\um. Likewise, \cite{rampazzo2013} composed an archival atlas of 91 ETGs with full \textit{Spitzer} IRS 5--38\um\ nuclear spectra. Of the 91, 58 possess IRS spectra which deviate from a purely passively evolving early-type template. Of these, 43 exhibit PAH emission (required for later comparison). And of these 43 sources, we find 16 with cross-matches in \cite{temi2009} with detections all three MIPS bands. TIR luminosities are calculated using the three-filter prescription of \cite{dalehelou2002}.

This subset of early type galaxies is more infrared-bright than typical, passive ETGs. Of galaxies in the \cite{temi2009} sample, the sources detected in all three MIPS bands possess a geometric mean L$_{\TIR}$/L$_{B} = 0.1$\ --- $>$10$\times$\ higher than the L$_{\TIR}$/L$_{B} \leqslant 0.01$\ for sources with limits on 70\um\ and/or 160\um. For reference, only 40\% of the \cite{temi2009} parent sample of ETGs possessed detections in all three MIPS bands --- reflecting the passive, dust-poor nature of the majority of ETGs. Most of these IR-bright ETGs are thought to be the remnants of minor mergers or interactions \citep{davis2015}. Supplementing these 16 IR-bright ETGs with the seven ETGs from SINGS, we thus compose a sample of 23 dusty ETGs for comparison against our E+A sample. 

In addition to our composite ETG comparison sample, we make use of the ATLAS$^{\rm 3D}$\ sample --- a multiwavelength, volume-limited survey of 260 ETGs out to 42 Mpc \citep{cappellari2011}. The molecular gas content of the ATLAS$^{\rm 3D}$\ sample is described in \cite{young2011}, while a recent paper by \cite{lapham2017} details the FIR cooling-line properties of a small subset of the sample.

\section{Results}
\label{sec:results}

In this section, we first present global results from the infrared photometry for the sample --- such as dust masses and abundances, multiwavelength comparisons and extinction analysis, and spatial distribution. We then discuss results from the IRS and PACS spectroscopy, including PAH, nebular line, and H$_{2}$\ emission, ISM energetics, and cooling-line deficit. We then discuss time evolution of ISM properties, followed by an analysis of potential AGN activity. We conclude with an analysis of the star formation properties of the sample, as well as comparisons to galaxy star forming relations. In the following section (\S\,\ref{sec:discuss}), we discuss the significance and implications of these results regarding the prior and future evolution of these galaxies.

\begin{deluxetable}{lppppp}
\tablecaption{DL07 SED Fit Parameters\label{tab:dust}}
\tablecolumns{6}
\tabletypesize{\scriptsize}
\tablehead{%
\colhead{Galaxy} &
\colhead{$\mathrm{\log M_{Dust}}$} &
\colhead{$q_{\PAH}$} &
\colhead{} &
\colhead{$\gamma$} &
\colhead{$\mathrm{\log L_{\TIR}}$} \\
%%%
\colhead{(SDSS)} &
\colhead{(M$_{\odot}$)} &
\colhead{(\%)} &
\colhead{$U_{\mathrm{min}}$} &
\colhead{(\%)} &
\colhead{(L$_{\odot}$)} \\
%%%
\colhead{(1)} & 
\colhead{(2)} & 
\colhead{(3)} & 
\colhead{(4)} & 
\colhead{(5)} & 
\colhead{(6)} \\
}
\startdata
0336\_469\_51999&$6.74 \pm 0.14$&4.5&2.5&3.26&$9.51 \pm 0.013$\\
0379\_579\_51789&$7.13 \pm 0.17$&1.8&2.5&1.99&$9.74 \pm 0.007$\\
0413\_238\_51929&$6.26 \pm 0.09$&4.7&8.0&6.11&$9.58 \pm 0.005$\\
0480\_580\_51989&$8.09 \pm 0.27$&0.4&10.0&0.00&$11.18 \pm 0.011$\\
0570\_537\_52266&$7.00 \pm 0.16$&5.5&1.2&14.90&$9.70 \pm 0.008$\\
0598\_170\_52316&$6.08 \pm 1.29$&4.7&5.0&19.14&$9.69 \pm 0.657$\\
0623\_207\_52051&$7.32 \pm 0.20$&5.5&0.7&0.37&$9.39 \pm 0.008$\\
0637\_584\_52174&$7.57 \pm 0.21$&6.5&3.0&0.81&$10.24 \pm 0.007$\\
0656\_404\_52148&$5.42 \pm 0.01$&4.8&15.0&1.63&$8.91 \pm 0.006$\\
0755\_042\_52235&$6.80 \pm 0.13$&7.5&5.0&5.34&$9.96 \pm 0.007$\\
0756\_424\_52577&$7.47 \pm 0.20$&3.6&3.0&1.64&$10.13 \pm 0.007$\\
0815\_586\_52374&$7.41 \pm 0.20$&8.5&1.2&0.48&$9.72 \pm 0.004$\\
0870\_208\_52325&$7.02 \pm 0.73$&0.8&1.5&12.37&$9.98 \pm 0.140$\\
0951\_128\_52398&$5.29 \pm 0.05$&5.0&15.0&1.95&$8.77 \pm 0.005$\\
0962\_212\_52620&$6.67 \pm 0.12$&2.7&20.0&1.37&$10.12 \pm 0.005$\\
0986\_468\_52443&$6.35 \pm 0.09$&4.7&15.0&1.82&$9.88 \pm 0.006$\\
1001\_048\_52670&$6.50 \pm 0.11$&2.9&4.0&3.33&$9.41 \pm 0.008$\\
1003\_087\_52641&$6.39 \pm 0.10$&6.9&15.0&3.46&$9.96 \pm 0.010$\\
1039\_042\_52707&$5.59 \pm 0.02$&2.4&30.0&2.23&$9.32 \pm 0.006$\\
1170\_189\_52756&$6.65 \pm 0.12$&1.9&12.0&0.89&$9.87 \pm 0.005$\\
1279\_362\_52736&$6.94 \pm 0.14$&4.6&6.0&0.08&$9.83 \pm 0.006$\\
1352\_610\_52819&$6.53 \pm 0.10$&6.5&7.0&2.40&$9.68 \pm 0.007$\\
1604\_161\_53078&$6.70 \pm 0.13$&7.3&2.0&10.36&$9.71 \pm 0.007$\\
1616\_071\_53169&$6.38 \pm 0.10$&1.8&4.0&0.66&$9.07 \pm 0.008$\\
1853\_070\_53566&$6.02 \pm 0.06$&7.3&10.0&0.80&$9.28 \pm 0.008$\\
1927\_584\_53321&$5.80 \pm 0.04$&5.3&10.0&1.94&$8.96 \pm 0.005$\\
2001\_473\_53493&$7.56 \pm 0.21$&0.3&25.0&0.00&$11.08 \pm 0.012$\\
2276\_444\_53712&$7.70 \pm 0.23$&2.7&10.0&0.38&$10.84 \pm 0.015$\\
2360\_167\_53728&$6.86 \pm 0.90$&0.0&30.0&1.97&$10.51 \pm 0.090$\\
2365\_624\_53739&$7.93 \pm 0.24$&6.2&4.0&0.95&$10.72 \pm 0.007$\\
2376\_454\_53770&$7.93 \pm 0.25$&2.1&6.0&1.18&$10.89 \pm 0.007$\\
2750\_018\_54242&$6.30 \pm 0.10$&4.4&30.0&5.77&$10.23 \pm 0.011$\\
2777\_258\_54554&$8.70 \pm 0.33$&0.6&15.0&0.00&$10.86 \pm 0.015$\\
\enddata
\tablecomments{(1) Galaxy ID using SDSS notation. \\
(2) Derived dust mass. \\
(3) Derived PAH mass abundance. \\
(4) Lower cutoff for starlight intensity scale factor $U$. \\
(5) Fraction of dust mass in regions with $U > U_{\mathrm{min}}$. \\
(6) Total integrated 3--1100\um\ luminosity.}
\end{deluxetable}

\subsection{Infrared SEDs}
\label{sec:seds}
The infrared spectral energy distributions (SEDs) of galaxies are dominated by stellar emission in the near-infrared (NIR; $\sim$0.8--5\um), and reprocessed emission from interstellar dust grains in the mid- and far-infrared (MIR, FIR; $\sim$3--1100\um). 

In Figure~\ref{fig:fir_col-col} we show starlight-subtracted infrared color-color diagrams for the E+A sample, compared to the KINGFISH sample of nearby star-forming galaxies \citep{dale2014}. In general, the E+As display infrared colors consistent with the (wide) range of KINGFISH galaxies. Many of the E+As have bluer ($\sim$5$\times$) 8\um/24\um\ color than KINGFISH sources at a similar 70\um/160\um\ color, indicating dominant PAH emission in the MIR. Additionally, $\sim$50\% of the \textit{Spitzer} sub-sample are non-detections at 500\um\ and the rest have 100\um/500\um\ $>$\ 20, both indicative of warm dust peaks.

\begin{figure*}[t]
\centering
\leavevmode
\includegraphics[width={0.98\linewidth}]{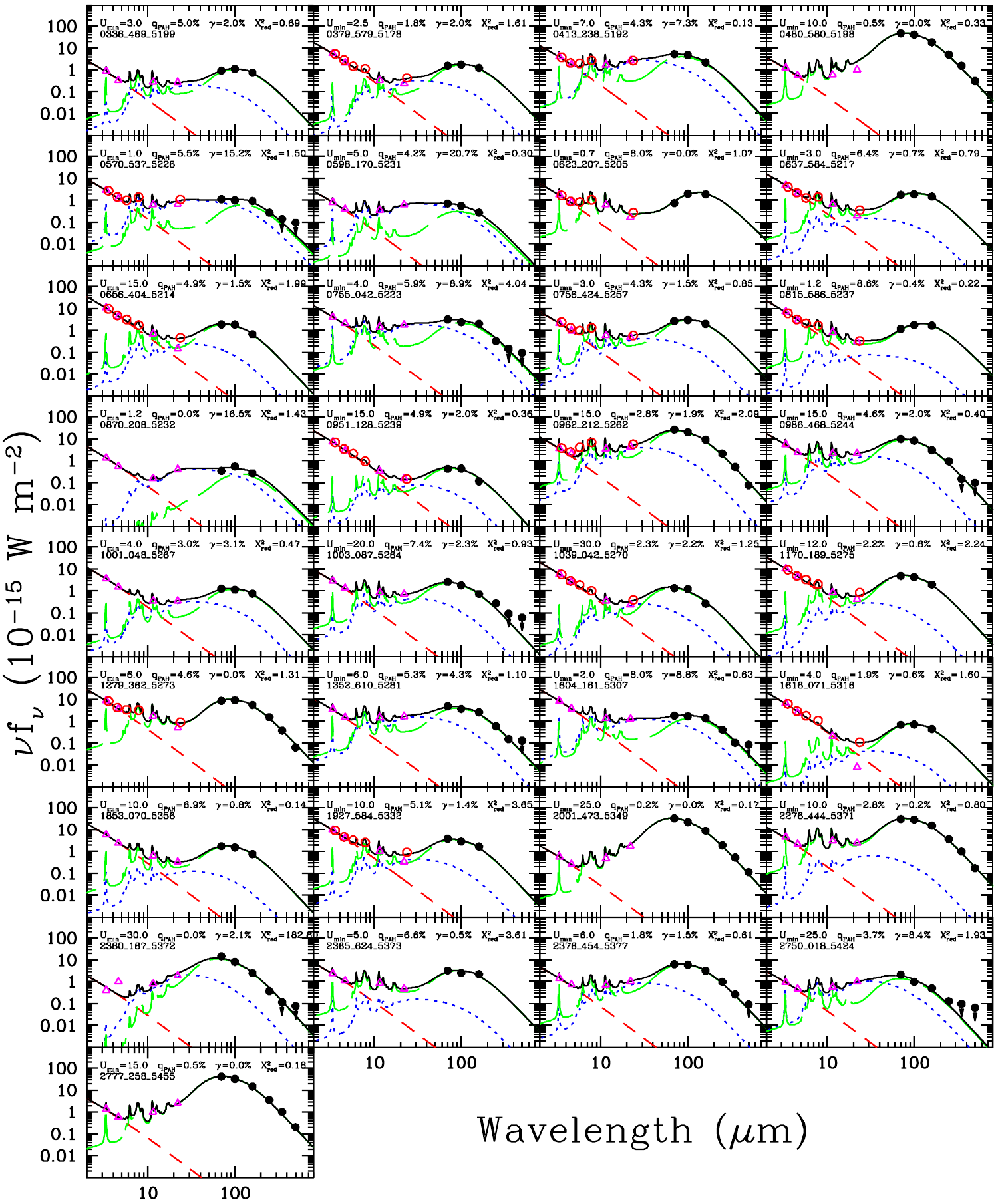}
\caption{Globally-integrated infrared/sub-millimeter spectral energy distributions for the sample, sorted by SDSS plate ID. The following symbols are utilized: filled black circles (\textit{Herschel}), open magenta triangles (WISE), and open red circles (\textit{Spitzer}). Arrows indicate 5$\sigma$ upper limits. The solid curve is the sum of a 5000\,K stellar blackbody (red dashed) along with \cite{draineli2007} models of dust emission from PDRs (blue dotted; $U>U_{\rm min}$) and the diffuse interstellar medium (green long-dashed; $U=U_{\rm min}$). The fitted parameters from these model fits are listed within each panel along with the reduced $\chi^2$.
}
\label{fig:sed}
\end{figure*}

\begin{figure}[t]
\centering
\leavevmode
\includegraphics[width={0.95\linewidth}]{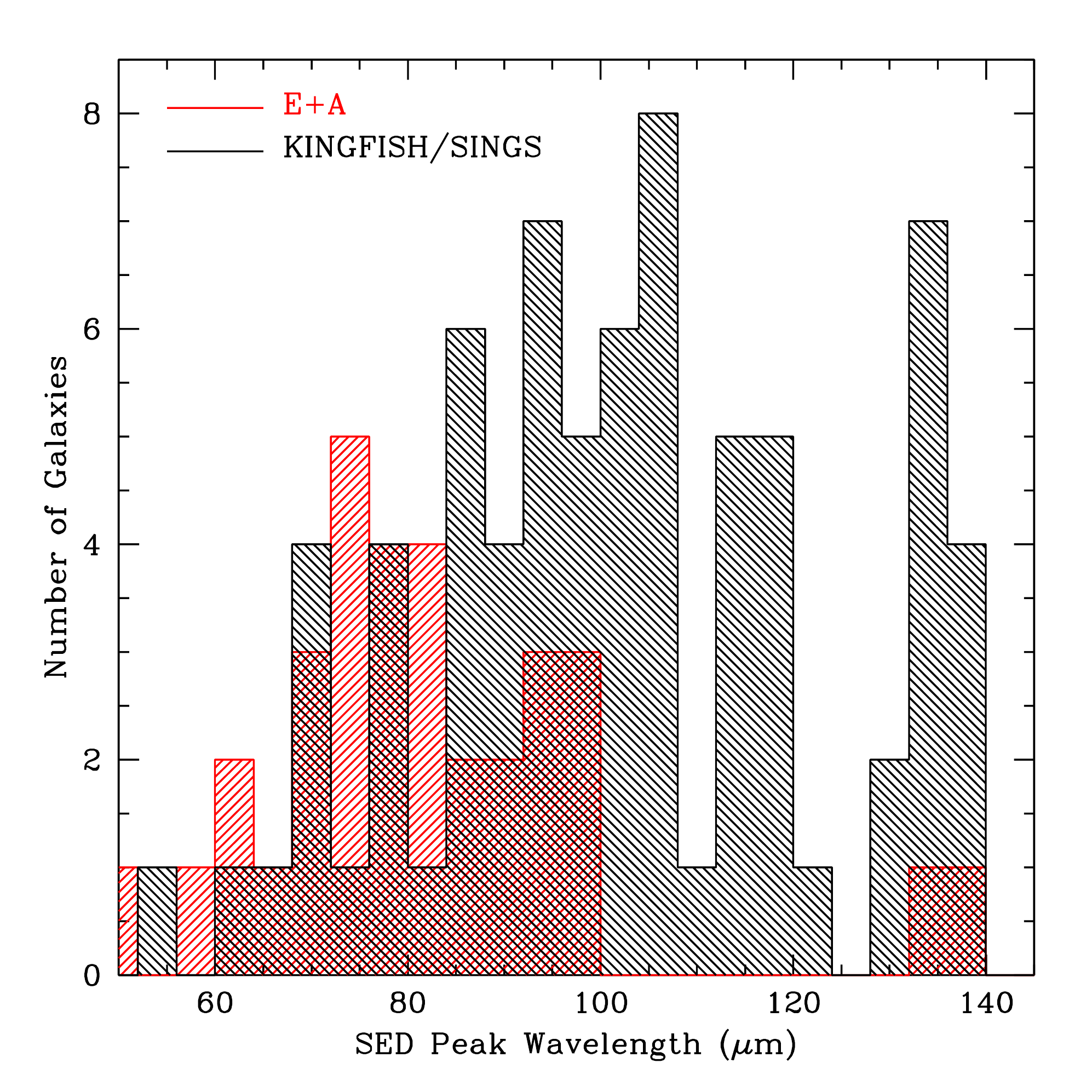}
\caption{Comparison histograms of the SED peak wavelength for the E+A and matched KINGFISH/SINGS samples. Note that the 13 SINGS/KINGFISH sources with peaks $\sim$130\um\ are low-metallicity dwarf galaxies. The E+A sample has infrared SED's that peak at $\sim$70\um, much warmer than the KINGFISH/SINGS mean of $\sim$100\um.}
\label{fig:peak}
\end{figure}

Our SED modeling approach is explained in detail in \S\,\ref{sec:sed}.  Figure \ref{fig:sed} shows the SED fit for each galaxy and Table \ref{tab:dust} gives the dust model parameters, such as dust mass, polycyclic aromatic hydrocarbon (PAH) mass abundance, and derived radiation field intensity. The peak of the IR SED corresponds roughly to an effective dust temperature, and warmer grains imply a higher intensity radiation field. Figure \ref{fig:peak} shows the SED peak wavelength for the E+As relative to the KINGFISH/SINGS matched sample. The E+A sample's IR SEDs peak, on average, at $\sim$70--75\um\ --- significantly warmer than the $\sim$100\um\ peak of normal galaxies in the SINGS/KINGFISH samples, and comparable to (U)LIRGs in the GOALS sample \citep{u2012}. 

\subsection{Dust Masses, PAH Abundances, and DGR}
\label{sec:dmass}
Modeling of the E+A sample's infrared SEDs reveal that all sources possess significant dust reservoirs. Derived dust masses reach as high as $5\times 10^{8} $\,M$_{\odot}$, comparable to the SINGS sample. However, the modeled PAH mass fractions (q$_{\PAH}$) range from 0--8.5\%, with a median value of 4.6\%, substantially larger than the SINGS median of 3.2\%. Indeed, nearly half (16/33) of the E+A SEDs yield a best-fit q$_{\PAH} > 4.6\%$, the \textit{highest} value for SINGS. 

\begin{figure}[t]
\leavevmode
\centering
\includegraphics[width={0.95\linewidth}]{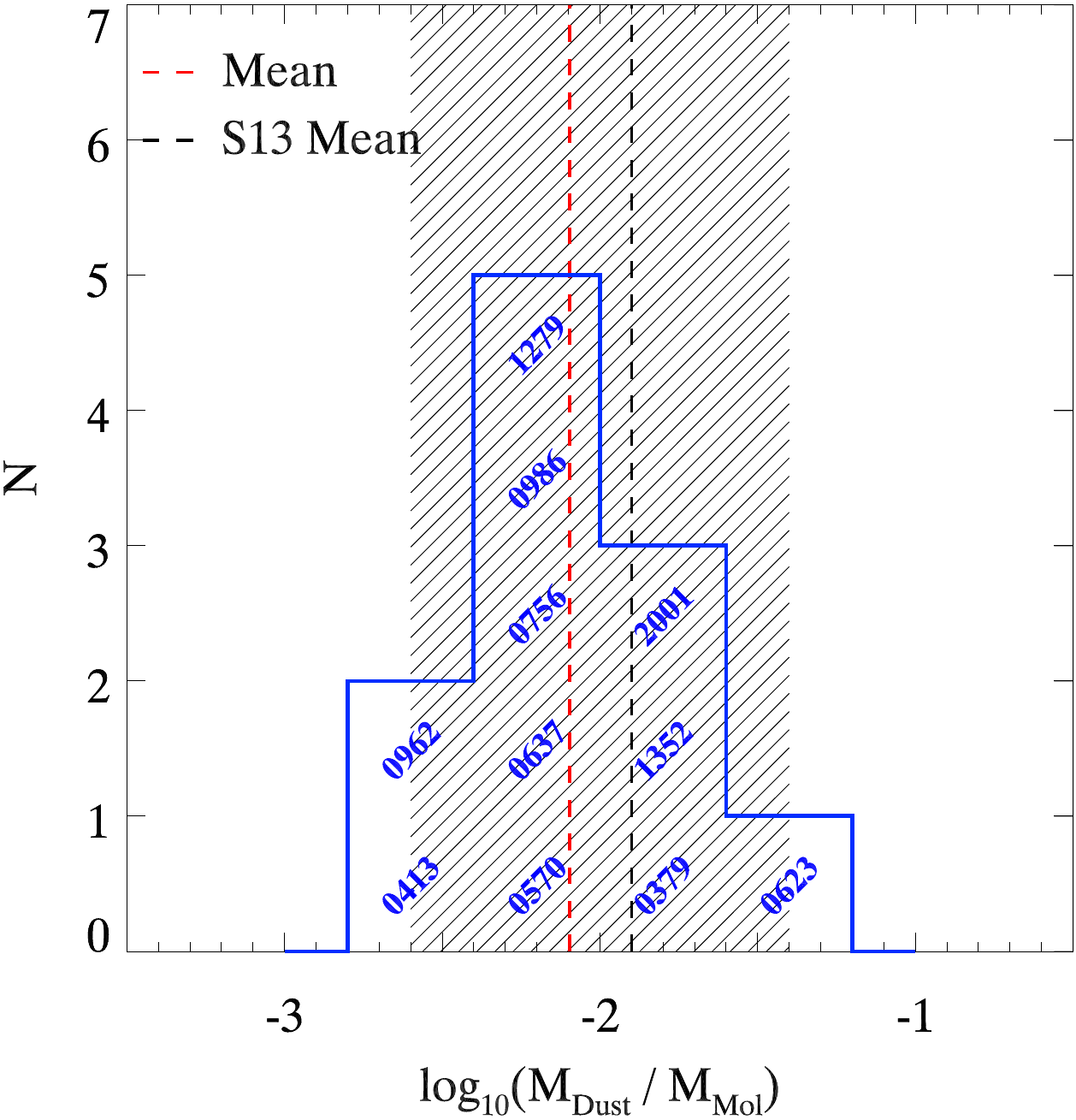}
\caption{A histogram of the DMGR for the 11 E+As with CO intensities detected at $\geqslant$3$\sigma$\ and modeled dust masses known to better than 50\%. The SDSS plate IDs are shown for the galaxies in each bin. The red dashed line denotes the sample mean, while the black dashed line is the KINGFISH/THINGS mean DGR from \cite{sandstrom2013}. The hatched region shows the full range of DGR from \cite{sandstrom2013}. Overall, the E+As’ DMGRs are consistent with the DGRs observed in nearby star-forming galaxies.}
\label{fig:dgr}
\end{figure}

Estimates of the molecular gas content of this sample were obtained from CO observations with IRAM and SMT, detailed in FYZ15. An $\alpha_{\mathrm{CO}} = 4$\ M$_{\odot}$\,(K km s$^{-1}$\ pc$^{2}$)$^{-1}$\ was used (the Milky Way value; see \S\,\ref{sec:aditya_h2_mass} for $\alpha_{\mathrm{CO}}$\ definition and discussion). Combining these measurements with our estimates of the dust mass, we examined the dust-to-molecular gas ratio (DMGR = M$_{\mathrm{Dust}}$/M$_{\mathrm{Mol}}$). Only 17 E+As ($\sim$50\%) were detected in CO at the $>$3$\sigma$\ level. Of those 17 galaxies, $\langle \log_{10}$(DMGR)$\rangle = -1.99$. In Figure \ref{fig:dgr}, we show the distribution of DMGR for the CO-detected sub-sample and compare to the dust-to-total gas ratios (DGR = M$_{\mathrm{Dust}}$/[M$_{\mathrm{Mol}}$ + M$_{\mathrm{H\,I}}$]) of KINGFISH/THINGS galaxies \citep{sandstrom2013}. The E+As possess a DMGR mean within 40\% of the \cite{sandstrom2013} DGR mean, and span a similar range. This suggests that either their atomic mass fractions are low, or that their intrinsic DGR values are substantially higher than that of normal galaxies. One source (0623) lies above the DGR range from \cite{sandstrom2013}, suggesting that this particular source could possess a higher atomic fraction. We discuss additional possibilities for this source in \S\,\ref{sec:aditya_h2_mass}. However, some studies have found a broader range of DGR among nearby galaxies than \cite{sandstrom2013} (e.g., \citealt{remy-ruyer2014}).

\begin{figure*}[t]
\centering
\leavevmode
\includegraphics[width={0.8\linewidth}]{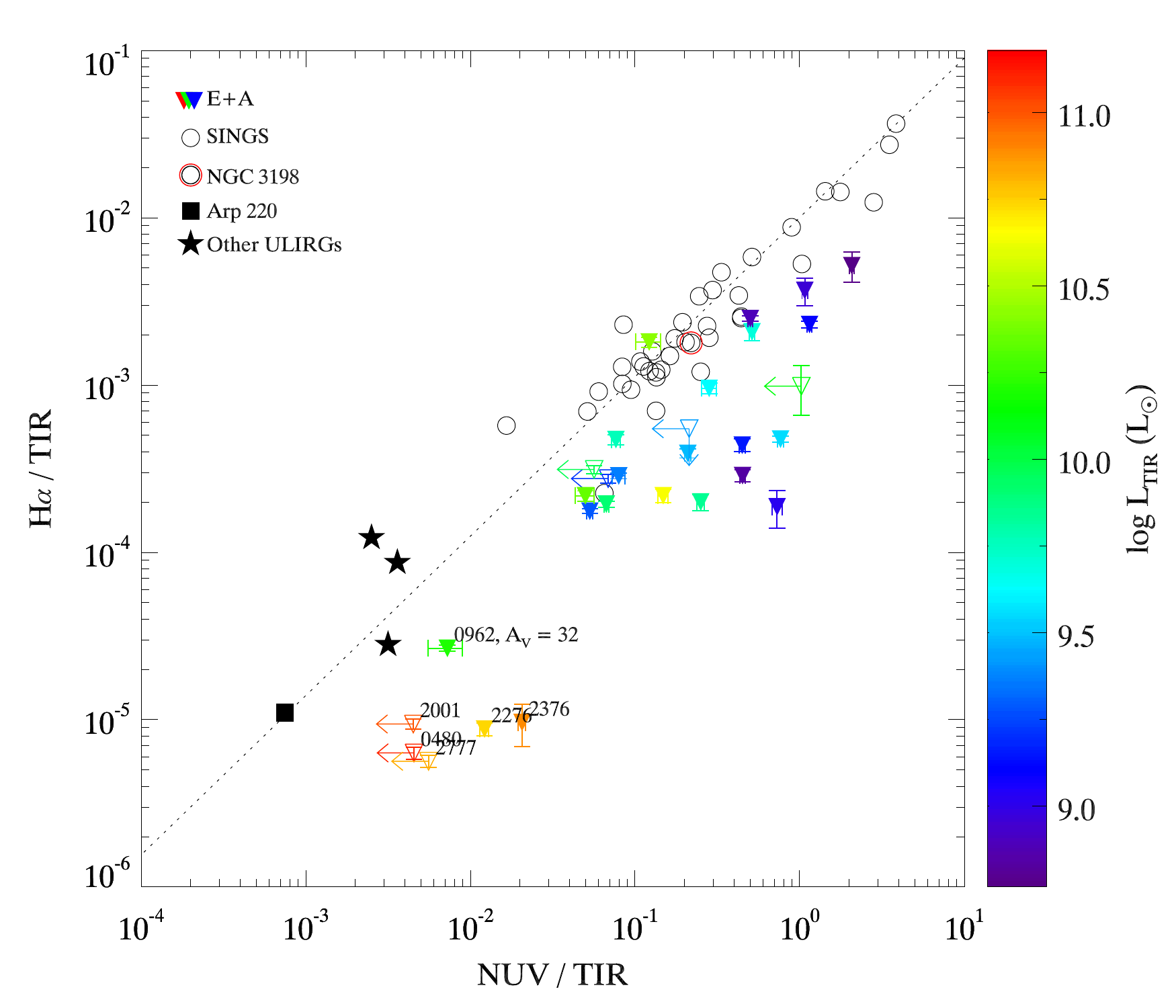}
\caption{Aperture-corrected H$\alpha$-to-TIR luminosity ratio, plotted as a function of the NUV-to-TIR luminosity ratio for the E+A sample (downward triangles), color-coded by TIR luminosity (red is most luminous). H$\alpha$\ and NUV have both been corrected for galactic extinction. Downward arrows denote upper limits for the H$\alpha$/TIR ratio, while leftward arrows denote upper limits for the NUV/TIR ratio. The SINGS/KINGFISH sample is shown as open circles, with NGC 3198 (a source with detected 9.7\um\ silicate absorption; SDD07) labeled with a second, concentric red circle. Arp 220 (black filled square) and several other ULIRGs (black filled stars; UGC 05101, UGC 08696, IRAS 08572+3915) are also plotted for reference (H$\alpha$, \citealt{moustakas2006}; TIR, \citealt{armus2009}; NUV, \citealt{brown2014}). The dotted line is an attenuation curve ranging from $\tau_{V} = $\ 0--10, assuming a \cite{charlot&fall2000} extinction law.  Only the slope of this curve is meaningful, given the anti-correlated impacts of attenuation on H$\alpha$/UV vs. IR emission. The trend with increasing extinction is remarkably consistent from the star-forming to ULIRG comparison samples, while E+A's lie consistently below it. The six most attenuated sources, all but one of which are 22\um-selected Herschel-only sources, are labeled for reference.  The single highly-obscured \textit{Spitzer} E+A source 0962\_212\_52620 is labeled with extinction measured directly from its 9.7\um\ silicate absorption (A$_{V}\!\simeq32$; see \S\,\ref{sec:specdecomp}). 
}
\label{fig:halpha-nuv}
\end{figure*}

\subsection{UV/Optical-to-Infrared Comparisons}
\label{sec:halpha-nuv}
As described in \S\,\ref{sec:sample-selection}, the E+As were selected to have very weak optical emission lines. However, the derived TIR luminosities are very comparable to galaxies in the SINGS sample, with both samples possessing medians of L$_{\TIR}\simeq7\times$10$^{9}$\,L$_{\odot}$. Using the SDSS photometry and the \cite{cook2014} $ugri - UBVR_{\mathrm{C}}$\ transformations, we compute a geometric mean TIR-to-$B$-band luminosity ratio L$_{\TIR}$/L$_{B} = 1$, with a full range of 0.044--31. This wide range is consistent with galaxies in the SINGS sample (SDD07), but even the highest values are significantly lower than those found in ULIRGs, which can reach L$_{\TIR}$/L$_{B} \sim 100$\ \citep[e.g., Arp 220;][]{armus2009,veron-cetty2010}.

In Figure \ref{fig:halpha-nuv} we plot the E+A MPA-JHU H$\alpha$\ line luminosities \citep{aihara2011}, relative to TIR, as a function of GALEX NUV/TIR and compare them to the SINGS sample. The measured H$\alpha$\ emission listed in Table \ref{tab:sample} is derived from the 3\arcsec\ SDSS fiber; many sources are larger than this 3\arcsec\ aperture, necessitating an aperture correction. H$\alpha$\ fluxes for sources which are fully resolved at 8\um\ (FWHM $\gtrsim$\ 3\arcsec)  were aperture-corrected from the 3\arcsec\ SDSS fiber by scaling the 8\um\ flux density within the fiber to the global value.  Size estimates for Herschel-only sources were obtained by adopting the mean 8\um-to-R$_{90}$\ size ratio for the \textit{Spitzer} sources (see \S\,\ref{sec:compact}). For \textit{Herschel} sources with projected 8\um\ FWHM $>$\ 3\arcsec\ (only 2), a linear fit to the 3\arcsec-to-global flux ratios as a function of 8\um\ size for the resolved \textit{Spitzer} sources was used to derive a correction. Of the 8 sources which are resolved (or projected to be resolved) at 8\um\ and, thus, require an aperture correction, the corrections are generally modest, from 3--5. A single source does require a larger correction of 9.1. Performing a similar analysis on the 8\um\ images of several compact SINGS sources with SDSS coverage yields H$\alpha$\ fluxes within a factor of two of the globally-obtained values.

\begin{figure*}[t]
\centering
\leavevmode
\includegraphics[width={0.8\linewidth}]{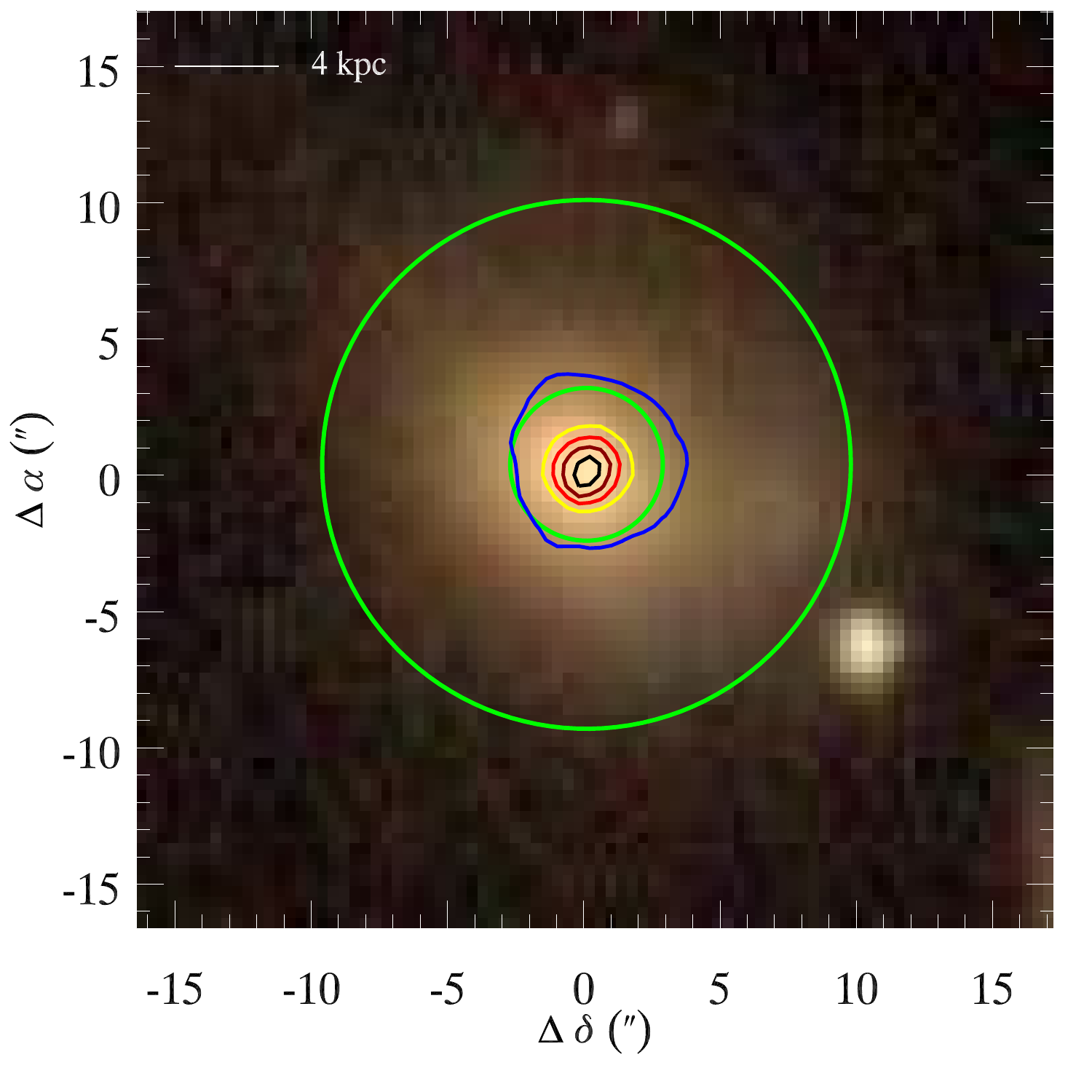}
\caption{An SDSS three-color image ($gri$) of 0379\_579\_51789 (North is right; East is up). Two green circles are overlayed, with radii equal to 0379's $r$-band Petrosian 50\% and 90\% radii, respectively (R$_{50} \sim 2.8\arcsec$/R$_{90} \sim 9.7\arcsec$, as derived by SDSS). Also shown are 0379's IRAC 8\um\ surface brightness isocontours, at 1, 3, 4, 5, and 6 MJy sr$^{-1}$. The 8\um\ FWHM is 3.8\arcsec, corresponding to a physical size of $\sim$4 kpc at the distance of 0379, which approximately overlaps the 4 MJy sr$^{-1}$~(yellow) isocontour. The 8\um\ 50\% diameter is only 2.2\arcsec, approximately overlapping the second innermost (dark red) contour.
}
\label{fig:8um-size}
\end{figure*}

Both SINGS and E+A sources are spread along the direction of a \cite{charlot&fall2000} reddening curve,
\begin{equation}
\begin{split}
T_{\lambda}^{\mathrm{ISM}} \propto e^{-\tau_{\lambda}}, \\
\tau_{\lambda} \propto \lambda^{-0.7}, 
\end{split}
\end{equation}
where $T_{\lambda}^{\mathrm{ISM}}$\ is the interstellar transmission function and $\tau_{\lambda}$\ is the optical depth, at a given wavelength. We have adopted an emission-line (e.g., H$\alpha$) optical depth of $\tau_{\lambda}^{\mathrm{H}\alpha} = 2\tau_{\lambda}^{\mathrm{ISM}}$, as is typically assumed due to the elevated extinction seen in H\,II regions where the emission originates. We find a wide range of extinction impact\footnote{Increasing dust column both removes H$\alpha$ and NUV power, and contributes additional infrared emission, making the slope of the extinction curve valid, but amplifying absolute extinction values.}. The ULIRGs (including Arp 220) fall at the lower left, along the extinction curve sequence from the normal galaxies. Notably, however, the E+As are offset from this sequence by a factor of $\sim$5, consistent with a robust young, but aging stellar population and relative dearth of ongoing star formation, resulting in more NUV than H-ionizing emission. Several of the IR-selected Herschel sources appear to possess high internal extinction, as they lie much further along the reddening curve than the SINGS sample. This raises the possibility of a ``skin'' effect in this small, dustier subset, in which the bulk of the stellar population could be hidden by substantial dust columns (see \S\,\ref{sec:discuss} for further discussion).

\subsection{Spatial Extent}
\label{sec:compact}
To examine the spatial extent of the MIR emission, we fit elliptical Gaussian functions to the IRAC 8\um\ images of each galaxy. The resulting 50\% enclosed flux radii (taken as the geometric mean between the Gaussian 50\% radius along two elliptical axes) reveal compact emitting regions --- on average 3--4$\times$\ smaller than their corresponding R$_{50}$\ optical sizes in $r$-band. Four of the galaxies are unresolved with IRAC, and, except for one source (0623\_207\_52051), the rest are very modestly resolved with FWHMs less than twice the size of the IRAC PSF ($\lesssim 6$\arcsec). In Figure~\ref{fig:8um-size}, we show the SDSS $gri$~image of 0379\_579\_51789, overlayed with 8\um\ surface brightness contours. As is typical of the sample, the 8\um\ emission is contained within a region $\sim$3$\times$\ smaller than the corresponding optical emission.

Using angular sizes derived from the redshift-dependent luminosity distances, we compare the physical 8\um\ FWHMs for the E+As to the 8\um\ FWHMs for the GOALS sample \citep{diaz-santos2010}. We estimate a mean 8\um\ FWHM of 2.8 kpc for the E+As, with a full range of 1.1--5.6 kpc. Of the 15 \textit{Spitzer} sources, four are more compact than the most compact GOALS ULIRGs (d$\sim$1.5 kpc) and four of the remaining 11 are as or more compact than the most compact GOALS LIRGs \citep[d$\sim$2.6 kpc][]{diaz-santos2010}. Overall, only three sources possess FWHM $>$4 kpc. Five of the sources are unresolved at 8\um\ --- in two of which this corresponds to a physical size $>$3 kpc --- and thus may be even more compact than estimated. 

Their compactness at 8\um\ seems to be consistent with the high optical central surface brightness of many E+As. \citeauthor{yang2006} (\citeyear{yang2006}, \citeyear{yang2008}) discovered a sample of E+As possessing compact blue cores, with typical sizes $< 1.4$\ kpc. Though the presence of significant dust columns may confuse such optical classification in the majority of this sample, as well as others, the consistency of 8\um\ sizes across the sample suggests that central compactness is likely a typical characteristic of E+As.

\begin{figure*}[t]
\centering
\leavevmode
\includegraphics[width={0.8\linewidth}]{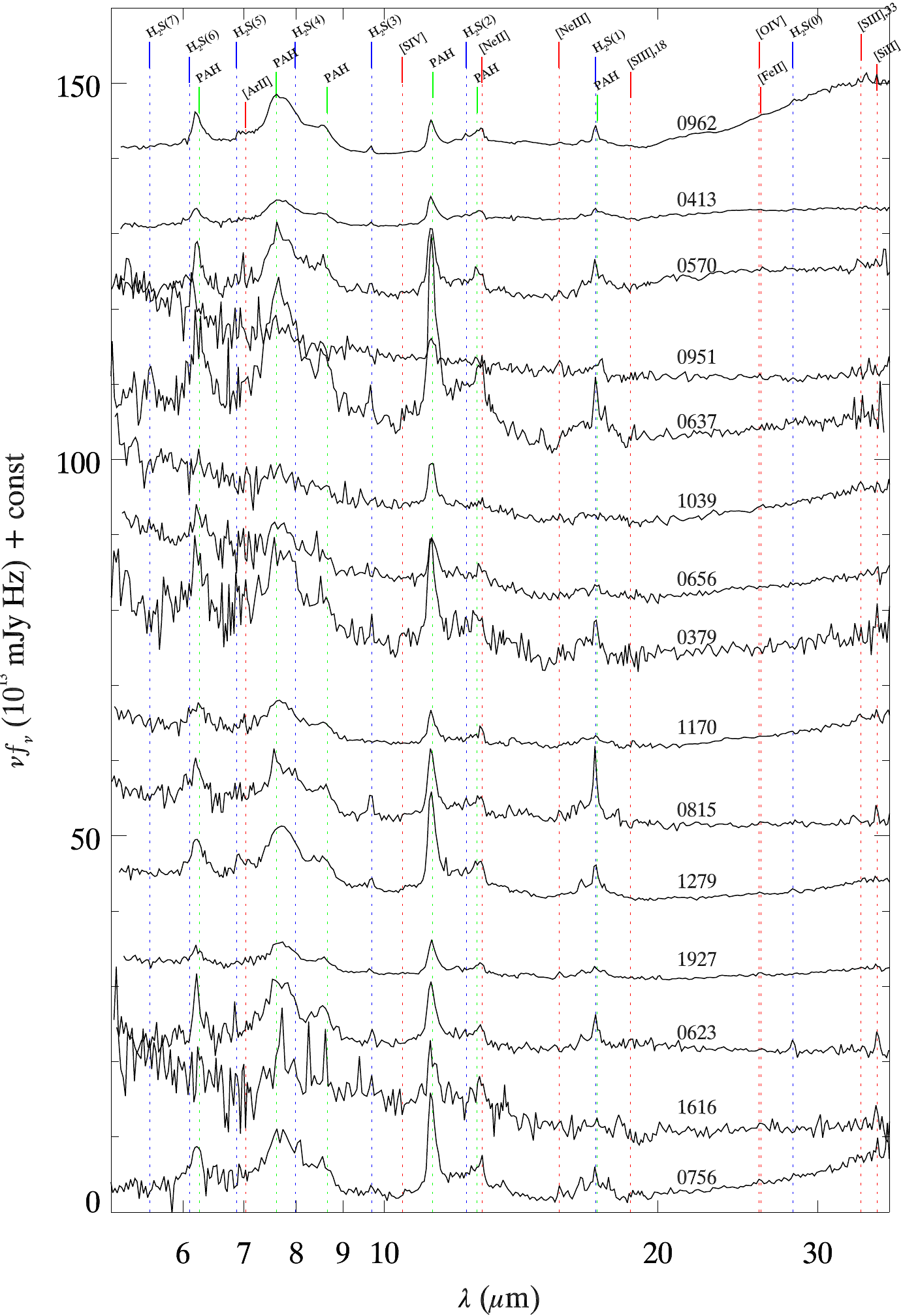}
\caption{Spitzer IRS rest-frame spectra of 15 E+As, sorted arbitrarily for maximum visibility. Green lines show the positions of primary PAH features, while blue lines show the positions of relevant H$_{2}$\ rotational lines and red show fine-structure emission lines. The plate number of each object is shown on the right. The spectra are PAH-dominated in all cases, but are correspondingly weak in nebular emission lines.}
\label{fig:spectra}
\end{figure*}

\begin{figure}[t]
\leavevmode
\centering
\includegraphics[width={0.95\linewidth}]{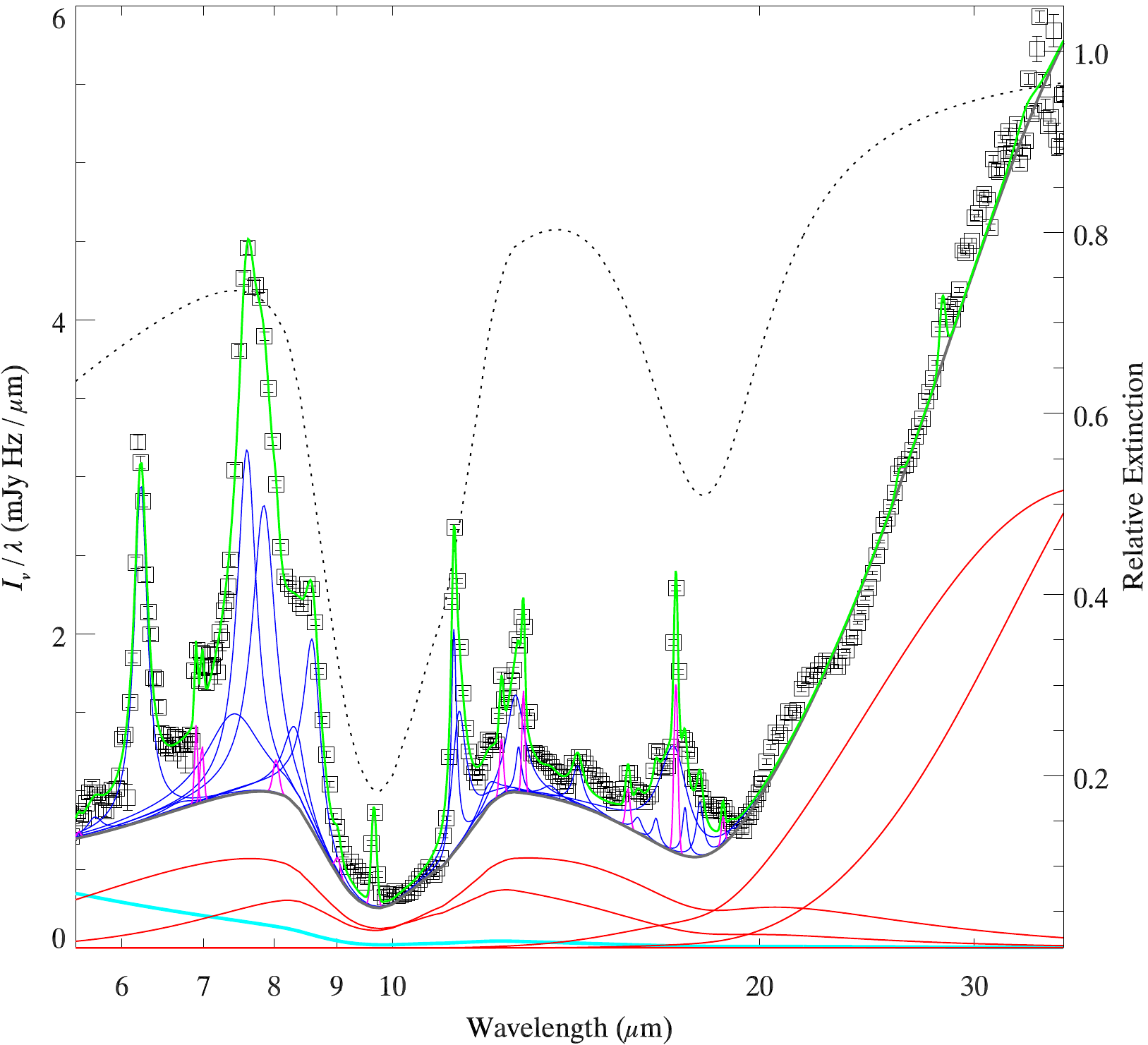}
\caption{PAHFIT decomposition of 0962\_212\_52620. Nebular and H$_{2}$\ rotational lines are shown in magenta and dust emission features in blue. The stellar continuum is shown as a cyan curve, the dust continua as red curves, and the total fit as a green curve. The attenuation profile is shown as a dotted line, with the relative extinction shown on the right-hand axis.}
\label{fig:pf962}
\end{figure}

\subsection{Spectral Properties}
Relatively unaffected by dust attenuation, the mid-infrared (MIR) spectrum of galaxies provides unique insight into the properties of their ISM. Using the \emph{Spitzer/IRS} spectra obtained for our 15 galaxy sub-sample, we investigate physical properties of the dust and gas, and an extinction-independent analysis of residual star formation (see \S\,\ref{sec:sfr}). The full sample of E+A spectra are shown in Figure~\ref{fig:spectra}, with various spectral features labeled. PAH emission is strongly detected in the entire sample, along with unusually bright H$_2$ rotational emission lines, but the nebular emission lines are considerably fainter than found in typical star-forming galaxies, as discussed below.

\subsubsection{PAHFIT Decomposition \& Silicate Opacity}
\label{sec:specdecomp}
The IRS spectra were decomposed using the spectral decomposition model PAHFIT\footnote{\color{blue}{http://tir.astro.utoledo.edu/jdsmith/research/pahfit.php}}, described in SDD07. PAHFIT treats IRS spectra as linear combinations of four distinct components: a blackbody stellar continuum, warm dust blackbody continua, fine-structure emission features, and dust (PAH) emission features. Spectra are decomposed by minimizing the $\chi^2$\ of the fit resulting from iterative addition of the four components. Fine structure lines are fit using Gaussian profiles and the PAH emission features with Drude profiles. Figure \ref{fig:pf962} shows the decomposition of 0962\_212\_52620, one of the nearest and brightest sources.

PAHFIT also internally fits and corrects emission features for extinction. Extinction in the MIR is dominated by two distinctive, broad silicate features: a deep feature at 9.7\um\ and a shallower feature at 18\um. PAHFIT offers the choice of several extinction curves (e.g., \citealt{chiar&tielens2006}; \citealt{kemper-vriend-tielens2004}) and a well-mixed or foreground screen geometry.  

In contrast to the rest of our \textit{Spitzer} sub-sample, 0962\_212\_52620 displays significant 9.7\um\ and 18\um\ silicate absorption, most visible in Fig.~\ref{fig:pf962} in the declining continuum between 15 and 20\um, and sharp upturn thereafter. Assuming a screen geometry, PAHFIT returns a 9.7\um\ silicate optical depth $\tau_{9.7} = 1.77$. Using the ratio of \textit{V}-band extinction-to-silicate optical depth $A_{V}/\tau_{9.7} = 18$\ for the local ISM \citep{roche&aitken1984,rieke&lebofsky1985}, this corresponds to $A_{V} = 32$. For comparison, the ULIRG Arp 220 has been estimated to possess $A_\mathrm{V}\!\gtrsim\!100$\footnote{Local extinction may be higher or lower depending on clumpiness of the ISM.}, using screen extinction models and the observed 9.7\um\ abosorption \citep{smith1989,spoon2006}. In \S\,\ref{sec:agn} we discuss the possibility of a buried AGN in this source. All other sources exhibit no detectable silicate absorption, corresponding to global extinction $A_\mathrm{V}\lesssim 3$. 

\subsubsection{PAH and Nebular Line Emission}
\label{sec:pah}
Polycyclic aromatic hydrocarbons (PAHs) are small, carbonaceous dust grains found in galaxies' ISM across a wide range of  environments.  PAHs are thought to act as the principal heating sources of the neutral ISM, via photoelectrons liberated from them by the ISRF. Due to their small size ($< 16\angstrom$; \citealt{draineli2007}), PAHs possess very low heat capacities and, thus, absorption of a single UV photon heats them to high, non-equilibrium temperatures. Consequently, the resulting emission also manifests as individual transition bands. As such, PAHs are more sensitive to changes in the ISRF. See the \cite{tielens2008} review for an in-depth discussion of astrophysical PAH chemistry and emission processes.

PAH emission has been studied in great detail in nearby galaxies (e.g., SINGS, SDD07; \citealt{draine2007}) and has been found to be fairly ubiquitous, even at high redshift \citep[e.g.][]{teplitz2007}, across a wide range of galaxy evolutionary stages. Due to their modest preference for UV photons \citep{crocker2013}, some have sought to calibrate PAH emission as a star formation indicator (\citealt{peeters2004}; \citealt{wu2005}) --- the grain fluorescence attributed to photodissociation regions (PDRs) bordering H\,II regions \citep{povich2007}. Conversely, \cite{lidraine2002_1} showed that PAH emission need not be elicited by UV emission, casting doubt on their usefulness as star-formation tracers. Supporting this, PAH emission has been seen in many nearby ETGs with visible dust lanes and seemingly small amounts of ongoing star formation \citep{temi2007,vega2010,rampazzo2013}. 

\begin{figure}[!ht]
\leavevmode
\centering
\includegraphics[width={0.95\linewidth}]{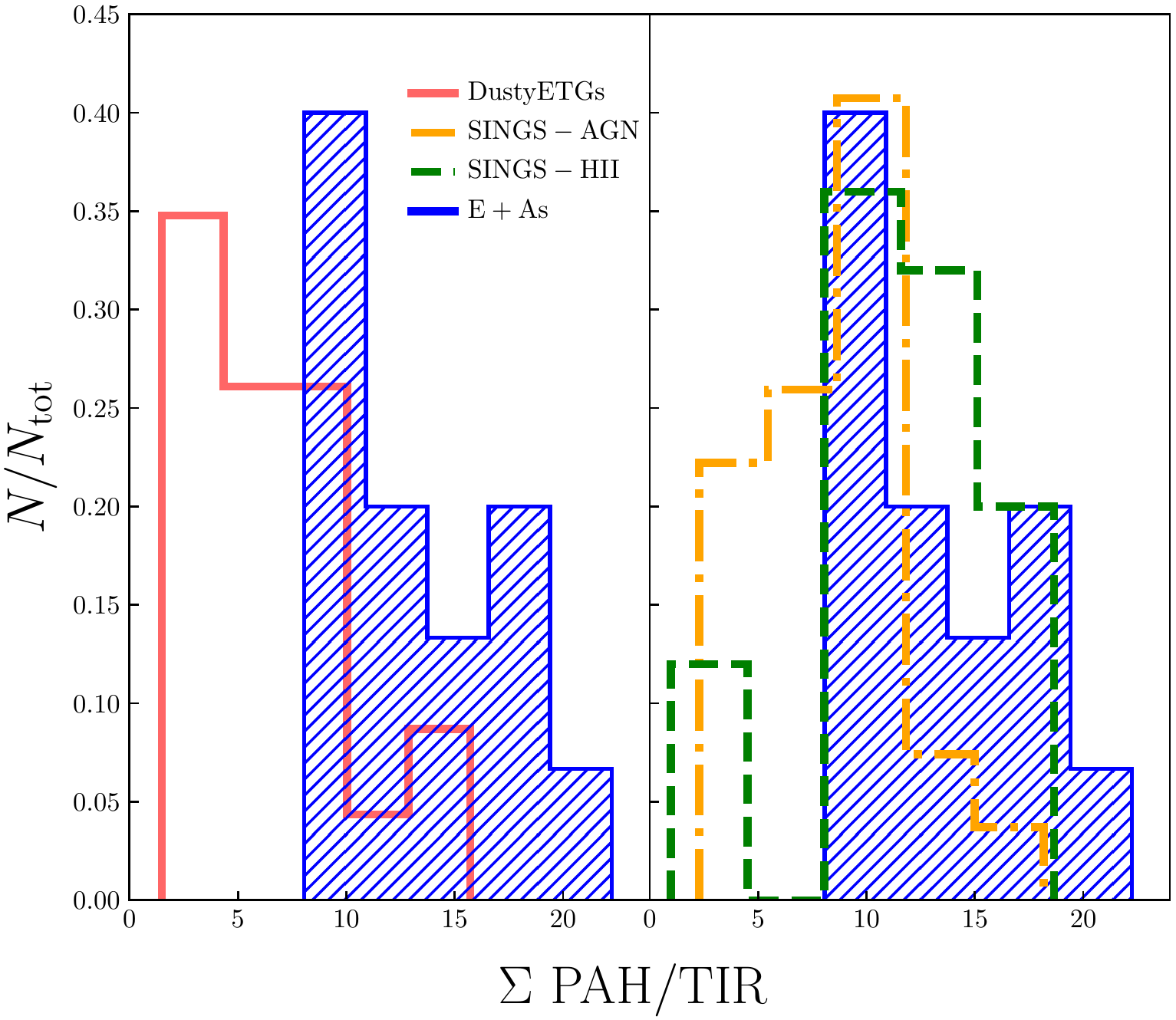}
\caption{Histograms of total PAH-to-TIR luminosity for the E+A sample. \uline{Left}: PAH/TIR for the E+As (blue) compared to the dusty ETGs (red). The E+As are, on average, 3$\times$\ as fractionally PAH luminous as the ETGs. \uline{Right}: PAH/TIR for the E+As compared to the SINGS AGN (orange) and H\,II (green) sub-samples, individually. The E+As are most similar to the SINGS H\,II-dominated galaxies, but display, on average, brighter PAH emission than either comparison sample. 
}
\label{fig:pah-24}
\end{figure}

PAH emission features dominate the spectra of all 15 of the \textit{Spitzer}-observed E+As. Indeed, the vast majority of the sub-sample (13/15) possess all of the primary features identified by SDD07 in nearby star-forming galaxies --- at 6.2\um, 7.7\um, 8.6\um, 11.3\um, 12.7\um, and 17\um. In Figure \ref{fig:pah-24} we show the distribution of E+A total integrated PAH emission strengths, relative to TIR luminosity. As a population, the E+As display much more dominant PAH emission than the dusty ETG sample or AGN-hosting star-forming galaxies, and are most consistent with that of the SINGS H\,II-dominated sources.

Indeed, the total fractional contribution of PAH emission to the infrared is high, with all but one source displaying PAH/TIR $\geqslant$\ 10\%. The geometric mean of 13.0\% is noticeably higher than the SINGS value of 8.7\%, with one source (1616\_071\_53169) displaying an exceptional PAH/TIR = $22.2 \pm 2.1$\% --- if confirmed, this is the highest known fractional PAH luminosity of any galaxy. 

\begin{figure*}[t]
\centering
\leavevmode
\includegraphics[width={0.8\linewidth}]{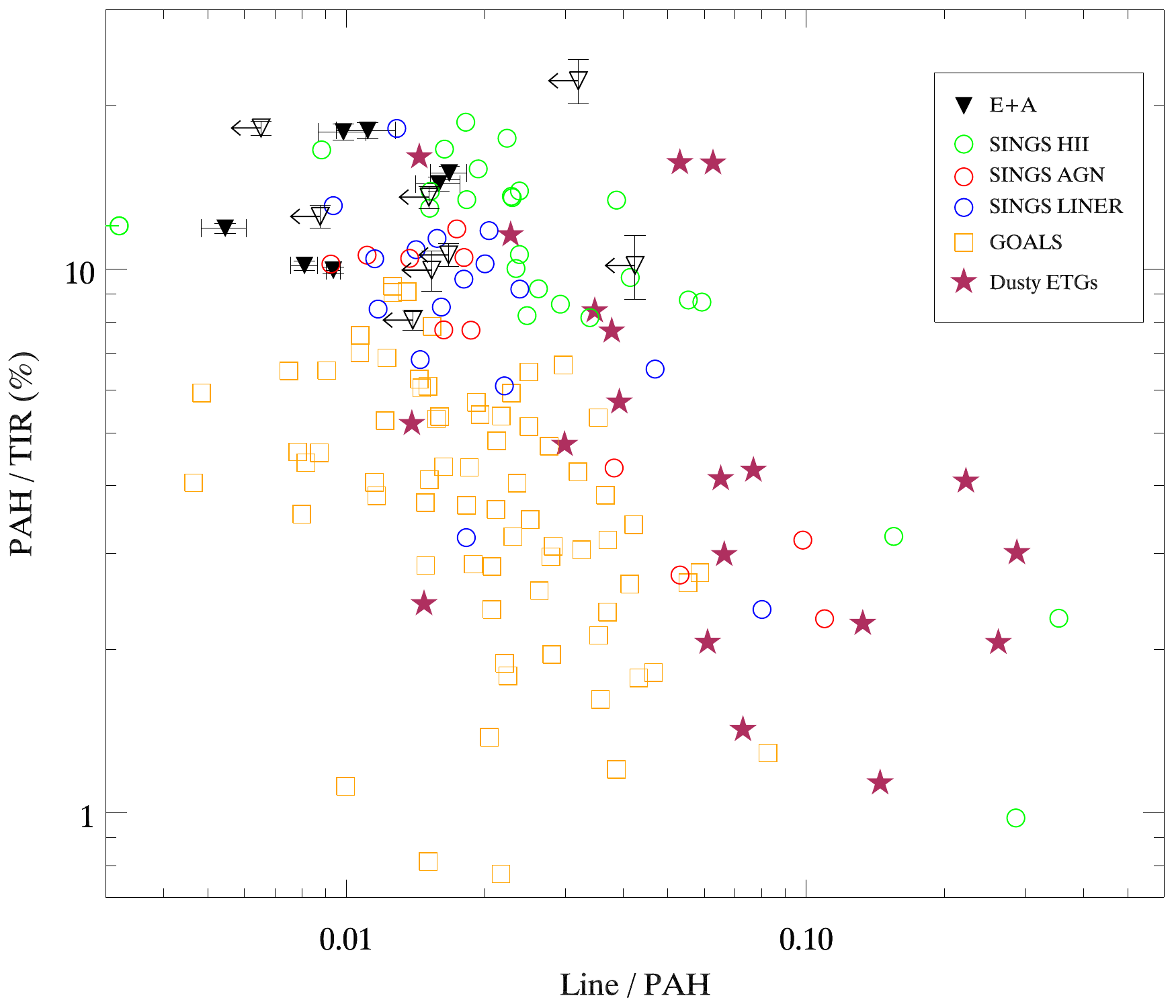}
\caption{Fractional PAH luminosity, relative to TIR, plotted as a function of total line (\neII\ + \neIII\ + \sIV\ + \sIII\,18\um) emission, relative to PAH. E+As are shown as black downward triangles, with open symbols and arrows denoting upper limits. The SINGS sample is shown as open circles (SDD07), with HII nuclei in green, Seyfert nuclei in red, and LINERs in blue. The GOALS sample is shown as open orange squares \citep{armus2009,inami2013,stierwalt2014}. As shown here (and as can also be intuited from their spectra), the E+As possess very bright PAH emission, but exceptionally weak relative nebular line emission --- distinct from all other comparison samples.}
\label{fig:pahtir-linepah}
\end{figure*}

\begin{figure*}[t]
\centering
\leavevmode
\includegraphics[width={0.95\linewidth}]{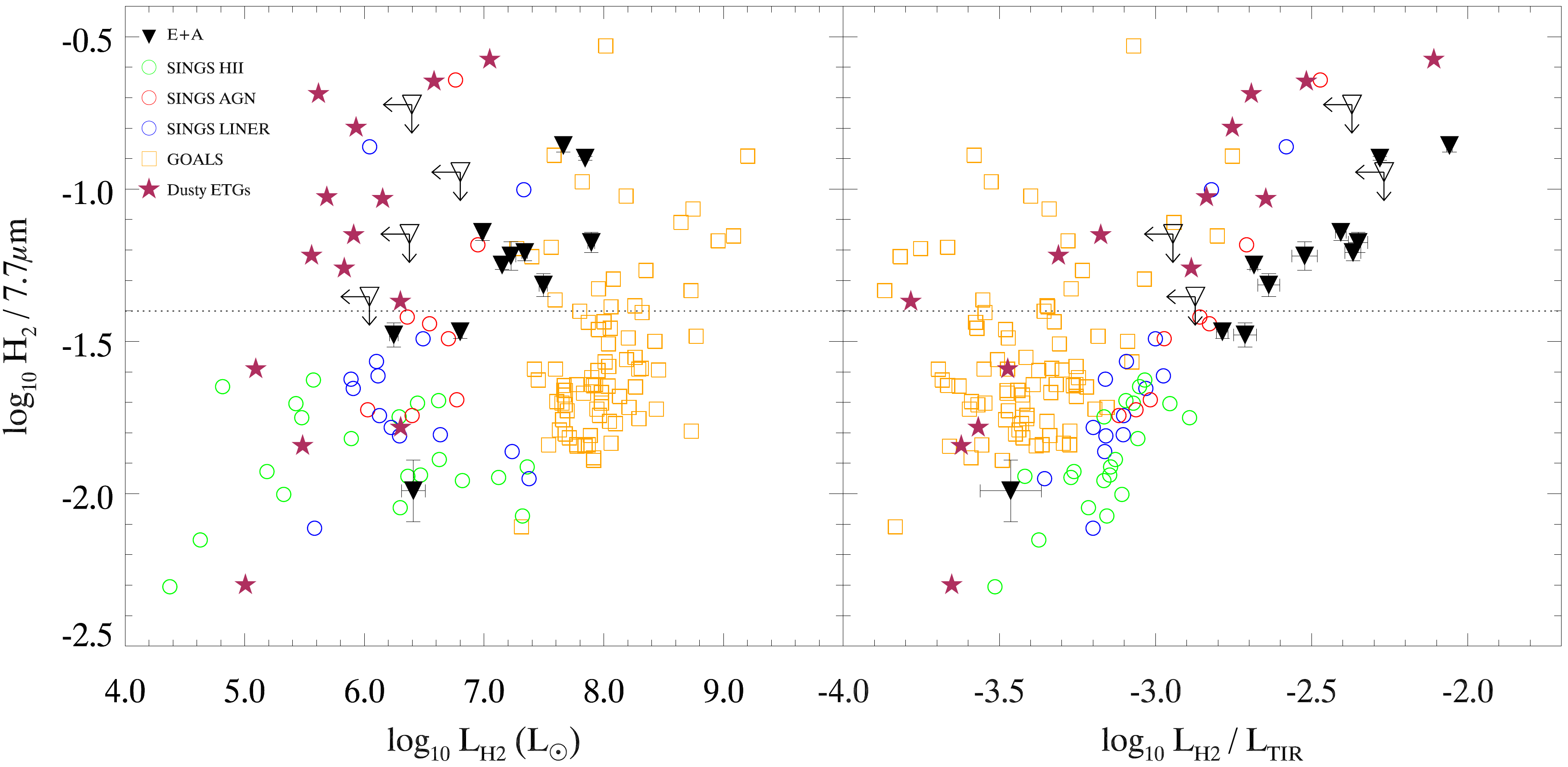}
\caption{\uline{Left}: The ratio of total H$_{2}$\,S(0--3) emission to 7.7\um\ PAH emission, as a function of H$_{2}$\ luminosity. Arrows denote 4$\sigma$\ upper limits. Orange open squares denote the GOALS sample \citep{stierwalt2014}, filled magenta squares denote the dusty ETGs, and the green, red, and blue open circles denote SINGS HII, AGN, and LINER nuclei, respectively. The ETGs have been corrected for 7.7\um\ depletion using the geometric mean 7.7\um/11.3\um\ for the E+A sample. The E+As are shown as filled downward triangles. The dotted line is the \cite{ogle2010} H$_{2}$/7.7\um\ turbulent-heating (e.g., MOHEG) threshold. The E+As are overall more luminous than either the dusty ETGs or SINGS, but less luminous than GOALS sources. \uline{Right}: Same as the left figure, but plotted as a function of H$_{2}$/TIR on the x-axis. The majority of E+As display significantly stronger H$_{2}$\ emission, relative to TIR, than the majority of either SINGS or GOALS, and are most consistent with the dusty ETGs. In the majority of E+As, an unusually high 0.1--1\% of their TIR luminosities appears to be from H$_{2}$\ rotational emission.}
\label{fig:h2_goals}
\end{figure*}

The MIR forbidden line transitions of numerous ionized species provide relatively extinction-independent probes of the radiation field. In Figure \ref{fig:pahtir-linepah} we plot PAH/TIR as a function of the combined nebular emission from \neII, \neIII, \sIV, and \sIII\,18\um, relative to PAH. The E+As largely occupy a distinct parameter space compared to the SINGS, GOALS, and dusty ETG samples, with weaker line emission, but higher PAH/TIR. In the seven E+As which possess a $>$5$\sigma$\ sum, the geometric mean Line/TIR is $\sim$3$\times$\ lower than for the SINGS H\,II dominated sample, yet their geometric mean PAH/TIR of 13.0\% is significantly higher than the 9.6\% of the H\,II sources. GOALS sources typically possess significantly lower PAH/TIR, but relative line luminosities more comparable to the E+As. The dusty ETGs live largely at the other extreme, with many possessing line-bright spectra but overall weak PAH emission --- indicative of star-formation or shock-dominated environments which are more hostile to the survival of small grains. 

Thus, the ISM energetics of E+A galaxies are unusual --- PAH emission is even more dominant than in normal star-forming galaxies, but nebular line emission is considerably weaker. This suggests that, in the E+As, the primary PAH heating mechanism is less well-coupled to the primary line emission mechanism. We discuss this further in \S\,\ref{sec:discuss}.

\subsection{Properties of Rotational H$_{2}$ Emission}
\label{sec:h2}
As a symmetric diatomic molecule, H$_{2}$~possesses no permanent dipole moment. However, at temperatures $\gtrsim$80\,K, its quadrupole moment allows for emission via pure rotational transitions. Excitation mechanisms range from star formation (UV pumping in PDRs; \citealt{hollenbach&tielens1997}), to AGN (X-rays; \citealt{draine&woods1992}), cosmic rays \citep{dalgarno1999}, and turbulent/shock heating \citep{shull&hollenbach1978}. H$_{2}$\ rotational emission is seen, to some degree, in most nearby star-forming galaxies (SDD07). All galaxies in our E+A sample display some amount of H$_{2}$\ rotational emission. We discuss potential excitation mechanisms, as well as emission-based temperature and mass estimates below.  

\begin{figure*}
\centering
\leavevmode
\includegraphics[width={0.95\linewidth}]{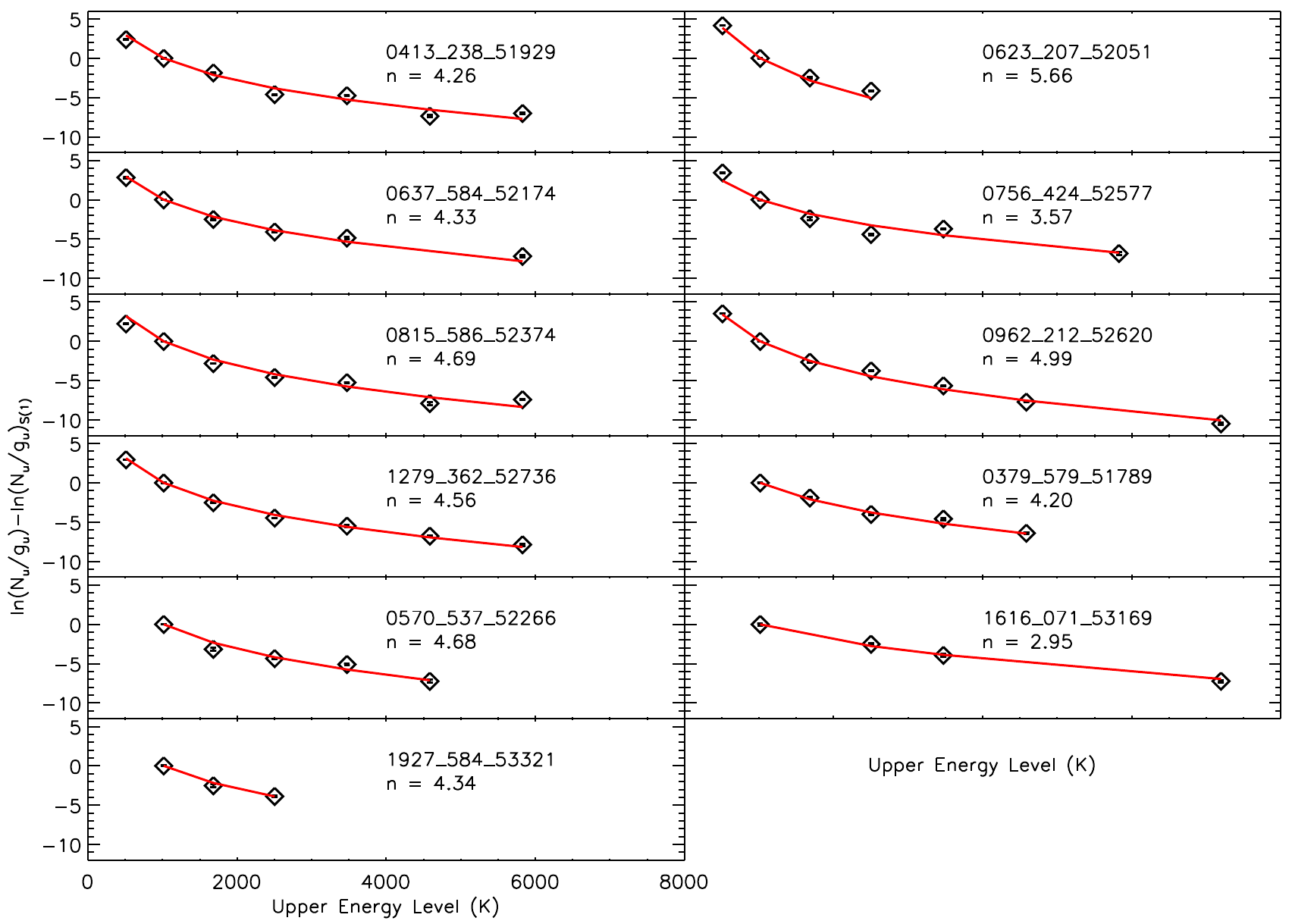}
\caption{Power law model fits (red solid) to the observed H$_{2}$\ rotational line ratios (black diamonds) for our 11 selected galaxies. Several sources display nearly the full compliment of MIR H$_{2}$\ rotational lines. The relatively shallow power law slopes indicate that in many cases the molecular gas is quite warm. Uncertainties are shown, though in most cases they fall within the symbols.}
\label{fig:excitation_diagrams}
\end{figure*}

\subsubsection{Excitation Mechanisms}
\label{sec:h2_excitation}
\cite{stierwalt2014} showed that most (U)LIRGs in the GOALS sample host exceptionally strong H$_{2}$\ emission compared to normal star-forming galaxies, likely due to turbulent shock heating rather than UV excitation --- the dominant mechanism in normal galaxies. They find that the emission ratio of H$_{2}$\ to the 7.7\um\ PAH feature has a strong positive correlation with merger state, with the latest-stage mergers also displaying the highest H$_{2}$/7.7\um. Much of the power in the 7.7\um\ PAH feature is thought to arise from the photo-excitation of ionized grains \citep{li&draine2001}. Thus, H$_{2}$/7.7\um\ traces the importance of non-radiative heating to the observed H$_{2}$\ emission (see \S\,\ref{sec:firdeficit}). In Figure \ref{fig:h2_goals}, we show the sum of the H$_{2}$\,S(0--3) luminosities relative to the 7.7\um\ PAH luminosity, as a function of both the H$_{2}$\ luminosity and H$_{2}$/TIR. The E+As exhibit a slightly larger than order-of-magnitude range in H$_{2}$/PAH luminosity. A majority (10/15) of the E+As display far higher H$_{2}$/7.7\um\ ratios than are seen in star-forming galaxies, which have a geometric mean H$_{2}$/7.7\um\ $= 0.013$\ (SDD07) --- 4$\times$\ lower than the E+A value of 0.053. The E+As are, however, quite consistent with the most luminous GOALS sources \citep{stierwalt2014}, but at substantially lower total luminosity. The E+As are also strong outliers in H$_{2}$/TIR, with values consistently higher than GOALS or SINGS sources, but very comparable to a number of the dusty ETGs. 

Of the 23 galaxies in the dusty ETG sample, eight have bright H$_{2}$\ emission, relative to both their TIR and PAH luminosities. Most of the ETGs show indications of artificially high H$_{2}$/7.7\um\ ratios due to small-grain depletion by AGN or a soft exciting radiation field, as evidenced by 7.7/11.3 PAH band ratios significantly below 1 (see \S\,\ref{sec:evo} \& \ref{sec:agn}). This effect is corrected by deriving a 7.7\um\ using the geometric mean 7.7/11.3 for the E+A sample (little-to-no depletion; 7.7/11.3 = 2.8). Even after this correction, the eight ETGs show high H$_{2}$\ relative to both 7.7\um\ and TIR --- though note that TIR luminosities are much fainter than the E+As ($\sim$10--100$\times$\ lower). It is interesting to note that these dusty ETGs all possess either nuclear dust rings or sporadic dust patches, as well as strong shells or tidal streams, both suggestive of relatively recent gas-rich mergers/interactions \citep{rampazzo2013} --- unlike ``traditional'' dust-poor pure ellipticals.

Much recent work has focused on identifying contributions to H$_{2}$\ emission not associated with the radiation field, particularly in cases where these non-radiative mechanisms dominate. \cite{ogle2010} discovered a class of H$_{2}$-emitting radio galaxies, which they dubbed molecular hydrogen emission-line galaxies (MOHEGs), all with exceptional H$_{2}$/7.7\um\ $\geqslant$\ 0.04 --- an empirically-derived radiative heating limit determined by comparing to the SINGS star-forming sample. Galaxies with H$_{2}$/7.7\um\ above the \cite{ogle2010} threshold of $0.04$\ are, then, assumed to be dominated by H$_{2}$\ non-radiative H$_{2}$\ excitation mechanisms. Of the 15 \textit{Spitzer}-observed E+As, eight sources have detected H$_{2}$\ lines which satisfy the H$_{2}$/7.7\um\ $\geqslant$\ 0.04 MOHEG criterion. 

\begin{figure*}[t]
\centering
\leavevmode
\includegraphics[width={0.95\linewidth}]{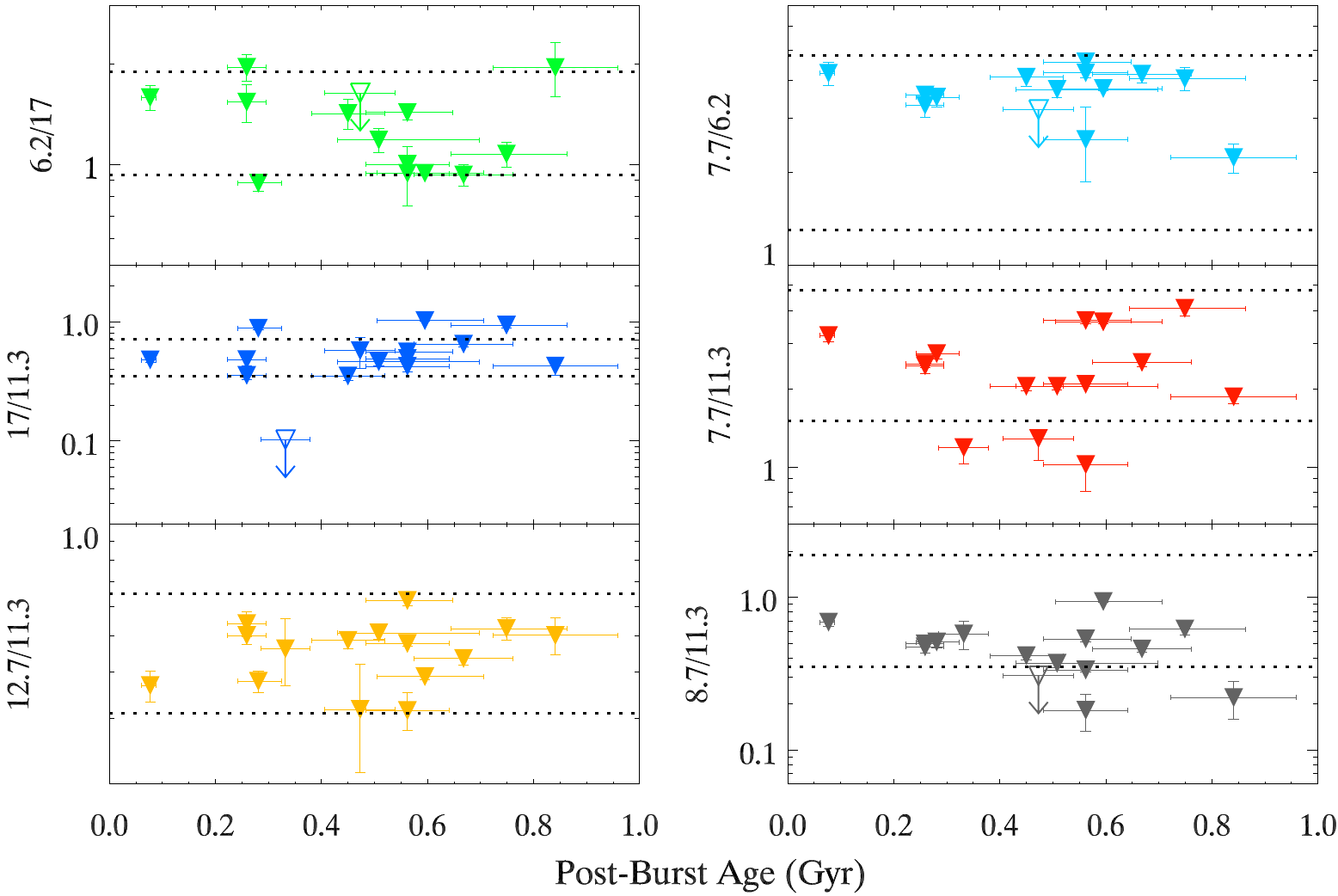}
\caption{Luminosity ratios of the 6 primary PAH features, plotted as a function of post-burst age (French et al. 2017, submitted). The dashed lines correspond to the SINGS sample 10-90\% range for the given ratio, from SDD07. The E+As appear to possess ratios in good agreement with SINGS.
}
\label{fig:bands}
\end{figure*}

\cite{cluver2013} detected a sample of MOHEGs in Hickson Compact Groups (HCGs), all with significant intragroup interactions. Several galaxies in their sample have confirmed strong AGN activity and, thus, may exhibit artificially high H$_{2}$/7.7\um\ values due to PAH grain depletion. However, they determine that the H$_{2}$\ emission present in the vast majority of their sample is driven by shock heating due to ram pressure from the ongoing intragroup interactions. The PAH band ratios of the E+A sample (see \S\,\ref{sec:evo}, Figure \ref{fig:bands}) show little evidence for preferential small-grain destruction, suggesting that shock (or turbulent) heating of the molecular gas is a viable mechanism for producing the observed emission.  These shocks are thought to be magnetohydrodynamic (MHD) and intrinsically low-velocity in nature, and particularly efficient at cooling via H$_{2}$\ rotational lines \citep{draine1983}.\footnote{While in many systems low-velocity molecular shocks and fast, radiative shocks may be connected via a turbulent cascade (e.g., Stephan's Quintet; Phil Appleton, private communication), sustained turbulent heating, resulting in significant localized gas dispersions (e.g., $\gtrsim$\,20 km\,s$^{-1}$), is all that is required to excite H$_{2}$\ rotational emission.}

It seems that for the majority of the E+As, turbulent heating is the most viable option for producing the observed strong H$_{2}$\ emission, due to consistently high H$_{2}$/7.7\um\ ratios and little evidence for small-grain depletion. Indeed, with H$_{2}$\ contributing $\sim$1\% of TIR, 0815\_586\_52374 possesses one of the highest known fractional H$_2$ emission in galaxies --- rivaled only by the intragroup shock ridges between interacting galaxies, such as those in Stephan’s Quintet \citep{cluver2010,appleton2017}, or very IR-faint post-merger ETGs. Two potential origins for the turbulent heating are (1) the effects of a recent galaxy interaction/merger, and (2) radio-mode (jet) feedback from a SMBH. Each of these mechanisms will be discussed in greater detail in \S\,\ref{sec:discuss}.

\subsubsection{Estimating Mass and Temperature}
\label{sec:aditya_h2_mass}
In the cold, dense interiors of molecular clouds, conditions do not allow for the excitation of H$_{2}$'s quadrupole transitions, leaving the bulk of the molecular gas mass essentially dark. The only reliable method of tracing this gas is via rotational transitions of the carbon monoxide (CO) molecule, which \textit{do} emit at the $\sim$10 K temperatures present deep in molecular clouds. Converting the observed CO brightness to a molecular gas mass requires a CO-to-H$_{2}$\ conversion factor, $\alpha_{\textrm{CO}}$, 
\begin{equation}
\mathrm{%
M_{H_{2}} (M_{\odot}) = \alpha_{CO}\,L_{CO}^{\prime}
}, 
\end{equation}
where $\alpha_{\textrm{CO}}$\ is in units of M$_{\odot}$\,(K km s$^{-1}$\ pc$^{2}$)$^{-1}$.

\cite{togi&smith2016}, by adopting a power-law temperature distribution, developed a method to fit and extrapolate the H$_{2}$\ excitation diagrams to derive estimates of the total H$_{2}$\ mass, independent of CO. The slope of the fitted power-law is directly related to how much of the H$_{2}$\ is warm (above a typical ``cold'' threshold of $\sim$50 K), which in turn is related to the conditions present in the ISM. We adopt this method for the E+A sample and compare to the estimates from CO. 

For robust estimates, sources must possess at least three H$_{2}$\ rotational lines with S/N $\geqslant$\ 5, one of which must be the 17.04\um\ S(1) line. The H$_{2}$\ column density is assumed to be distributed via the power-law,
\begin{equation}
dN \propto T^{-n} dT,
\end{equation}
where $dN$\ is the number of H$_{2}$\ molecules in the temperature range $T$\ to $T+dT$. The model takes three parameters --- the upper ($T_{u}$) and lower ($T_{l}$) temperature limits, as well as the power law index, $n$. $T_{l}$\ is the lowest temperature found in the range of possible temperatures required to explain the observed excitation. Using the power law index required to match the MIR rotational line fluxes and extrapolating down to $T_{l}=50$\,K, the estimated H$_{2}$\ mass is in good agreement with the CO-derived estimates for nearby galaxies \citep{togi&smith2016}. As the power law model is extrapolated down to the $T_{l}$\ ``temperature floor'', the assumption of 50\,K is perhaps the most significant model uncertainty. For example, it is possible that $T_{l}$\ could be elevated in very turbulent systems, leading to an over-estimation of H$_{2}$\ mass. 

\begin{deluxetable}{rttt}
\tablecaption{\label{tab:h2}Power Law H$_{2}$\ Model Results}
\tablecolumns{4}
\tabletypesize{\small}
\tablehead{%
\colhead{} &
\colhead{M$_{\mathrm{Mol}}$(Model)} &
\colhead{} &
\colhead{} \\
%%%
\colhead{Galaxy} &
\colhead{($10^{8}$\,M$_{\odot}$)} &
\colhead{$n$} &
\colhead{$\mathrm{\frac{M_{\mathrm{Mol}}(Model)}{M_{Mol}(CO)}}$} \\
%%%
\colhead{(1)} & 
\colhead{(2)} & 
\colhead{(3)} & 
\colhead{(4)} \\
}
\startdata
0379\_579\_51789 & 7.5 & $4.20 \pm 4.20$ & 0.62 \\
0413\_238\_51929 & $3.63 \pm 0.63$ & $4.26 \pm 0.04$ & 0.71 \\
0570\_537\_52266 & 25.0 & $4.68 \pm 4.68$ & 2.04 \\
0623\_207\_52051 & $27.5 \pm 18.8$ & $5.66 \pm 0.02$ & 7.9 \\
0637\_584\_52174 & $43.8 \pm 7.8$ & $4.33 \pm 0.04$ & 0.88 \\
0756\_424\_52577 & $6.0 \pm 1.5$ & $3.57 \pm 0.06$ & 0.10 \\
0815\_586\_52374 & $53.8 \pm 3.8$ & $4.69 \pm 0.02$ & $>$3.9 \\
0962\_212\_52620 & $65.0 \pm 2.4$ & $4.99 \pm 0.01$ & 3.7 \\
1279\_362\_52736 & $12.0 \pm 1.0$ & $4.56 \pm 0.02$ & 0.87 \\
1616\_071\_53169 & 0.25 & $2.95 \pm 2.95$ & $>$0.06 \\
1927\_584\_53321 & 0.96 & $4.34 \pm 4.34$ & $>$1.75 \\
\enddata
\tablecomments{(1) Galaxy ID, (2) power law model-derived H$_{2}$\ mass (in units of $M_{\odot}$), (3) power law index of best-fit model, (4) ratio of power law-derived to CO-derived molecular gas mass. CO estimates are taken from FYZ15 (see Table \ref{tab:sample}), assuming $\alpha_{\mathrm{CO}} = 4$.}
\end{deluxetable}

Of the 15 \textit{Spitzer} sources, 11 possess at least 3 H$_{2}$\ lines detected at 5$\sigma$. Figure \ref{fig:excitation_diagrams} shows the excitation diagrams for the 11 selected galaxies. We compare the model estimates to the FYZ15 CO-derived masses, for which an $\alpha_{\textrm{CO}} = 4$\,M$_{\odot}$\ (K km s$^{-1}$\ pc$^{2}$)$^{-1}$\ was assumed. 

\begin{figure*}[t]
\centering
\leavevmode
\includegraphics[width={0.95\linewidth}]{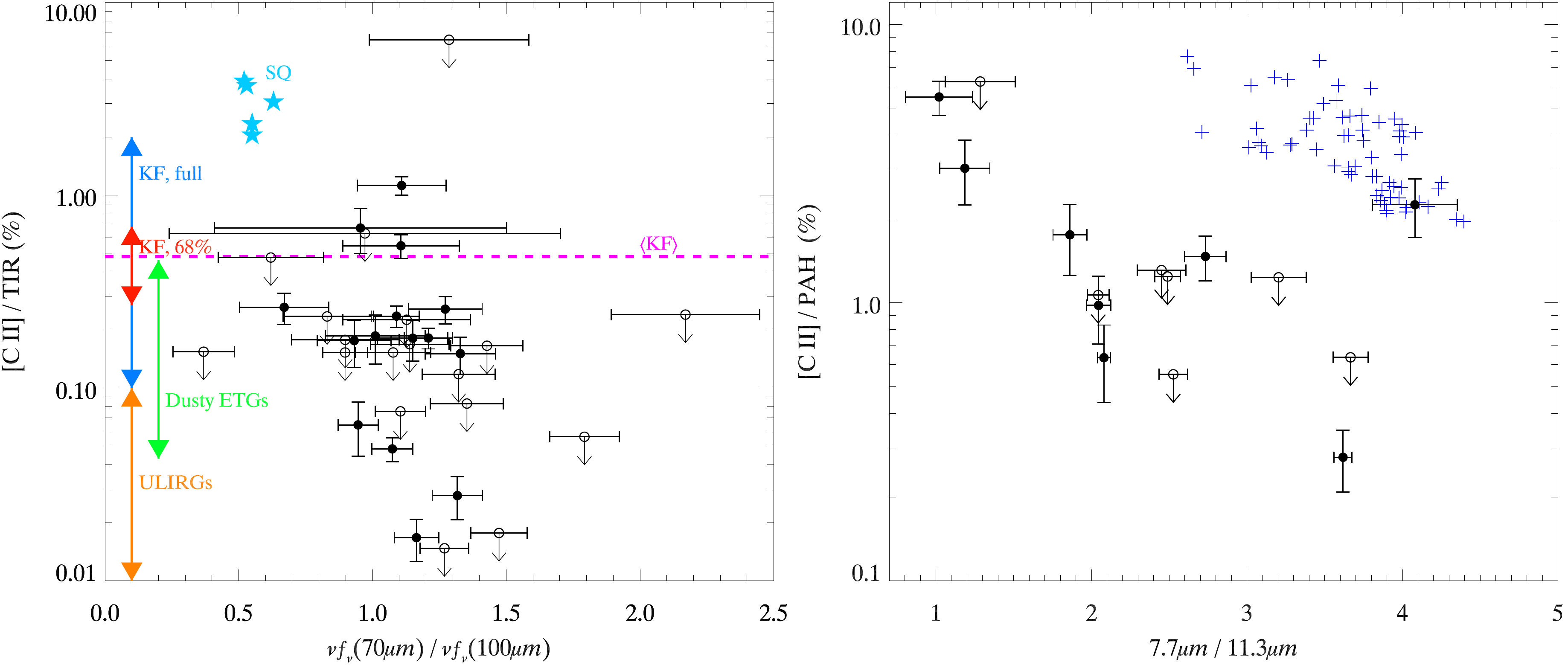}
\caption{\uline{Left}: The \cII\ deficit (\cII/ TIR) plotted as a function of 70\um/100\um\ color. The E+As are shown as black, filled circles, where open circles with downward arrows denote 3$\sigma$\ upper limits. The double-headed arrows denote the ranges of various comparison samples: the KINGFISH \citep{smith2017} full range (blue), 68\% range (red), and mean (magenta), dusty ETGs (green), and GOALS ULIRGs (orange; 
\citealt{diaz-santos2013}, assuming a multiplicative factor of 2 for FIR-to-TIR conversion). We also show individual regions of the inter-galaxy shock ridge in Stephan's Quintet (light blue stars; \citealt{appleton2013}). \uline{Right}: \cII-to-total PAH emission as a function of the 7.7\um/11.3\um\ PAH band ratio. The E+As follow the same schema as the left figure. The blue plus symbols correspond to resolved regions within the star-forming galaxies NGC 1097 and NGC 4559 \citep{croxall2012}. The E+As seem to posses significantly deeper \cII\ deficits than most normal, star-forming galaxies, relative to both TIR and PAH emission.}
\label{fig:deficit}
\end{figure*}

The resulting masses are given in Table \ref{tab:h2}, along with the ratio of model-to-CO-based masses. The power-law model masses are in good agreement with the estimates from CO, to within a typical factor of 2--4 (similar to the model uncertainties), for all but 3 sources. Cases where the model estimate is quite low (e.g., 0756) could be explained by a decreased $\alpha_{\textrm{CO}}$, as is typically assumed for turbulent conditions in ULIRG molecular clouds. However, cases where the model estimate is high are harder to explain (e.g., 0623, 0815). The most likely explanation is that these sources have elevated H$_{2}$\ temperature floors above 50\,K. Indeed, this seems to be supported by these two sources' high H$_{2}$/7.7\um\ ratios (both among the top three in the sample), indicating particularly turbulent ISM. The derived power-law slopes are shallower than those found for SINGS galaxies \citep{togi&smith2016} and comparable to those of ULIRGs (Togi, private communication) and turbulent systems such as Stephan's Quintet \citep{appleton2017}, indicating that much of the gas is warm. 

\subsection{FIR Line Deficit}
\label{sec:firdeficit}
The primary heating source of the diffuse ISM is thought to be the photoelectrons liberated from small dust grains, such as PAHs, by ionizing radiation. The photoelectric heating efficiency of the gas is defined as the ratio of gas heating from photoelectrons to total dust heating from the absorption of UV and optical photons \citep{mochizuki2004}. The gas then cools via emission from collisionally-excited infrared forbidden lines such as \cII, \nII, \oI, and \siII. Of these lines, the \cII\ 158\um\ and \oI\ 63\um\ lines dominate in low and high-density environments, respectively \citep{wolfire2003}. These two transitions are thought to be responsible for $>$90\% of neutral gas cooling \citep{kaufman2006}. \cite{malhotra2001} showed that the photoelectric heating efficiency decreases in galaxies with higher dust temperatures --- a measure of ISRF intensity. When the ISRF intensity is high enough such that the PAH photoionization rate is elevated above the electron capture rate, the ionization potential increases, thus decreasing photoelectric yields \citep{croxall2012}. As the cooling-line emission is directly proportional to the photoelectric efficiency, a decrease in photoelectric efficiency results in a deficit of cooling-line emission.

\oI\ is only detected (very modestly; above 3$\sigma$) in three sources, while 17/33 sources are detected in \cII. In Figure \ref{fig:deficit} we show the \cII/TIR ratio for our sample, as a function of 70\um/100\um\ FIR color (a measure of the dust temperature). Most galaxies in our sample show a significant deficit in \cII\ emission, indicative of inefficient gas heating. There appears to be a modest negative trend with 70\um/100\um\ color. Though the E+A sample displays a wide range of deficits with many upper limits, 85\% lie below the KINGFISH mean of 0.48\%. A significant fraction of the sample ($\sim$25--30\%, including upper limits) falls below the full KINGFISH range, overlapping with GOALS ULIRGs. Many of the E+As are, however, very consistent with the range of deficits seen in dusty ETGs from the ATLAS$^{\rm 3D}$\ sample \citep{lapham2017}, which are largely intermediate between the KINGFISH and GOALS ULIRG samples. Several sources display \textit{severe} deficits (upper limits at $< 0.02$\%) --- comparable to GOALS ULIRGs and some high-redshift sub-millimeter galaxies \citep[see][]{diaz-santos2013,smith2017}. 

The E+A sample is inconsistent with the surplus of (shock-excited) \cII\ emission seen in the individual shock regions of Stephan's Quintet (SQ; \citealt{appleton2013}) --- which are colliding at 1000 km\,s$^{-1}$. This is interesting, as the E+As' H$_{2}$\ rotational emission (see \S\,\ref{sec:h2_excitation}) approaches the emission seen in SQ. This suggests that the turbulent heating found in E+As is altogether different from the fast shocks seen in SQ (which couple the \cII\ and H$_{2}$\ cooling via a turbulent cascade) and is, instead, a primarily low-velocity phenomenon. 

We also plot the ratio of \cII/PAH emission, as a function of the 7.7\um/11.3\um~PAH band ratio. The 7.7\um~feature is thought to originate from ionized grains and the 11.3\um\ feature from neutral grains. Thus, the 7.7\um/11.3\um\ traces the grains' ionization state. In this sample, we see a strong decrease in the photoelectric efficiency (\cII/PAH) as the grains become more ionized (7.7\um/11.3\um), similar to trends seen in nearby star-forming galaxies \citep{croxall2012}. However, the post-starbursts lie \textit{systematically} $\sim$10$\times$\ below the trend for nearby star-forming galaxies, indicating particularly low photoelectric heating. 

Compared to resolved regions within NGC 1097 \& 4559 \citep{croxall2012}, the three sources detected in \oI\ are all deep outliers in the cooling-line ratio (\cII/\oI) --- as low as 0.04 in one case. The total line luminosity relative to TIR in these cases --- (\cII\ + \oI)/TIR  --- is high, with the two \cII\ detections possessing (\cII\ + \oI)/TIR $>$\ 0.01. Additionally, both of those sources possess deep \cII\ deficits, below 0.1\%. These characteristics --- deep \cII\ deficit and dominant \oI\ --- are typically seen in ULIRGs, with dense-gas dominated ISM. Such dominant \oI\ is expected in MHD-driven molecular shocks, where carbon is largely neutral \citep{draine1983} --- consistent with the properties of H$_{2}$\ emission discussed in \S\,\ref{sec:h2_excitation}.

\subsection{Time Evolution of the ISM}
\label{sec:evo}
The derivation of the E+As' post-burst ages involves the decomposition of their UV-optical SEDs and detailed stellar population modeling, described in an upcoming paper by French et al. (2017, submitted). Many other works have investigated the evolution of the ISM in PSB galaxies (e.g., colors, AGN activity, metallicity; \citealt{yesuf2014}, \citealt{alatalo2014}, \citealt{alatalo2017}). Here we examine relationships between the properties of PAH emission and ISM content of the E+A sample with post-burst age.  

As described in \S~\ref{sec:pah}, PAH emission is closely linked to the ISRF. The primary emission features (6.2, 7.7, 8.6, 11.3, 12.7, and 17 \um) are thought to arise from grains of different characteristic sizes and ionization states \citep{draineli2007}. As the stellar populations of post-burst galaxies age, the radiation field heating dust grains should become softer. Thus, we might expect to observe a time evolution of E+As' dust emission properties with post-burst age. In Figure \ref{fig:bands}, we show luminosity ratios of the primary PAH bands, each as a function of post-burst age. Surprisingly, all ratios remain nearly flat with age and most lie comfortably within the SINGS 10--90\% ranges. When discounting the three highest-value sources, there appears to be a \textit{very} slight negative trend with age in the shortest-wavelength bands, similar to the trend observed in \cite{roseboom2009}. 

\begin{figure}[t]
\leavevmode
\centering
\includegraphics[width={0.95\linewidth}]{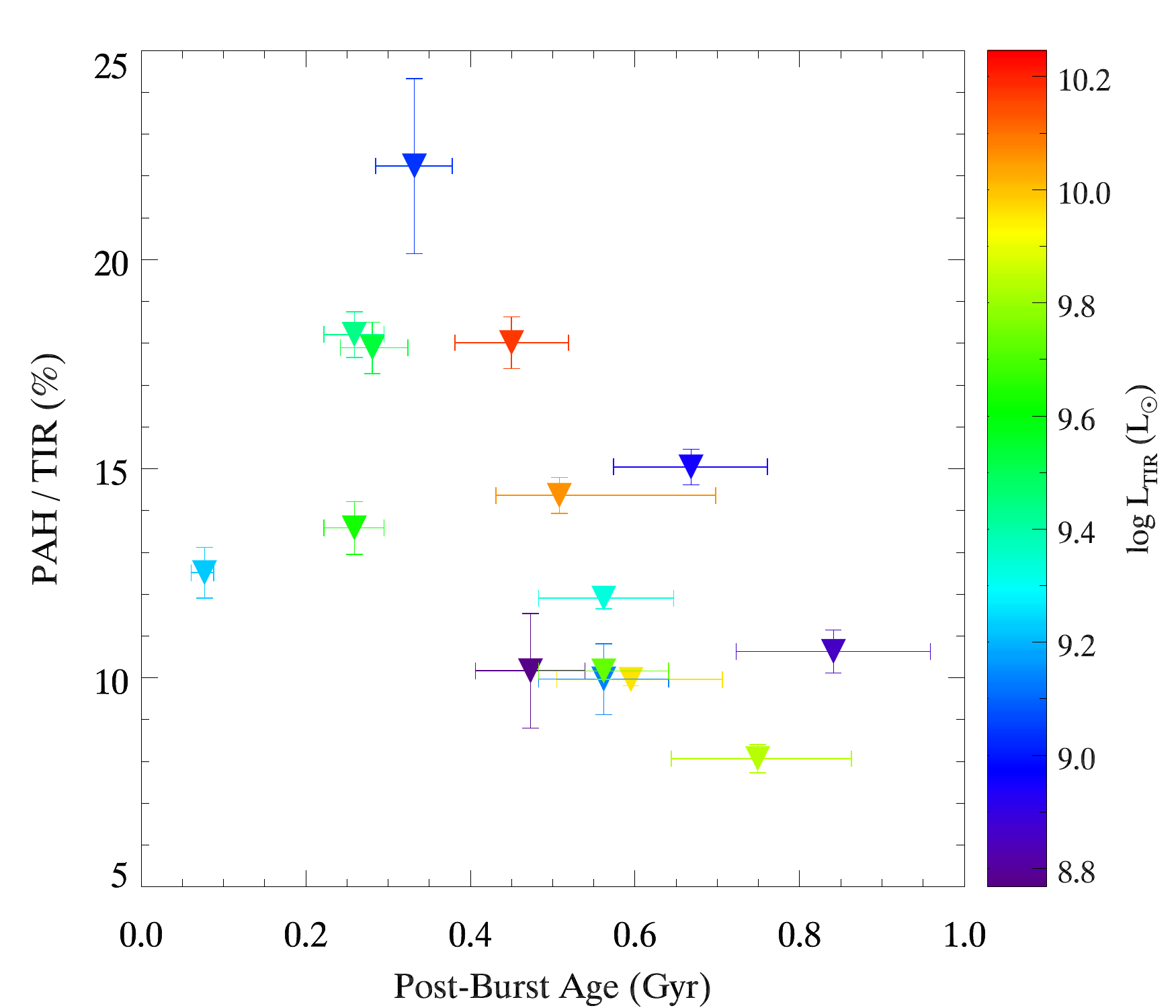}
\caption{Total PAH/TIR, plotted as a function of post-burst age (French et al. 2017, submitted). Sources are color-coded by TIR luminosity. A generally decreasing trend is seen, with the oldest sources displaying the lowest PAH emission, relative to TIR.
}
\label{fig:pahtir_age}
\end{figure}

In Figure \ref{fig:pahtir_age}, we plot fractional PAH emission (PAH/TIR) as a function of post-burst age. We observe a generally strong negative trend of PAH/TIR with burst age, but with an apparent peak occurring at $\sim$400 Myr. Suppression of PAH/TIR is expected in the presence of grain destruction --- PAHs are very small and much more easily destroyed than larger grains, while larger grains contribute the bulk of the infrared emission. Thus, PAHs will contribute fractionally less to TIR in the presence of grain destruction. Interestingly, it seems that the effects of grain destruction accumulate with age in the E+As. In fact, the oldest E+As approach the PAH/TIR $<$\ 10\% seen in the dusty ETG sample.

It has been suggested that a time delay should exist between star formation and PAH emission. \cite{galliano2008} proposed that following a burst of star formation, PAH emission should be absent for the first $\sim 400$\ Myr, until PAH production begins in the atmospheres of carbon-rich asymptotic giant branch (AGB) stars, and use this to explain the relative weakness of PAH emission in low-metallicity starbursting dwarf galaxies. While we do see some evidence for a peak in PAH/TIR at ages of $\sim$400 Myr, more recent work by \cite{sandstrom2010} showed no correlation between the distribution of PAH emission and AGB stars in the Small Magellanic Cloud (SMC). They do observe a strong correlation between the distribution of PAH and CO emission, potentially indicating that PAH grain growth occurs predominantly in the cold, dense interiors of molecular clouds. \cite{zhukovska2016} (and references therein) also argue that galaxies' dust production in general is dominated by interstellar grain growth in the cold neutral medium, rather than stellar-related production (e.g., winds \& supernovae). They find that in a typical star-forming galaxy, grain destruction by supernovae and growth in the ISM should reach a ``steady-state'' in $\sim$140 Myr (after the onset of hydrodynamic simulations). They do suggest that carbonaceous seed grains supplied by AGB stars, on which gas-phase metals from the ISM could accumulate, could accelerate dust regrowth during early galactic evolution. 

\begin{figure*}[t]
\centering
\leavevmode
\includegraphics[width={0.95\linewidth}]{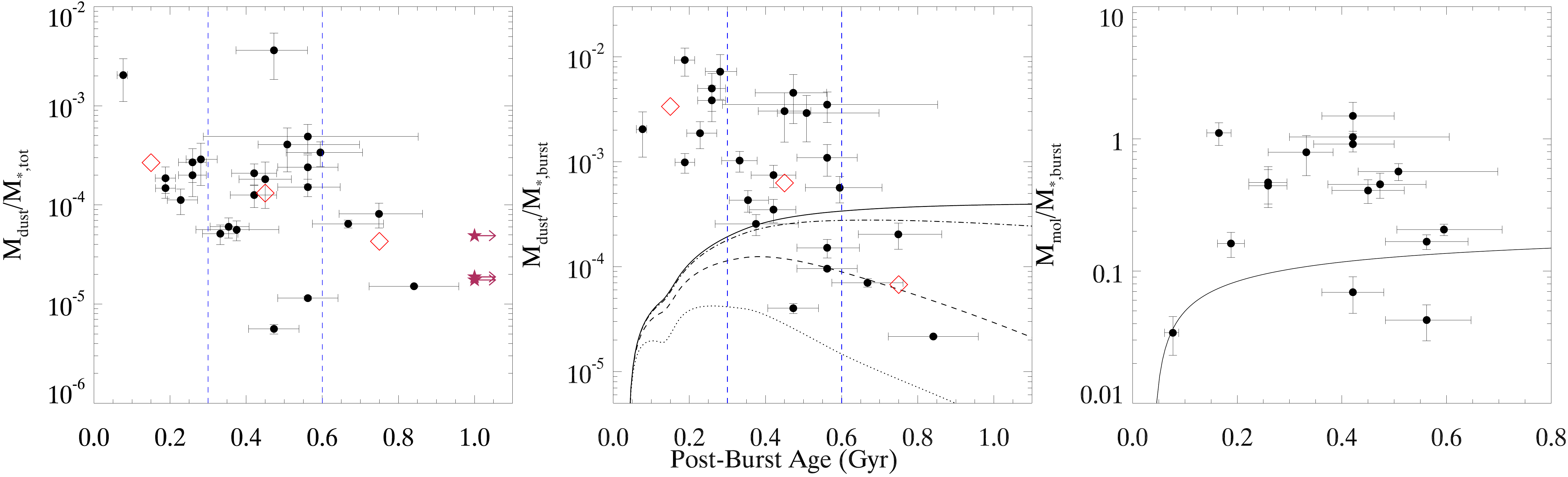}
\caption{\uline{Left}: Ratio of dust mass per unit total stellar mass, as a function of time since the burst (post-burst age; French et al. 2017, submitted). Only sources with M$_{\mathrm{dust}}$/M$_{\mathrm{*,burst}}$\ $\geqslant$\ 2$\sigma$\ are shown, where $\sigma$\ is the total propagated uncertainty from each quantity. The red diamonds are the median M$_{\mathrm{dust}}$/M$_{\mathrm{*,tot}}$\ values in three age bins: 0--300 Myr, 300--600 Myr, and $>$600 Myr, the locations of which are denoted by the blue dashed lines. The three magenta stars are ETGs in SINGS with M$_{*,tot} > 10^{10} M_{\odot}$\ and well-measured dust masses. Their stellar populations are assumed to be old, though uncertain, denoted by the right-facing arrows. \uline{Center}: Dust mass per unit burst stellar mass, as a function of post-burst age. Only sources with M$_{\mathrm{dust}}$/M$_{\mathrm{*,burst}}$\ $\geqslant$\ 2$\sigma$\ are shown. The red diamonds are the median M$_{\mathrm{dust}}$/M$_{\mathrm{*,burst}}$\ values in the three age bins. The four curves are the theoretical AGB dust models of \cite{zhukovskaPhD} (solid -- no destruction, dash-dot -- low-density, dashed -- 10$^6$\,K, dotted -- 1.5$\times 10^7$\,K; see \S\,\ref{sec:evo}). The declining trend is even more visible here. \uline{Right}: The ratio of observed molecular gas mass per unit burst stellar mass, also plotted as a function of post-burst age. Only sources detected in CO ($\geqslant$3$\sigma$) and with M$_{\mathrm{Mol}}$/M$_{\mathrm{*,burst}}$\ $\geqslant$\ 2.5$\sigma$\ are shown. The solid curve is an updated theoretical model from \cite{zhukovskaPhD}, showing the total injection of gas from stellar evolution, for the modeled burst, over time. For reference, by $\sim$300 Myr the formed stellar population has returned $\sim$10\% of its stellar mass to the ISM. As the vast majority of E+As lie above the curve, they appear to possess significantly higher gas masses than can be explained by stellar recycling alone.}
\label{fig:agbrec}
\end{figure*}

In Figure \ref{fig:agbrec}, we plot the observed dust mass and molecular gas mass for the sample, each normalized to the modeled stellar mass produced in the burst, as a function of post-burst age. To assess the contribution of dust produced solely from the aging stellar population, we employ theoretical dust yield models for single stellar populations, including AGB stars. The models are calculated for solar metallicity as described in \cite{zhukovskaPhD}, assuming a \cite{kroupa2001} IMF. Dust yields for AGB stars are taken from \cite{zhukovska2008}. Models of dust destruction in hot gas are also included. In conditions found in early-type galaxies, dust grains are efficiently destroyed by thermal sputtering on the timescale $\tau_{\rm spu} = 10^{5}$\ $(1 + (10^6{\, \rm K}/T_{\rm gas})^3)$\ $n_{\rm e}^{-1} {\rm \,yr\,}$\ for typical 0.1\um-sized grains \citep{dwek&arendt1992}. For simplicity, we consider grain sputtering in a hot gas with the fixed density and temperature for three cases: (1) $n_{\rm H} = 0.002$\,cm$^{-3}$, $T_{\rm e} = 1.5 \times 10^7$\,K, (2): $n_{\rm H} = 0.001$\,cm$^{-3}$, $T_{\rm e} = 10^6$\,K, and (3) a lower density gas with $n_{\rm H} = 7\times 10^{-5}$\,cm$^{-3}$, $T_{\rm e} = 1.5\times 10^7$\,K, similar to the outer regions of elliptical galaxies \citep{mathews&brighenti2003}. Dust is rapidly destroyed in the ISM in the first two cases on the timescales of 50 and 200 Myr, respectively, and on the longer timescale $\tau_{\rm spu} =1.5$\ Gyr in the third case. Possible dust destruction by blast waves from type Ia SNe is neglected.

We observe a steep negative trend of dust-to-burst stellar mass with age (M$_{\rm dust}$/[$f_{\rm *,burst}$\ M$_*$]), similar in slope to the two high-density sputtering models. Like the trend with PAH/TIR, this implies that the effects of dust destruction continue to accumulate in E+As, rather than reaching the steady-state predicted by \cite{zhukovska2016}, and potentially points to sputtering in a hot, low-density medium. It should be noted that older post-burst populations are more likely to have originated from longer, stronger bursts \citep{snyder2011}, thus creating an intrinsic anti-correlation between age and 1/$f_{\rm *,burst}$\ (see French et al. 2017, submittd) even disregarding M$_{\rm dust}$. However, this effect is much weaker than the observed trend. Indeed, the trend remains even when considering M$_{\rm dust}$/M$_{\rm *,tot}$, though with higher scatter. All sources younger than 400 Myr possess considerably higher measured dust masses than can be explained by recycled input from AGB stars. Older sources, however, lie closer to the curve. However, all sources possess more dust than can be explained by AGB star input when accounting for destruction using either of the two high-density sputtering models. 

In terms of the gas produced by stellar recycling, due to limited H\,I observations, we do not have total gas mass. But using molecular mass as a lower gas mass limit, we note that, although there is little observable trend in the molecular gas-to-burst stellar mass with age, the majority of sources lie significantly above the \cite{zhukovskaPhD} model. For these sources, it seems that at least a significant fraction of the ISM must be preexisting, rather then regrown via stellar recycling. This has interesting implications for the evolution of these galaxies, discussed further in \S\,\ref{sec:discuss}. There are several sources, however, which fall below the model prediction, implying that in these few cases the ISM may, indeed, have been regrown by gas injection from the burst stellar population. 
 
\subsection{AGN Diagnostics}
\label{sec:agn}
This sample was optically selected to possess weak-to-nonexistent \oIII\ emission, thus attempting to eliminate sources with strong AGN. Additionally, eight of the E+As have existing \textit{Chandra} coverage. \cite{depropis2014} used the X-ray luminosities to place limits on AGN luminosity in these eight sources at $<$\,0.1\% of the Eddington luminosity --- consistent with our \oIII\ selection. However, the sample possesses significant dust reservoirs and a range of extinction, introducing the possibility of obscured AGN which could contribute significantly to the ISM energetics --- as have been found in other PSB studies \citep[e.g.,][]{ko2013,yesuf2014,alatalo2017}. Additionally, in \S\,\ref{sec:h2_excitation} we revealed that the majority of galaxies in the sample display strong, shock-powered H$_{2}$\ rotational emission. AGN feedback has been proposed as a potential mechanism for eliciting similar emission in some early-type galaxies, motivating a closer study in the E+A sample. In this section we investigate various AGN indicators to probe the possibility of both obscured and radio-mode AGN. We find evidence for one deeply embedded AGN, and some evidence for excess radio activity in an additional three sources. In the vast majority of sources (29/33), however, we find no compelling evidence of the presence of an AGN. 

\subsubsection{Probing Obscured AGN}
\label{sec:agn_obscured}
Several high-ionization forbidden transitions exist in the MIR and serve as extinction-insensitive probes of the presence of AGN activity: mainly \neV\ at 14.32\um\ and \oIV\ at 25.96\um. \neV\ is absent in all of our IRS spectra, ruling out the presence of strong AGN. However, at low luminosities \neV\ often becomes blended with the 14.3\um\ PAH feature during decomposition, making it difficult to rule out the presence of low-luminosity AGN (SDD07).

\oIV\ is more reliable at low luminosities, however it is also blended, this time with \feII\ (SDD07). \cite{satyapal2004} calibrated the ratio of \oIV/\neII\ as a LINER/AGN indicator. In the single source where both lines are detected (0756\_424\_52577), $\log_{10}$(\oIV/\neII) $= -0.55$. If the detected line is entirely \oIV, this ratio could be consistent with either very low-level AGN or LINER activity. However, the presence of a luminous, buried AGN is highly unlikely. 

0962\_212\_52620, the source with detected silicate absorption, is not detected in \oIV. To more robustly probe the presence of a buried AGN, we use the classification scheme of \cite{spoon2007}: combining the apparent silicate strength with the 6.2\um\ equivalent width (EQW). Silicate strength is defined as, 
\begin{equation}
S_{\mathrm{Sil}} = \ln \frac{f_{\mathrm{obs}}(9.7\mathrm{\micron})}{f_{\mathrm{cont}}(9.7\mathrm{\micron})}, 
\end{equation}
where $f_{\mathrm{obs}}$\ and $f_{\mathrm{cont}}$~are the observed 9.7\um\ flux density and the local MIR continuum flux density, evaluated at 9.7\um, respectively. We estimate $S_{\mathrm{Sil}} \approx -1.5$\ for 0962, combined with a 6.2\um\ EQW of $0.78$\um. These place 0962 in \cite{spoon2007} class 2C --- populated primarily by PAH-dominated ULIRG spectra. Thus, we find no evidence of a buried AGN in any of the \textit{Spitzer} sources.  

Another potential PAH-based AGN indicator can be found in the 7.7\um/11.3\um\ PAH band ratio, which is tied to ionization state and grain size distribution. SDD07 observed a strong suppression in 7.7/11.3 with increasing \neIII/\neII, which was later confirmed in larger surveys (e.g., \citealt{diamond-stanic2010}). The majority of SINGS galaxies with 7.7/11.3 $<$\ 3 are either optically confirmed AGN or LINER types. The majority of the E+As have 7.7/11.3 $<$\ 3, consistent with their optical and few reliable \oIV/\neII\ LINER classifications. Three sources possess 7.7/11.3 $<$\ 1.5, below the lowest 10\% of the SINGS sample. However, in this weak suppression regime, LINERs and true AGN are virtually indistinguishable. When combined with the modest \oIV\ detections and, more often, non-detections, the 7.7/11.3 ratios do not provide any further evidence of strong buried AGN and, instead, further support their optical LINER classification. 

\begin{figure}[t]
\centering
\leavevmode
\includegraphics[width={0.9\linewidth}]{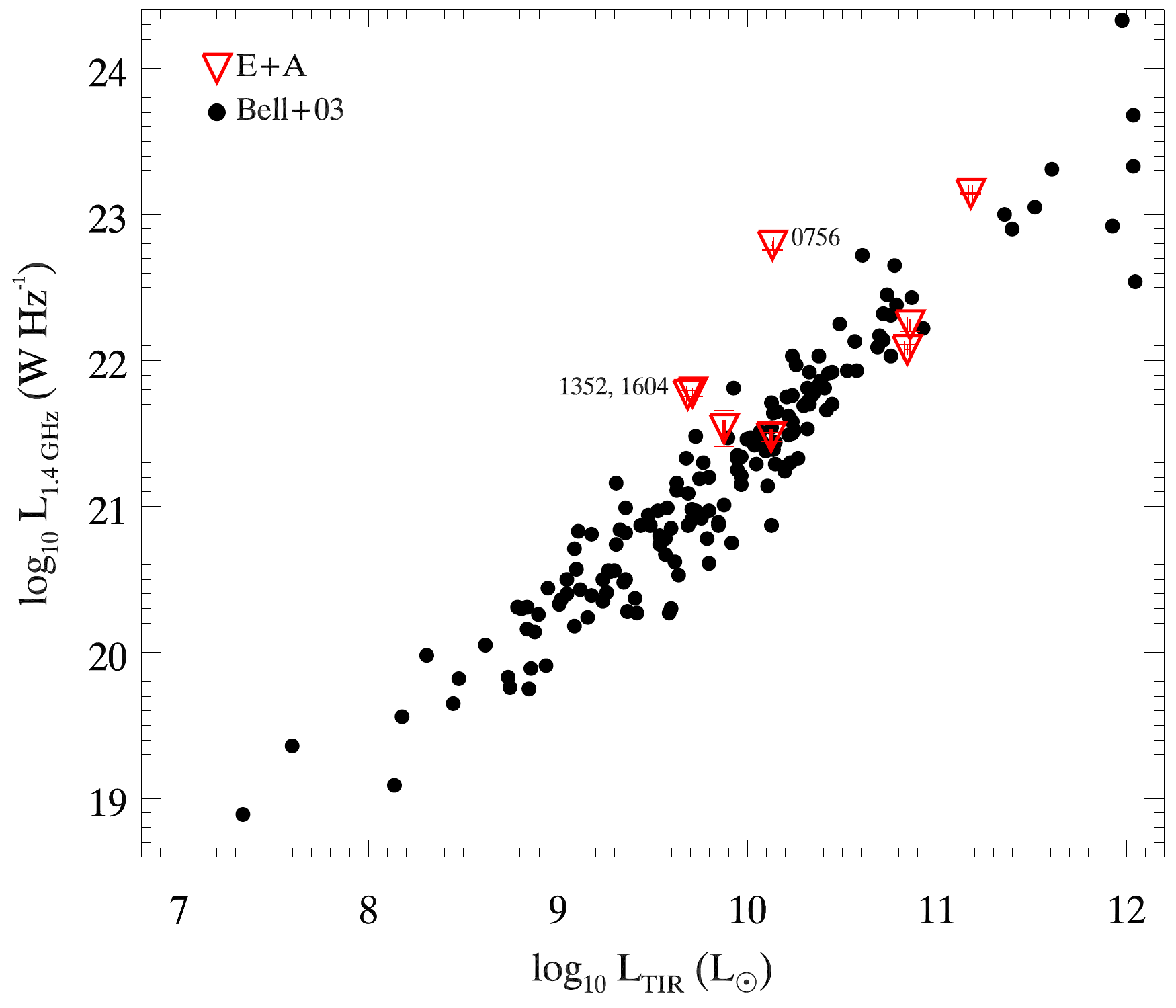}
\caption{The radio-infrared correlation. For the eight E+As possessing FIRST and/or NVSS coverage, their 1.4 GHz luminosity density is plotted against their TIR luminosity (blue triangles). The diverse sample of star-forming galaxies from \cite{bell2003} is plotted for comparison (black circles). The three sources with apparent 1.4 GHz excess are labeled.}
\label{fig:radio-ir}
\end{figure} 

Though not observed with \textit{Spitzer} IRS, one \textit{Herschel}-only source (2360\_167\_53728) displays extreme WISE colors of $[3.4] - [4.6] = 1.8$\ and $[4.6] - [12] = 2.8$, which place it at the top of the WISE AGN selection region from \cite{jarrett2011}. Its WISE photometry is best matched by the NIR/MIR spectrum of the ULIRG IRAS 08572+3915 (Jarrett, private communication) --- an obscured AGN-host. Indeed, as seen in Fig.~\ref{fig:sed}, the rise from 3.4\um\ to 4.6\um\ is poorly fit by the \cite{draineli2007} starlight + dust SED model, potentially indicating the presence of a very hot dust component peaking at $\sim$5--10\um. However, the TIR luminosity of 3.3$\times$10$^{10}$\,L$_{\odot}$\ is 50$\times$\ lower than that of IRAS 08572+3915, making it a truly peculiar source. 

\subsubsection{Radio Activity}
\label{sec:radio}
Six of the E+As were detected in the Very Large Array (VLA) Faint Images of the Sky at Twenty-Centimeters (FIRST) survey \citep{becker1995}, and another two in the NRAO VLA Sky Survey (NVSS; \citealt{condon1998}). Using the observed 1.4 GHz flux densities, we compare the E+As to the radio-infrared correlation --- an empirical relationship between star-forming galaxies' infrared and radio luminosities \citep{dejong1985,helou1985}. In star-forming systems, the radio emission is thought to arise from non-thermal synchrotron emission caused by the acceleration of cosmic rays in supernova remnants \citep{condon1992}.

It is worth first noting that the detection rate of the E+As in FIRST is significantly lower than for normal star-forming galaxies --- their radio luminosities are intrinsically lower. Among a sample of 258 SDSS DR14-selected star-forming galaxies --- L$_{\rm H\alpha} \gtrsim 10^{7} L_{\odot}$, or SFR $\gtrsim$\ 1 $M_{\odot}$\,yr$^{-1}$\ \citep{kennicutt1994} --- with redshifts $0.049 < z 0.051$, the FIRST detection rate is $\sim$35\%. In comparison, the E+A detection rate is only 18\% (6/33) --- significantly lower. 

We plot the eight E+As' radio and TIR luminosities in Figure \ref{fig:radio-ir}, against the star-forming comparison sample of \cite{bell2003}. Four of the E+As lie within the locus of the star-forming comparison sample. Three sources (0756\_424\_52577, 1352\_610\_52819, and 1604\_161\_53078), however, appear to be offset --- 0756 significantly so. 

Radio excess is commonly interpreted as evidence for the presence of a radio AGN. Both sources detected in FIRST (1352, 1604) are unresolved, with a resolution of $\sim$5\arcsec. For 1352, this is comparable to the optical size. However, 1604's optical size is 12\farcs5 --- much larger than the unresolved 1.4 GHz emission. Though a true central point source cannot be determined, AGN activity remains a plausible explanation. 

0756's position relative to the radio-IR correlation shows a significant excess at 1.4 GHz, a generally unequivocal indication that a radio AGN is present in the system. However, this particular galaxy was detected in NVSS, which possesses a much larger beam size --- FWHM $\simeq$\ 45\arcsec. SDSS imaging shows that this galaxy is interacting with another galaxy (not targeted by SDSS), with a separation significantly smaller than the 45\arcsec\ beam, making it possible that some of the observed emission is associated with the companion. 0756 is also the only source with detected \oIV, as discussed in the previous section.

\subsection{Star Formation}
\label{sec:sfr}
The stellar population modeling effort detailed in French et al. (2017, submitted) reveals that the E+As possessed, on average, peak SFRs in excess of 100\,M$_{\odot}$\,yr$^{-1}$, during the height of their past starbursts. Limits from H$\alpha$\ suggest that these galaxies possess current SFRs which are $>$100$\times$\ lower than their former peaks. However, the sample was selected to possess very weak H$\alpha$, introducing a potential bias. To robustly assess the current star-forming properties and, thus, post-starburst nature of the sample, we adopt three different infrared SFR calibrations, using a mixture of photometric and emission line-based tracers, each discussed below. SFRs for each galaxy, using each tracer, are given in Table \ref{tab:sfr}. Many of these SFRs may be upper limits, however, due to the presence of an unusual, old starlight-dominated radiation field. See \S\,\ref{sec:sfr_lim} and \ref{sec:discuss} for further discussion of these limitations. 

\subsubsection{SFR Calculations}
\label{sec:sfr_calc}
We adopt the TIR SFR calibration of \cite{murphy2011}, 
\begin{equation}
\mathrm{%
SFR(M_{\sun}\,yr^{-1}) = 3.88 \times 10^{-44} L_{IR}(erg\,s^{-1})
},
\end{equation}
adopting a factor of L$_{\TIR}$/L$_{\IR}$\ = 1.06 to convert from 3–1100\um\ TIR to 8–1000\um\ IR luminosity. 

\begin{deluxetable}{rtttt}
\tablecaption{Global Star Formation Rates\label{tab:sfr}}
\tablecolumns{5}
\tabletypesize{\scriptsize}
\tablehead{%
\colhead{Galaxy} &
\colhead{TIR} &
\colhead{12\um} &
\colhead{\cII\,158\um} &
\colhead{\neII + \neIII} \\
%%%
\colhead{(SDSS)} &
\colhead{(M$_{\odot}$\,yr$^{-1}$)} &
\colhead{(M$_{\odot}$\,yr$^{-1}$)} &
\colhead{(M$_{\odot}$\,yr$^{-1}$)} &
\colhead{(M$_{\odot}$\,yr$^{-1}$)} \\
%%%
\colhead{(1)} & 
\colhead{(2)} & 
\colhead{(3)} & 
\colhead{(4)} &
\colhead{(5)}
}
\startdata
0336\_469\_51999&0.46&1.20&$<$~0.59&\ldots\\
0379\_579\_51789&0.78&1.70&$<$~0.77&$<$~0.35\\
0413\_238\_51929&0.54&0.93&$<$~0.23&0.17 $\pm$ 0.03\\
0480\_580\_51989&21.15&2.40&1.89 $\pm$ 0.47&\ldots\\
0570\_537\_52266&0.71&2.51&\ldots&$<$~0.09\\
0598\_170\_52316&0.69&2.63&5.11 $\pm$ 0.48&\ldots\\
0623\_207\_52051&0.35&0.81&$<$~0.30&$<$~0.14\\
0637\_584\_52174&2.47&5.89&2.32 $\pm$ 0.64&1.77 $\pm$ 0.27\\
0656\_404\_52148&0.12&0.16&0.12 $\pm$ 0.03&$<$~0.08\\
0755\_042\_52235&1.29&2.88&$<$~0.82&\ldots\\
0756\_424\_52577&1.90&2.51&\ldots&1.71 $\pm$ 0.19\\
0815\_586\_52374&0.74&1.23&1.04 $\pm$ 0.19&0.56 $\pm$ 0.10\\
0870\_208\_52325&1.36&2.69&$<$~3.27&\ldots\\
0951\_128\_52398&0.08&0.71&\ldots&$<$~0.03\\
0962\_212\_52620&1.87&1.78&0.29 $\pm$ 0.07&0.67 $\pm$ 0.02\\
0986\_468\_52443&1.06&1.41&1.04 $\pm$ 0.13&\ldots\\
1001\_048\_52670&0.36&1.29&2.15 $\pm$ 0.23&\ldots\\
1003\_087\_52641&1.28&3.98&\ldots&\ldots\\
1039\_042\_52707&0.29&0.85&0.87 $\pm$ 0.12&$<$~0.04\\
1170\_189\_52756&1.04&0.76&1.02 $\pm$ 0.24&$<$~0.07\\
1279\_362\_52736&0.96&1.15&0.34 $\pm$ 0.11&0.30 $\pm$ 0.03\\
1352\_610\_52819&0.68&1.29&0.56 $\pm$ 0.12&\ldots\\
1604\_161\_53078&0.72&1.66&$<$~0.60&\ldots\\
1616\_071\_53169&0.16&0.17&0.61 $\pm$ 0.16&$<$~0.62\\
1853\_070\_53566&0.27&1.05&\ldots&\ldots\\
1927\_584\_53321&0.13&0.23&$<$~0.06&0.13 $\pm$ 0.01\\
2001\_473\_53493&16.83&4.47&$<$~1.61&\ldots\\
2276\_444\_53712&9.76&6.03&2.51 $\pm$ 0.34&\ldots\\
2360\_167\_53728&4.56&2.82&1.29 $\pm$ 0.37&\ldots\\
2365\_624\_53739&7.39&13.18&9.85 $\pm$ 1.60&\ldots\\
2376\_454\_53770&10.98&7.94&13.38 $\pm$ 1.68&\ldots\\
2750\_018\_54242&2.39&6.31&$<$~3.05&\ldots\\
2777\_258\_54554&10.11&2.24&\ldots&\ldots\\
\enddata
\tablecomments{(1) Galaxy ID. \\
(2) TIR-based SFR from the \cite{murphy2011} calibration. \\
(3) WISE 12\um-based SFR from Cluver et al. (2017, in preparation). \\
(4) \cII-based SFR from the \cite{delooze2011} calibration. \\
(5) Neon-based SFR from the \cite{ho2007} calibration.}
\end{deluxetable}

Additionally, we adopt the neon-based calibration of \cite{ho2007}, which estimates the SFR from MIR forbidden transitions of ionized neon (\neII\ at 12.8\um\ and \neIII\ at 15.6\um). The calibration is:  
\begin{multline}
\mathrm{%
SFR(M_{\sun}\,yr^{-1}) = 4.34 \times 10^{-41} \left[\frac{L_{\neII+\neIII} (erg\,s^{-1})}{\textit{f}_{+} + 1.67\textit{f}_{+2}} \right]
},
\end{multline}
where L$_{\rm \neII+\neIII}$\ is the sum of integrated \neII\ and \neIII\ line luminosities, and $f_{+}$\ and $f_{+2}$\ are the fractions of singly and doubly ionized neon, respectively. 

Using the CLOUDY photoionization code \citep{cloudy}, we simulated model H\,II regions\footnote{CLOUDY models emission from H\,II regions only and does not include shock emission} using realistic temperatures and densities, and adopted ionization fractions typical for Neon --- $f_{+} = 0.17-0.53$\ and $f_{+2} = 0.47-0.83$. It should be noted that the range of ionization fractions is small and thus, due to their linear contribution to the \cite{ho2007} calibration, they are not a dominant source of uncertainty in the SFRs. 

Integrated line fluxes are obtained from PAHFIT, which we find can overestimate the significance of the \neII\ line at low luminosities, due to the line's proximity to the broad 12.7\um\ PAH emission feature. Therefore, for most sources we predicate ``detections'' of the \neII\ and \neIII\ sum on the detection of \neIII\ at $\geqslant$5$\sigma$, followed by eye-inspection of the spectra. In two cases (0637 \& 0756) \neIII\ is not detected, but \neII\ is detectable by eye (also with high S/N), resulting in a S/N $> 5$~for the sum. The neon SFR calibration was determined to be self-consistent with the TIR calibration (when using identical IMFs) in the star-forming SINGS sample. 

Lastly, we adopt the \cII-based SFR calibration of \cite{delooze2011}:  
\begin{equation} 
\mathrm{%
SFR(M_{\sun}\,yr^{-1}) = \frac{\left[L_{\cII} (erg\,s^{-1}) \right]^{0.983}}{1.028 \times 10^{40}}
}.
\end{equation}
Upper limits are assessed at the 3$\sigma$\ level. 

\subsubsection{SFR Considerations \& Limitations} 
\label{sec:sfr_lim}
Recent work by \cite{hayward2014} has shown that PSBs can remain dust-obscured, with UV emission from the post-burst population leading to a boost in infrared emission unrelated to star formation for up to 1.5 Gyr after the burst. Their work suggests that TIR may overestimate the SFR of PSBs by up to two orders of magnitude, in some cases. Additionally, the adopted \cite{murphy2011} relation was calibrated on galaxies with steady star formation. Thus, if the E+As are, truly, post-burst, all SFR estimates will be expected to be upper limits. Indeed, in sources where Neon is detected, SFR(\TIR) is typically twice as high as the corresponding SFR(\neII+\neIII) and ranges from 1.1--3.2$\times$\ higher. 

\begin{figure*}[t]
\centering
\leavevmode
\includegraphics[width={0.8\linewidth}]{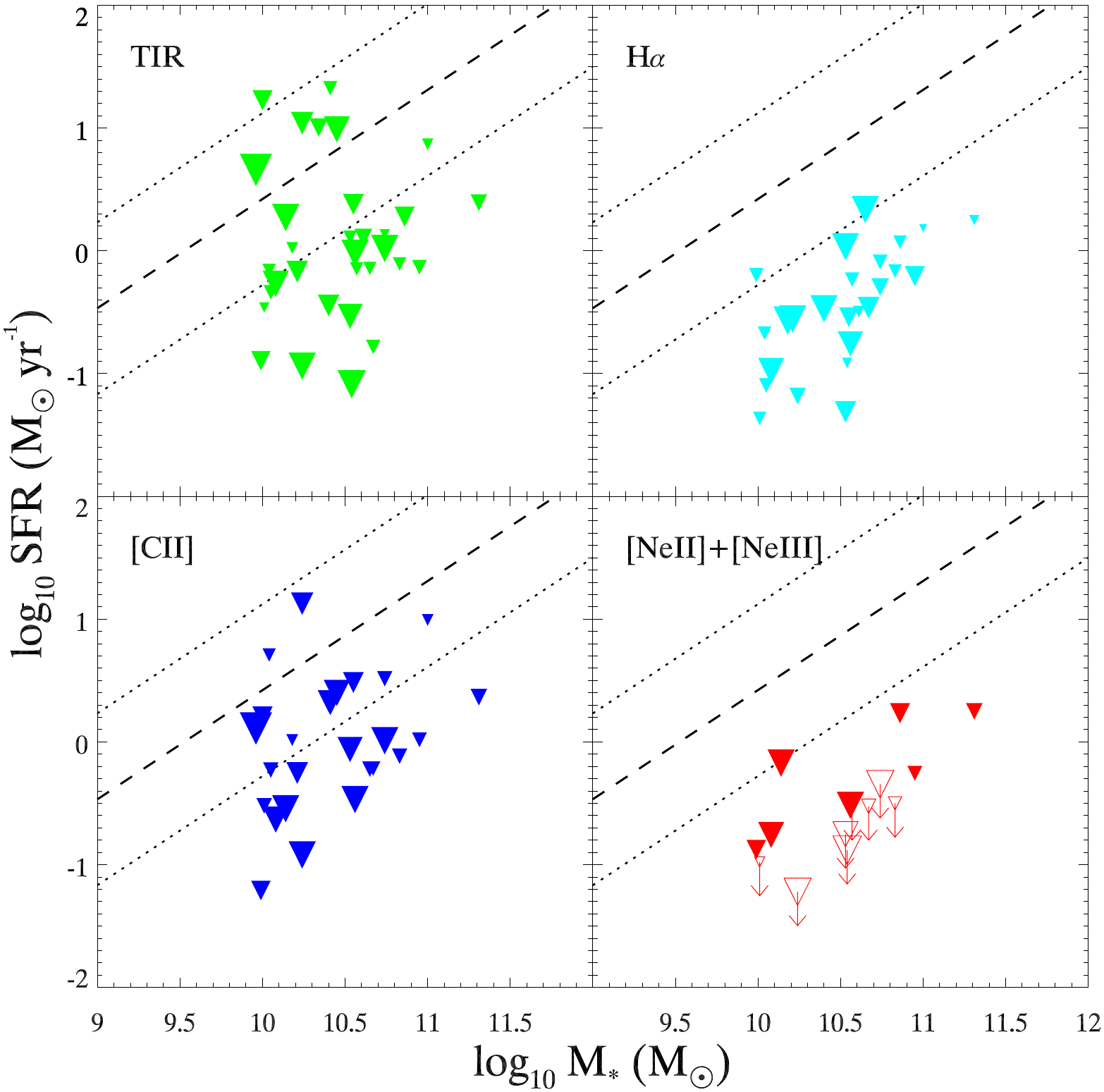}
\caption{SFR rate vs. stellar mass. For the E+A sample, we plot each of our derived SFRs in a separate panel: TIR (upper left; green), \cII\ (bottom left; blue), and Neon (bottom right; red). We also include H$\alpha$\ (FYZ15, upper right; cyan). Symbol size is scaled linearly with post-burst age (oldest have largest symbols). SFR limits are denoted by downward arrows. The dashed and dotted lines show the WISE best-fit and full range, respectively, to late-type galaxies in the GAMA G12 field \citep{jarrett2017} --- the star-forming main sequence. H$\alpha$\ SFRs have been excluded for the seven sources with high inferred extinction. Though there is considerable scatter among different tracers, the E+As overwhelmingly lie below the star-forming main sequence.} 
\label{fig:sf-ms}
\end{figure*}

Additionally, \cite{smith2017} suggest that in dense stellar environments (such as stellar bulges), the energy density of UV-optical emission is high enough to heat dust grains without a significant contribution from the FUV (e.g., star formation). These systems display \cII\ deficits which do not reflect their star formation, due to a decoupling of infrared emission and gas heating. Most of the E+As have steep surface brightness profiles \citep[e.g.,][]{yang2008,abramson2013}, indicative of dense central stellar environments, and most display \cII\ deficits similar to or even deeper than those found in the densest central bulges of the KINGFISH sample. Together these could indicate that SFRs derived from TIR and \cII\ emission are generally overestimates. In \S\,\ref{sec:discuss}, we discuss this in greater detail. 

PAH emission, as an oft-used SF indicator, was excluded on the basis of the E+A's potentially unusual radiation field. Indeed, \neII-to-total PAH emission (\neII/PAH) is 10$\times$\ lower, on average, than the SINGS sample. Additionally, individual PAH feature-based calibrations, such as those of \cite{hernnan-caballero2009}, yield SFRs which are $>$10$\times$\ higher, on average, than those based on TIR. We do, however, include SFRs based on WISE 12\um\ \citep{cluver2017}, which were calibrated using TIR. These are still an average of 3$\times$\ higher than the neon SFRs. Thus, in these sources, PAH-based SFRs are completely unreliable --- consistent with the findings of \cite{lidraine2002_1}.

Though only a subset of the 15 \textit{Spitzer} sources are detected in \neII\ and \neIII, it is perhaps the most robust of the three calibrations. However, caution is still advised. Most of the sample can be classified as LINERs via their optical emission line ratios. This was confirmed in \S\,\ref{sec:agn} using the \neII\ and \oIV\ lines. Given the weakness of even the detected lines, it is feasible that in these sources, some of the nebular emission could be attributable to a low-level nuclear source or to extended LINER (LIER) emission from evolved stars, as seen in many quiescent galaxies \citep{yan2006,herpich2016,belfiore2016}.

\begin{figure*}[t]
\centering
\leavevmode
\includegraphics[width={0.95\linewidth}]{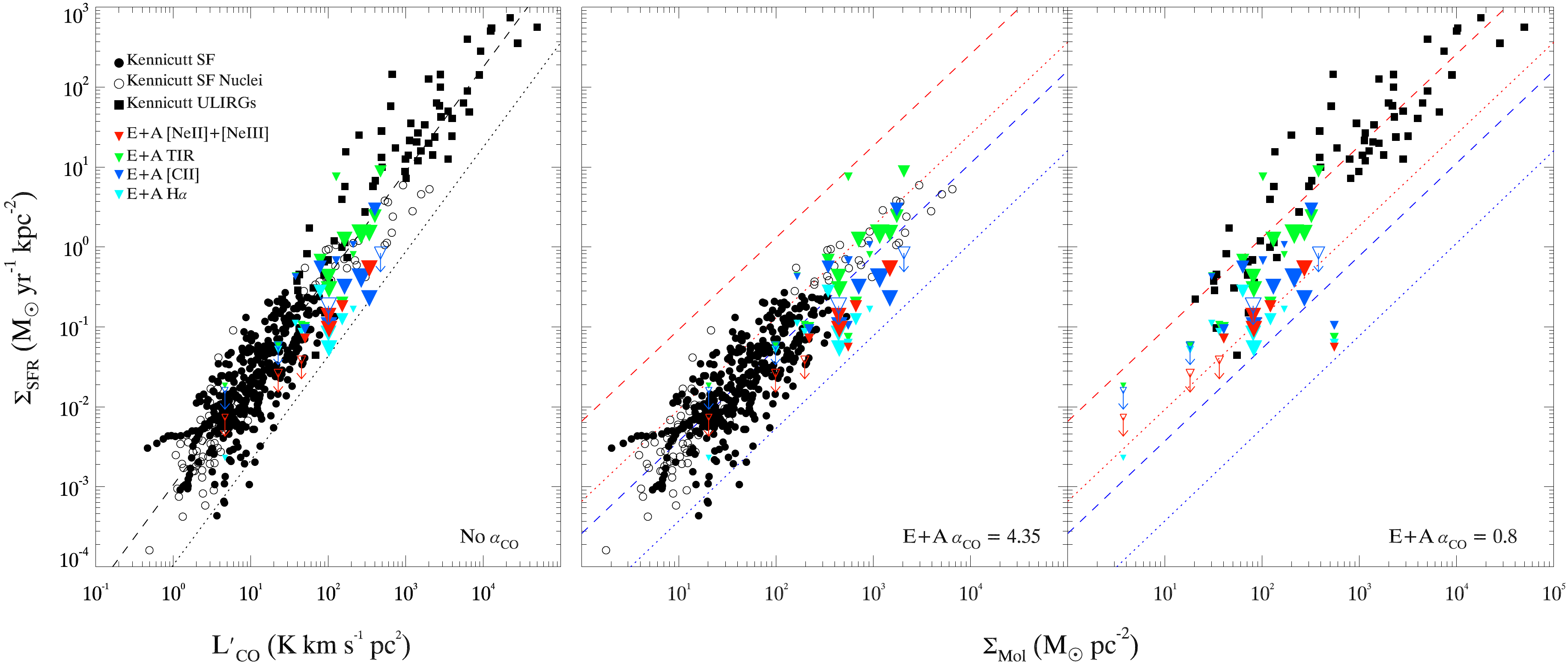}
\caption{The Kennicutt Schmidt star-formation diagram. SFR surface density is plotted as a function of CO surface brightness (left) and molecular gas surface density (center, right). Galaxies with CO limits have been excluded, as have $\Sigma_{\rm SFR}$(H$\alpha$) for the seven sources with significant inferred extinction. The one exception is 0815, which, though it is not detected in CO(1--0), does have a reliable molecular mass from fitting of its H$_{2}$\ temperature distribution (\S\,\ref{sec:h2}). The comparison sample is derived from the original sample of \cite{kennicutt1998}, composed of star-forming galaxies (black, filled circles) with $\alpha_{co} = 4.35$, star-forming galaxy nuclei (black, open circles) with $1 \le \alpha_{\textrm{CO}} \le 3.6$, and ULIRGs (black, filled squares) with $\alpha_{\textrm{CO}} = 0.8$. The E+As are plotted assuming $\alpha_{\textrm{CO}}$\ factors of: (Left) None, (Center): 4.35, (Right): 0.8. Power-law fits to the comparison samples are shown in each panel as dashed lines: SF galaxies (blue) and ULIRGs (red). Lines with 10$\times$\ lower normalization than each fit (but identical slopes) are shown as dotted lines, again with blue corresponding to SF galaxies and red to ULIRGs. Assuming $\alpha_{co} = 4.35$, the E+As are offset from SF galaxies by a factor of 5 when considering Neon SFRs. Assuming $\alpha_{co} = 0.8$, the E+As are offset from ULIRGs by a factor of 10 when considering Neon.} 
\label{fig:ks}
\end{figure*}

Additionally, classical shocks have been shown to efficiently produce nebular line emission, including \neII\ and \neIII. Though no classical shock indicators are seen in the optical emission lines, shock diagnostics are worth investigating due to the observed turbulent ISM as traced by H$_{2}$\ rotational emission. We do find that the observed \cII\ emission is inconsistent with production in fast shocks as are seen in other turbulent systems like Stephan's Quintet. To further limit the impact of shocks on ISM diagnostics such as nebular line emission, we compare the E+As' observed MIR nebular line ratios to the models of \cite{allen2008}. The \neII, \neIII, and \sIII\ lines are amply produced in fast shocks, and display distinct evolution of their emission ratios with shock velocity; low-velocity shocks produce these lines only weakly. We compute observed line ratios for the E+As and compare to the predicted ratios to examine any potential significant contribution from such low-velocity shocks. From \cite{allen2008}, the predicted ratios for a 100 km\,s$^{-1}$\ or 200 km\,s$^{-1}$\ shock are as follows: \neII/\neIII\ = 2, \neII/\sIII\ = 2.3, and \neII/\neIII\ = 1.9, \neII/\sIII\ = 2, respectively. At velocities $>$200 km\,s$^{-1}$\ \neIII\ and \sIII\ dominate over \neII. We can place limits on shocks for seven galaxies --- requiring at least one detected line and a limit on at least one of the other two lines. In 4/7 of the galaxies, we can stringently limit shock velocities to $\leqslant$\,200 km\,s$^{-1}$, while in the remaining three we can limit shocks even further, to $<$\,100 km\,s$^{-1}$. At these low velocities, shock production of \neII\ and \neIII\ is entirely negligible compared to production by the ionizing radiation field (see \citealt{allen2008}) and thus is not a dominant source of uncertainty in the derived SFRs. These velocity limits are in stark contrast to strongly shock-dominated systems, such as Stephan's Quintet (where the driving shock velocity is $\sim$1000 km\,s$^{-1}$), and are in good agreement with our \cII\ and H$_{2}$\ observations.  

SFRs for the sample were also computed by FYZ15, using the H$\alpha$\ luminosities from the MPA-JHU emission line analysis of the SDSS DR7 \citep[][see Table \ref{tab:sample} for fiber H$\alpha$\ fluxes]{aihara2011}. Of the seven sources with reliable neon-based SFRs, neon agrees with the aperture-corrected H$\alpha$\ SFRs of FYZ15 to within a factor of 2, on average. One source, 0962, shows a $\sim$10$\times$\ higher SFR in neon, compared to H$\alpha$, though still modest at $< 1$\,M$_{\odot}$\,yr$^{-1}$. This is the same source which possesses significant dust attenuation, as discussed in \S\,\ref{sec:specdecomp} and \S\,\ref{sec:agn}. 
 
\subsubsection{Efficiency of Star Formation in E+As --- Comparing to Star Formation Laws}
\label{sec:ks}
Galaxies star-forming properties are often placed in the context of empirical star formation laws. The star-forming main sequence is an empirical power-law relation between a star-forming disk galaxy's SFR and  stellar mass. Quiescent galaxies deviate from this relation, with early-types possessing far lower SFRs for their mass than a corresponding spiral galaxy. The Kennicutt-Schmidt Law (KS; \citealt{kennicutt1998}) is an empirical power-law relationship that exists between SFR and gas density in galaxies, expressed in terms of the SFR and either molecular or total gas (H\,I + H$_{2}$) surface densities ($\Sigma_{\textrm{SFR}}$, $\Sigma_{\textrm{Gas}}$). The relation is shown to hold (with relatively low scatter) across 6--7 orders of magnitude, indicating that gas density is a fundamental parameter in setting a galaxy's SFR --- likely a natural balance between feedback and gravitational collapse \citep{hopkins2013}. See the review by \cite{kennicutt&evans2012} for a more detailed discussion of observational results. 

In Figure \ref{fig:sf-ms}, we show SFRs for the E+A sample (using infrared tracers, as well as H$\alpha$) plotted as a function of stellar mass, and compare to the empirical relationship for star-forming galaxies (GAMA G12; \citealt{jarrett2017}). The E+As lie overwhelmingly below the relation for all tracers, with tracers such as \neII\ + \neIII\ and H$\alpha$\ showing the deepest suppression. In the vast majority of sources (all but the deepest outliers in Figure \ref{fig:halpha-nuv}) dust attenuation is very typical and, thus, the extinction-corrected H$\alpha$\ SFRs should be quite accurate --- as suggested by their consistency with Neon-derived SFRs (where detected).

The KS relation, alternatively, traces how efficiently a galaxy is forming stars given its existing molecular reservoirs --- independent of the galaxy's mass. Most early-type galaxies which are considered ``quiescent'' from their position compared to the star-forming main sequence do, in fact, closely follow the KS relation \citep{davis2014}. One of the limitations of the KS relation, however, is the implicit choice of a CO-to-H$_{2}$\ conversion factor (e.g., $\alpha_{\textrm{CO}}$, \S\,\ref{sec:h2}). We have little information about the $\alpha_{\textrm{CO}}$\ factor in the E+As and, thus, adopt separate conversion factors when considering different comparison samples. 

FYZ15 found significant offsets from the KS relation, using optical size estimates and SFR indicators. We re-examine these offsets in Figure \ref{fig:ks}, with our infrared size estimates and additional SFR tracers. We plot the E+A sample against the original \cite{kennicutt1998} sample, comparing to the star-forming (center) and ULIRG (right) sample separately. We follow our assumption from \S\,\ref{sec:halpha-nuv} that the spatial distribution of any ongoing star formation will likely be well-traced by the 8\um\ emission. This should also trace the molecular gas well, given the consistency of our DMGRs with typical DGRs (see \S\,\ref{sec:dmass}). Therefore, we adopt the 8\um\ FWHM sizes for computation of $\Sigma_{\textrm{SFR}}$\ and $\Sigma_{\textrm{Mol}}$. Though we argue that PAH (8\um) emission is a poor star formation indicator in E+As, the adopted 8\um\ sizes provide a stringent \textit{upper limit} on the extent of star formation.  

Compared to star-forming galaxies (using an appropriate $\alpha_{\textrm{CO}} = 4.35$), the E+As possess gas densities which are higher than 90\% of the star-forming comparison sample --- rivaled only by some of the gas-rich, star-forming nuclei. They also display a modest 3--5$\times$\ offset below the KS relation when considering the neon SFRs, and often $>$5$\times$\ when considering H$\alpha$ (as found in FYZ15) --- comparable to and exceeding the offsets observed in gas-rich ETGs \citep{davis2014}. The gas densities are very typical of ULIRGs (using an appropriate $\alpha_{\textrm{CO}} = 0.8$), however, in line with the lower 30\% of the comparison sample. The offset in $\Sigma_{\textrm{SFR}}$\ from ULIRGs is more severe, reaching $>$10$\times$\ in some cases. SFRs based on TIR display only a more modest offset, with two sources even lying above the relation (both \textit{Herschel} sources with scaled 8\um\ sizes). However, as discussed in \S\,\ref{sec:sfr_lim}, TIR likely provides only an upper limit in many cases and may overestimate the SFR substantially. The subset of sources with detected neon all display a $\sim$10$\times$\ offset in $\Sigma_{\textrm{SFR}}$\ compared to the ULIRGs. This is corroborated by equivalent offsets in H$\alpha$, most of which occur in sources unaffected by dust attenuation. 

In the small subset of 22\um-selected sources with high inferred dust attenuation (0480, 0962, 2001, 2276, 2360, 2376, 2777), H$\alpha$-based SFRs likely suffer from significant optical depth effects, even after correcting for extinction using $\tau_{\rm V}$\ derived from the stellar population fitting (e.g., Balmer decrement). In Figures \ref{fig:sf-ms} \& \ref{fig:ks} we therefore exclude SFR and $\Sigma_{\textrm{SFR}}$\ based on H$\alpha$\ for these seven sources. We find that, in these cases alone, differences between TIR and H$\alpha$-based SFRs are extreme --- with SFR(\TIR)/SFR(${\rm H\alpha}$) ranging from 25--275. These seven sources appear to form the spur of sources in FYZ15 which possess the most significant offsets from the KS law --- up to a factor of 100 in one case. Although it is now clear that such extreme offsets in H$\alpha$-inferred star formation efficiency are the result of the impact of dust obscuration, the single \textit{Spitzer} source among these seven, 0962, still possesses a 5--10$\times$\ offset in $\Sigma_{\textrm{SFR}}$\ using the dust-insensitive MIR neon line (for SF and ULIRG comparisons, respectively).  This measure is 3$\times$\ lower than SFR based on TIR, though it is still $\sim$10$\times$\ higher than that of H$\alpha$. We discuss the interpretation of this small subset of galaxies in \S\,\ref{sec:discuss}. 

\section{Discussion}
\label{sec:discuss}
Post-starburst galaxies are traditionally viewed as gas and dust poor --- having had the bulk of their ISM expelled by the event which nearly truncates their star formation on short timescales. In direct contrast to canonical timescale arguments for quenching via AGN-driven outflows, E+As have been found to host significant molecular reservoirs (FYZ15; \citealt{rowlands2015}). Here, we have revealed that these reservoirs are warm, dusty, and likely dominate the gas content --- forming an ISM with unusual emission properties and energetics. Below, we discuss the implications these characteristics have for the evolutionary origin and ultimate fate of post-burst systems.

\subsection{Examining the Post-Starburst Classification}
A comparison of the H$\alpha$, NUV, and TIR luminosities of the E+A sample reveals a  dust-attenuation sequence which spans \textit{three orders of magnitude} (see Figure \ref{fig:halpha-nuv}). A nearly-identical sequence is observed in star forming galaxies, extending down to ULIRGs such as Arp 220, with the vast majority of E+As displaying obscuration typical of normal star-forming galaxies (A$_{V} \lesssim 3$; SDD07). One of the \textit{Spitzer}/IRS sources (0962; \S\,\ref{sec:specdecomp}) displays significant silicate absorption ($\tau_{9.7\textrm{\um}} = 1.7$) and several of the WISE/22\um-selected \textit{Herschel}-only sources appear to be nearly as dust-obscured as Arp 220 (from the attenuation curve) --- highlighting a potential ``skin-effect'' regarding their classification. It should be noted that the E+As which were selected with the stringent WISE 22\um\ flux cut are, on average, dustier and more embedded. The potential for ``skin effect'' stellar populations is apparently small in a purely optically-selected E+A sample. Regardless of level of obscuration, the E+As display a $>$5$\times$\ \textit{systematic deficiency} of H$\alpha$\ emission (relative to TIR), compared to all other comparison galaxies --- consistent with their selection and young but aging stellar populations. For the deeply embedded E+As, there are two explanations for this deficiency: 
\begin{itemize}
\item A unique geometry is present, where significant ongoing star formation is completely dust-obscured behind a post-starburst ``skin'', which contributes the bulk of NUV and H$\alpha$\ emission. Though star-forming regions can be preferentially deeply embedded, it is difficult to reconcile the complete segregation of the clear post-burst ``skin'' stellar population from an obscured starbursting population.  
\item The ``skin'' is representative of, or related to, the rest of the stellar population, and thus the galaxies are truly H$\alpha$\ (and thus star-formation) deficient --- even though the bulk of the stars remain dust obscured. In support of this, these galaxies have TIR-based SFRs which are typically 5--10$\times$\ lower than their modeled peak burst SFRs. 
\end{itemize}
Due to the constancy of the H$\alpha$--NUV attenuation law across a wide variety of galaxy morphologies and evolutionary states, it seems unlikely that dust geometry can account for the observed deficiency. Though Arp 220 possesses one of the highest known global extinction values, with both its dual nuclei and many of its star-forming regions heavily embedded \citep{smith1989,scoville1991,sturm1996}, it does not deviate from the observed relation by more than a factor of two. Indeed, despite high general obscuration, UV-bright clusters are readily seen in the visible skin of most ULIRGs \citep{goldader2002}. 

Furthermore, \cite{wild2011} finds that in galaxies with very high silicate-inferred extinction (A$_{V} > 50$), Balmer decrement-derived extinction measurements do, indeed, trace silicate-based measurements, suggesting the visible stellar populations contain information about the embedded populations. Thus, it seems most likely that the E+A ``skin'' is also representative of the embedded stellar population, in the few cases with evidence for substantial dust attenuation. Some of these infrared-bright sources may, indeed, harbor obscured regions of ongoing star formation. However, their TIR-based SFRs (which are likely \emph{overestimates}; see \S\,\ref{sec:sfr_lim}) indicate that they have still experienced dramatic decreases in SFR since their starburst peaks.  Moreover, in the obscured IRS source 0962, the measured neon-based SFR, which can readily penetrate columns of $A_\mathrm{V}\sim100$ or greater, is nearly 200$\times$\ lower than the peak modeled burst SFR.  The burst mass fraction and burst age recovered from stellar SED modeling in the sample would not change if a portion of the stellar population with similar properties is obscured by dust in the optical/UV, but the inferred peak burst SFR would \emph{increase} if accounting for this hidden population, further highlighting the strongly truncated star formation history of these systems.

\subsection{Origin of the Spectral Properties}
The majority of sources are compact at 8\um\ (1/3 are unresolved), with over half possessing FWHM sizes $<3$\ kpc --- on average, 5-6$\times$\ more compact than in the optical. This suggests that the dust and molecular gas reservoirs are very centrally concentrated, as is found in starbursting (U)LIRGs, supporting a connection between E+As and progenitor starbursts. Further evidence of this connection is found in their IRS spectra. Strong PAH emission is ubiquitous across the sample, with a geometric mean PAH/TIR of 13.0\% --- significantly higher than the SINGS geometric mean of 8.7\% (SDD07) and considerably higher than the GOALS or dusty ETG samples. However, high-ionization lines are astonishingly weak. \neII\ 12.8\um, one of the brightest infrared fine-structure lines in star-forming galaxies, is weaker in every source, relative to both PAH and TIR emission, than the SINGS mean, with some deficient by factors of $>$5. The E+As are \textit{substantially} different than the dusty ETG population, most of which have been ``revived'' by a minor, gas-rich merger. That the nebular line deficiency persists into the MIR confirms that these sources are indeed only weakly star-forming, supporting their optical classification.

Though remarkably devoid of nebular lines, the E+As nevertheless display near-ubiquitous bright rotational H$_{2}$\ emission. They are strong outliers in H$_{2}$/TIR compared to both SINGS and GOALS sources, rivaled only by turbulent, shock-powered systems similar to Stephan's Quintet \citep{cluver2010,appleton2017}. Indeed, 8 of the 15 E+As lie above the MOHEG H$_{2}$/7.7\um\ limit of 0.04 \citep{ogle2010,cluver2013} --- also indicating that the H$_{2}$\ excitation is shock-driven. \cite{stierwalt2014} found that most galaxies in the GOALS sample which satisfied the MOHEG H$_{2}$/PAH criterion were experiencing, or had recently experienced, the most major mergers. Power-law fits of the H$_{2}$\ rotational excitation diagrams reveal that much of the H$_{2}$\ is warm, with H$_{2}$\ temperature power-law slopes that are, on average, shallower than those of star-forming galaxies and similar to those of ULIRGs and shock-powered systems such as Stephan's Quintet \citep{appleton2017}.

The \cII\ cooling-line deficit (\cII/TIR) in the E+A sample varies by two orders of magnitude, with 85\% of sources deeper than the KINGFISH average of 0.48\% \citep{smith2017}. \cite{smith2017} found that the \cII\ deficit in galaxies is overwhelmingly driven by a single parameter: the SFR density ($\Sigma_{\textrm{SFR}}$). The underlying mechanism is thought to be grain charging --- increases in far-UV intensity result in a higher grain ionization potential, decreasing photoelectric heating efficiency in the gas. Using the relation between \cII/TIR and $\Sigma_{\textrm{SFR}}$, given by \cite{smith2017}, we find that the $\Sigma_{\textrm{SFR}}$\ predicted from the equation is, on average, \textit{100$\times$\ higher} than the $\Sigma_{\textrm{SFR}}$\ calculated directly from TIR (which are shown in Figure \ref{fig:ks}). \cite{smith2017} posited that severe discrepancies such as these often arise in environments with very high stellar density, which possess high surface brightness, but lower per-photon energy than in star-forming regions (e.g., softer radiation field). In such a ``high-soft'' environment, the photoelectric coupling between the dust and gas is effectively broken. The poor correlation between $\Sigma_{\textrm{SFR}}$\ and \cII/TIR here is strong evidence in support of E+As possessing an aging post-starburst stellar population (soft) which is compact and centrally-concentrated (high surface brightness). A high stellar density region dominated by A-stars (as is assumed in E+As) could quite effectively produce such a high-soft radiation field. This could also help to explain the observed warm dust peaks --- at $\sim$70--75\um\ they are significantly warmer than the $\sim$100\um\ peaks found in star-forming galaxies. Additionally, the PAH mass fraction, q$_{\PAH}$, is likely affected by a high-soft radiation field. \cite{draine2014} found that q$_{\PAH}$\ was 2$\times$\ lower in the bulge of M31 when adopting a typical ISRF, rather than the true, old starlight-dominated radiation field of M31's bulge. Given the aging post-burst populations in the E+As, the q$_{\PAH}$\ values quoted in this paper are, thus, likely only lower limits. 

\subsection{Insight into the Evolutionary Path}
A strong negative trend is visible between fractional PAH luminosity (PAH/TIR) and modeled post-burst age. PAH grains are small and more easily destroyed. Thus, a suppression of PAH/TIR is consistent with the accumulating effects of small grain destruction as these sources age. Indeed, in quiescent ellipticals, PAH emission vanishes almost entirely (e.g., \citealt{bressan2006}). We also find a strong negative trend in dust mass (normalized to both total stellar mass and stellar mass created in the burst) with age, also likely indicative of the effects of dust destruction. This trend continues for several dusty ETGs in the SINGS sample which possess well-measured dust masses. In the models of \cite{zhukovska2016}, grain growth in the cold ISM reaches a steady state with grain destruction $\gtrsim$140 Myr following a burst of star formation. As star formation seems to have all but ceased, both PAH/TIR and the dust-to-burst mass ratio should remain $\sim$constant in sources older than $\sim$100 Myr. What, then, is causing the continued dust destruction implied by the declining PAH/TIR and dust-to-burst mass ratios? We suggest that this is evidence for an incipient transition to early-type, hot gas-dominated ISM.

Indeed, of galaxies with modeled burst ages greater than 900 Myr, the majority display distinctly early-type morphologies (see FYZ15). Additionally, PAH/TIR in these oldest systems is very comparable to the dusty ETG comparison sample --- most have PAH/TIR $<$\ 10\%. Though these dusty ETGs have likely been revived by a minor merger (and host confirmed star formation), their diminished PAH/TIR compared to normal star-forming galaxies suggests that they indeed possess ISM conditions which are harsher to small-grain survival. These results provide a window into the potential future evolution of the E+A systems and appear to support the classical view of E+As as the progenitors of gas and dust-depleted pure elliptical galaxies. In these situations the gas must be driven by an ongoing (if sporadic) source of turbulent energy, after the initial disrupting event, in order to maintain suppression of star formation. The mechanism responsible for the gas's continued turbulence is unknown, but we discuss potential mechanisms below. 

Despite the unique conditions present in the E+As, and the attendant limitations of traditional SFR indicators, we find that, although the E+As ISM are dense --- comparable to ULIRGs (which are molecular gas dominated; \citealt{iono2005}), when averaged over a typical $\sim$2 kpc scale --- and centrally-concentrated, they display a significant offset from both the star-forming main sequence and from the KS star formation relation. Both are indicative of low star-forming efficiencies, the former on a galaxy-wide scale and the latter on a local ISM scale. Traditional models of the star formation quenching process (typically some combination of AGN and stellar feedback), require the expulsion of galaxies' dust and gas on short timescales. Even if the gas does not escape the galaxy entirely, its distribution should be significantly altered. In contrast to this picture, most of the E+As in this sample possess dust and molecular gas masses substantially higher than can be explained by stellar recycling. Thus, in most cases, the observed dusty ISM must be preexisting, rather than regrown. Since we have no reliable method of estimating the gas or dust masses prior to or during the burst, it is very possible that a significant fraction of the ISM was removed. Indeed, the molecular gas-to-stellar mass ratios (with the stellar mass created in the burst subtracted) --- typically from 5--25\%) are significantly suppressed compared to (U)LIRGs in the GOALS sample \citep{larson2016} --- often by a factor of $>$5. While molecular gas could have been converted into the atomic phase, the consistency of the DMGRs with nearby galaxies' full DGRs suggests that these are molecular gas-dominated. Thus, their low molecular-to-stellar mass ratios supports the scenario that much of the ISM was removed in outflows, similar to currently active post-starbursts such as those in \cite{yesuf2017}. However, these results indicate that even in the most violent quenching events, a significant fraction of the ISM must remain behind, consistent with recent observations of both younger \citep{alatalo2016a} and high-redshift \citep{suess2017} post-starbursts.

The rapid decline of star formation ($\gtrsim$100$\times$) since the peak indicates the operation of a strong quenching mechanism sometime in the recent past, while current high molecular gas densities combined with low SFRs indicate that the gas must be continually turbulently-supported against collapse --- supported by the properties of H$_{2}$\ emission. Assuming that the ISM is relatively smoothly distributed --- a good assumption given the near-lack of star formation --- and with no additional energy input, turbulent energy should be dissipated on timescales much shorter than their post-burst ages \citep{scoville2017}. By what mechanism, then, is the large-scale turbulence maintained? Though we cannot answer this definitively, we consider the following mechanisms: 
\begin{itemize}
\item In the case of Arp 220, \cite{scoville2017} find that turbulence in the dual nuclear molecular gas disks is maintained simply by the continual time variation of the gravitational potential in the merging system. Such merger-induced gravitational torques were also posited as a mechanism for the star formation suppression observed in some HCGs \citep{alatalo2015b}. In E+As, if a merger was responsible for quenching of the starburst, significant evolution of the potential likely persists on timescales of order the dynamical time (hundreds of Myr) --- similar to many of the post-burst ages. Thus, it seems plausible that after star formation is essentially halted, time variation of the potential may, in many cases, provide the required mechanical energy necessary to maintain the observed turbulent signatures.
\item An alternative (or perhaps complimentary) mechanism is that of radio-mode AGN feedback. \cite{terrazas2016,terrazas2017} suggest that low-level AGN feedback is critical for maintaining star-forming quiescence in already quiescent galaxies --- a phenomenon seen in NGC 1266 \citep{alatalo2015}. Indeed, in three of the eight sources with existing 1.4 GHz radio continuum measurements, we find evidence for radio excess. Thus, periodic animation of the central SMBH, resulting in kinetic feedback, may be a plausible method for shocking the gas and suppressing star formation, and may not necessarily provide clear evidence of its presence in emission diagnostics. 
%\item \cite{conroy2015} has recently shown that late-time stellar heating by AGB star winds is a viable mechanism for preventing the cooling of hot gas in early-type galaxies, via shocking of the ejected AGB envelopes with the ambient medium. Given the E+As' dense central stellar environments, typical stellar velocity dispersions $>$100 km\,s$^{-1}$\ (from SDSS), and existing molecular reservoirs, interactions between AGB ejecta and the molecular gas could very likely result in shocks. Thus, this may be a plausible source of additional mechanical energy injection.  
\end{itemize}

HST imaging obtained for four of the eight confirmed MOHEG sources reveals bright, highly asymmetric tidal substructure in all cases, indicating that these sources have, indeed, all undergone recent major mergers/interactions. Though the greatest changes to the potential in a merger occur during final coalescence, significant fluctuations of the potential likely persist for hundreds of Myr afterwards \citep{bois2010}. If these fluctuations are still present in these E+As, it may indicate that, as in the case of Arp 220, time variation of the potential can account for much of the observed turbulence. Additionally, work by \cite{lanz2016} has shown that radio jets appear to drive the H$_{2}$\ emission in many early-type, radio-emitting MOHEGs, leading to star formation suppression characterized by an offset from the KS relation. Many of these galaxies do not display clear AGN indicators in the optical or mid-infrared, typical of purely radio AGN and similar to our E+As. Only one of the three galaxies in the sample with confirmed radio excess was observed with \textit{Spitzer}, and that source is, indeed, a MOHEG. This may also indicate that, in some E+As, low-duty cycle radio-mode feedback is a plausible source of continued energy injection.

\section{Conclusions}
We have presented an analysis of the infrared properties of a sample of 33 E+A post-starburst galaxies, combining infrared photometry (spanning 3--500\um) with full \textit{Spitzer} IRS spectroscopy and \textit{Herschel} PACS \cII\ and \oI\ cooling-line observations. A wide range of physical conditions in the ISM of these E+As are revealed for the first time, providing a direct probe of their formation mechanism and subsequent evolution. We find: 
\begin{enumerate}
\item Significant dust reservoirs (as high as 5$\times$10$^{8}$\,M$_{\odot}$), which, when combined with CO-based H$_{2}$\ masses (assuming $\alpha_{\textrm{CO}} = 4$, as in FYZ15), provide reasonable dust-to-molecular gas ratios of $0.0025\leqslant$\ DMGR $\leqslant 0.04$ --- consistent with the dust-to-total gas ratios of nearby galaxies, suggesting that the sample may be molecular gas-dominated.

\item Warm dust, with an average SED peak wavelength of 70--75\um. The modest UV budget, but high central stellar surface brightness suggests that the dust may be heated by an unusual high-intensity, softer (``high-soft'') radiation field created by the A-star dominated population. 

\item High PAH abundances. Half of the E+A sample possess q$_{\PAH} > 4.6$\% --- the highest value for nearby, star-forming galaxies in the SINGS sample. Additionally, if E+As are dominated by a high-soft radiation field, q$_{\PAH}$\ is likely higher still. 

\item Unusually strong PAH emission --- the mean PAH/TIR is 13.0\%, compared to SINGS' 8.7\% --- but overall weak nebular lines. Where detected, total line emission (\neII\ + \neIII\ + \sIV\ + \sIII\,18\um) is 3$\times$\ lower than in star-forming galaxies, relative to PAH emission, but most sources are not detected in any nebular lines, indicating that strong PAH emission can persist hundreds of Myr after the bulk of star formation has stopped in galaxies, and cautioning against the use of PAHs as direct star formation tracers. 

\item A small number of WISE 22\um-selected sources appear to be deeply dust-embedded, when considering attenuation curves for H$\alpha$\ vs. NUV emission. In these sources, even extinction-corrected H$\alpha$\ likely provides a significant underestimate of the SFR, and could lead to extremely discrepant apparent star forming efficiencies. However, these few obscured sources still display a steep decline in SFR from their past peaks, and current obscuration-tolerant SFRs are generally very modest, thus preserving their post-starburst classification.

\item Compact infrared cores, $< 1/3$\ of the optical extent. Physical sizes can approach ULIRG dimensions --- 8\um\ FWHM $<1.5$\ kpc. Their compactness at 8\um\ suggests compact star-forming progenitors, with gas densities higher than 90\% of normal star forming galaxies and very comparable to (U)LIRGs. 
  
\item Strong H$_{2}$\ pure rotational emission. 75\% of sources observed with \textit{Spitzer} display H$_{2}$\ emission considerably higher than seen in star-forming galaxies, with $>$50\% classifiable as Molecular Hydrogen-Emitting Galaxies (MOHEGs; H$_{2}$/7.7\um\ $\geqslant$\ 0.04). These galaxies have among the highest fractional H$_{2}$ luminosities (compared to TIR) of any globally integrated galaxy --- comparable to shock-dominated regions such as in Stephan's Quintet. The slopes of their H$_{2}$\ power-law temperature distributions are shallower than found in SINGS, and similar to ULIRGs and other turbulent systems. These all suggest that the H$_{2}$\ emission is powered by turbulent heating, though it appears to be a purely low-velocity phenomenon (as opposed to SQ). This is also supported by the dominance of the \oI\,63\um\ cooling-line over \cII\,158\um, in several sources --- predicted in molecular cloud shock models. 

\item Significant \cII\ deficits in most sources, with 85\% of sources falling below the KINGFISH mean of 0.48\%. These deficits are, however, inconsistent with being star-formation driven and, instead, seem to confirm the action of a predominantly ``high-soft'' radiation field.

\item Dust and molecular gas-to-stellar mass ratios which are typically much higher than can be explained by stellar recycling from the aging burst population. This indicates that the dust and gas reservoirs were not entirely expelled during burst truncation.

\item Negative trends of PAH/TIR and dust-to-stellar mass as a function of post-burst age, consistent with observations of dusty ETGs, indicating the effect of accumulating grain destruction and an incipient transition to virialized early-type gas conditions. 

\item Little evidence for significant AGN activity beyond potential low-level activity associated with their existing LINER classifications. The high-ionization emission line \oIV\ is only detected in one source (weakly), and PAH band ratios show no evidence of small-grain depletion. One galaxy (2360\_167\_53728) displays extreme AGN colors in the near-IR WISE bands, with evidence for a very hot dust component to its SED. The presumed AGN is likely deeply embedded. Three sources appear to display an excess at 1.4 GHz, relative to the radio-infrared correlation, potentially indicating the presence of radio AGN activity. One source, in particular, is clearly a radio galaxy --- the same source that is detected in \oIV. 

\item Low star-formation rates and efficiencies. Offsets from the star-forming main sequence are typically $>$10 and reach factors of $>$100 in some cases, for a variety of tracers, characterizing these galaxies as truly quiescent. Offsets from the KS relation exist for most sources and range from $\sim$3 to $>$10 for reliable tracers, depending on the comparison sample considered.  
\\
\end{enumerate}

These results paint a compelling picture: one of galaxies whose star formation was, indeed, rapidly truncated, and which are transitioning away from compact starbursting systems, but in which the gas and dust has \textit{not} been completely expelled and instead is supported against further star formation by turbulent or mechanical heating. The resulting dense, modestly-aged burst stellar populations appear to provide a ``high-soft'' radiation field, which dominates the unusual ISM energetics of galaxies seen after the fall from a prior star-forming peak.

\acknowledgments
We thank the anonymous referee for helpful comments which substantially improved this paper.

AS acknowledges support for this work by the National Science Foundation Graduate Research Fellowship Program under grant No.~DGE 1256260. Any opinions, findings, and conclusions or recommendations expressed in this material are those of the author(s) and do not necessarily reflect the views of the National Science Foundation. 

JDS acknowledges direct support for this project from the Research Corporation for Science Advancement through its Cottrell Scholars program, as well as visiting support from the Alexander von Humboldt Foundation.

KDF is supported by Hubble Fellowship Grant HST-HF2-51391.001-A, provided by NASA through a grant from the Space Telescope Science Institute, which is operated by the Association of Universities for Research in Astronomy, Incorporated, under NASA contract NAS5-26555

AIZ acknowledges support from NASA ADAP grant 09-ADP09-0073.

This work is based in part on observations made with \textit{Herschel}, a European Space Agency Cornerstone Mission with significant participation by NASA. Sup- port for this work was provided by NASA through an award issued by JPL/Caltech. This work made use of the NASA/IPAC Extragalactic Database (NED), operated by JPL/Caltech, under contract with NASA.

We thank Vanja \v{S}arkovi\'c for her work on the stellar recycling models used throughout this work, Tim Carleton for his assistance with IRS data reduction, and Eric Pellegrini, Tim Heckman, Katey Alatalo, Roberto Rampazzo, and Phil Appleton for useful discussions which improved this paper. We also thank Ned Wright, Peter Eisenhardt, Sara Petty, and the WISE extragalactic working group for pre-release access to WISE photometry for our \textit{Herschel}-selection.

\cleardoublepage
\appendix

\section{Photometric Tables and SDSS Spectra}
In Tables \ref{tab:spitzer-phot}, \ref{tab:herschel-phot}, and \ref{tab:wise-phot} we present the photometry for the sample from the \textit{Spitzer}, \textit{Herschel}, and WISE space observatories, respectively. \textit{Spitzer} photometric data only exist for the original sample of 15 sources, while \textit{Herschel} and WISE data are shown for the full sample of 33.

The \textit{Spitzer} photometry and \textit{Herschel} spectrophotometry is divided by the respective instruments: IRAC and MIPS for \textit{Spitzer}, PACS and SPIRE for \textit{Herschel}. Targets were chosen for \textit{Herschel} SPIRE coverage based on the 250\um\ flux density of their \textit{Spitzer} or WISE-extrapolated SEDs. In addition to the \textit{Spitzer} photometry, we also provide the slit-loss corrections for the IRS spectra, and the derived 8\um\ FWHMs used for analysis. 

In Figure \ref{fig:sdss_spec} we show show the SDSS spectrum for each object in the sample. Spectra were downloaded from the SDSS Data Release 12 (DR12) Science Archive Server (SAS; \href{https://dr12.sdss.org}{https://dr12.sdss.org}).

\newcolumntype{s}{!{\extracolsep{8pt}}c!{\extracolsep{0pt}}}
\floattable
\begin{deluxetable*}{rccccccsccc}
\rotate
\tablecaption{\textit{Spitzer} Photometric Data\label{tab:spitzer-phot}}
\tablecolumns{11}
\tabletypesize{\small}
\tablehead{%
\colhead{} &
\multicolumn{6}{c}{IRAC} &
\multicolumn{4}{c}{MIPS} \vspace{1mm}\\ \cline{2-7} \cline{8-11}
%%%
\vspace{-2mm} \\
\colhead{} &
\colhead{$f_{\nu}(3.6\textrm{\um})$} &
\colhead{$f_{\nu}(4.5\textrm{\um})$} &
\colhead{$f_{\nu}(5.8\textrm{\um})$} &
\colhead{$f_{\nu}(8.0\textrm{\um})$} &
\colhead{FWHM$_{8\textrm{\um}}$} &
\colhead{} &
\colhead{$f_{\nu}(24\textrm{\um})$} &
\colhead{$f_{\nu}(70\textrm{\um})$} &
\colhead{$f_{\nu}(160\textrm{\um})$} &
\colhead{} \\
%%%
\colhead{Galaxy} &
\colhead{(mJy)} &
\colhead{(mJy)} &
\colhead{(mJy)} &
\colhead{(mJy)} &
\colhead{(arcsec)} &
\colhead{$f_{SL}$} &
\colhead{(mJy)} &
\colhead{(mJy)} &
\colhead{(mJy)} &
\colhead{$f_{LL}$} \\
%%%
\colhead{(1)} & 
\colhead{(2)} & 
\colhead{(3)} & 
\colhead{(4)} & 
\colhead{(5)} & 
\colhead{(6)} & 
\colhead{(7)} & 
\colhead{(8)} & 
\colhead{(9)} & 
\colhead{(10)} & 
\colhead{(11)} \\ 
\vspace{-3mm}}
\startdata
0379\_579\_51789&6.52 $\pm$~0.90&4.28 $\pm$~0.59&2.83 $\pm$~0.39&3.03 $\pm$~0.39&3.80&0.57&3.38 $\pm$~0.27&$<$26.20&$<$26.20&0.40\\
0413\_238\_51929&4.52 $\pm$~0.62&3.15 $\pm$~0.44&3.86 $\pm$~0.54&7.23 $\pm$~0.95&2.62&0.66&21.62 $\pm$~0.89&79.17 $\pm$~6.51&72.12 $\pm$~12.23&0.55\\
0570\_537\_52266&3.22 $\pm$~0.45&2.14 $\pm$~0.30&1.91 $\pm$~0.27&3.63 $\pm$~0.48&3.07&0.63&8.61 $\pm$~0.40&15.37 $\pm$~2.47&$<$15.37&0.56\\
0623\_207\_52051&1.97 $\pm$~0.27&1.36 $\pm$~0.19&1.58 $\pm$~0.22&2.78 $\pm$~0.36&7.01&0.40&2.16 $\pm$~0.22&$<$20.74&$<$20.74&0.45\\
0637\_584\_52174&4.80 $\pm$~0.66&3.28 $\pm$~0.46&2.46 $\pm$~0.34&3.81 $\pm$~0.50&3.48&0.56&2.83 $\pm$~0.22&22.50 $\pm$~2.94&20.66 $\pm$~7.09&0.47\\
0656\_404\_52148&11.46 $\pm$~1.58&7.35 $\pm$~1.02&6.00 $\pm$~0.82&4.60 $\pm$~0.60&3.11&0.50&3.73 $\pm$~0.21&31.15 $\pm$~3.68&11.91 $\pm$~6.93&0.56\\
0756\_424\_52577&2.86 $\pm$~0.40&1.96 $\pm$~0.28&1.88 $\pm$~0.27&3.52 $\pm$~0.47&2.76&0.64&4.69 $\pm$~0.22&73.39 $\pm$~5.43&56.67 $\pm$~7.72&0.54\\
0815\_586\_52374&7.21 $\pm$~0.99&4.68 $\pm$~0.65&4.01 $\pm$~0.55&4.44 $\pm$~0.58&3.77&0.53&2.64 $\pm$~0.28&21.70 $\pm$~3.67&29.12 $\pm$~7.87&0.36\\
0951\_128\_52398&8.16 $\pm$~1.12&5.16 $\pm$~0.71&4.00 $\pm$~0.55&2.46 $\pm$~0.32&3.60&0.50&1.16 $\pm$~0.32&$<$29.44&$<$29.44&0.41\\
0962\_212\_52620&4.45 $\pm$~0.62&3.75 $\pm$~0.53&7.86 $\pm$~1.09&18.36 $\pm$~2.42&2.65&0.70&44.31 $\pm$~1.82&655.53 $\pm$~46.33&288.95 $\pm$~36.63&0.47\\
1039\_042\_52707&6.66 $\pm$~0.92&4.33 $\pm$~0.60&3.58 $\pm$~0.50&2.68 $\pm$~0.35&3.15&0.59&3.05 $\pm$~0.23&24.17 $\pm$~3.31&$<$24.17&0.46\\
1170\_189\_52756&10.99 $\pm$~1.50&7.38 $\pm$~1.02&6.30 $\pm$~0.85&5.33 $\pm$~0.69&2.78&0.52&6.66 $\pm$~0.43&103.61 $\pm$~8.47&52.89 $\pm$~11.07&0.42\\
1279\_362\_52736&9.71 $\pm$~1.34&6.24 $\pm$~0.87&6.32 $\pm$~0.87&8.35 $\pm$~1.08&3.61&0.53&6.99 $\pm$~0.38&133.61 $\pm$~9.77&138.11 $\pm$~19.35&0.48\\
1616\_071\_53169&7.36 $\pm$~1.01&4.38 $\pm$~0.60&3.38 $\pm$~0.46&2.83 $\pm$~0.37&4.28&0.35&0.87 $\pm$~0.15&$<$29.98&$<$29.98&0.40\\
1927\_584\_53321&10.92 $\pm$~1.51&6.92 $\pm$~0.97&6.26 $\pm$~0.87&6.72 $\pm$~0.88&3.56&0.54&7.22 $\pm$~0.37&46.51 $\pm$~4.38&28.30 $\pm$~7.95&0.41\\
\enddata
\tablecomments{(1) Galaxy ID. \\
(2)-(5) \textit{Spitzer} IRAC flux densities. \\
(6) IRAC 8\um\ FWHM. Computed as the geometric mean of the major and minor axes of 2-D elliptical Gaussian. \\
(7) Ratio of synthetic IRAC 8\um, computed from the IRS spectrum, to global 8\um; used for spectral scaling of SL. \\
(8)-(10) \textit{Spitzer} MIPS flux densities. \\
(11) Synthetic-to-global ratio for MIPS 24\um; used for spectral scaling of LL.}
\end{deluxetable*}

\newcolumntype{s}{!{\extracolsep{8pt}}c!{\extracolsep{0pt}}}
\floattable
\begin{deluxetable*}{rcccccscc}
\rotate
\tablecaption{\textit{Herschel} Photometry\label{tab:herschel-phot}}
\tablecolumns{9}
\tabletypesize{\scriptsize}
\tablehead{%
\colhead{} &
\multicolumn{5}{c}{PACS} &
\multicolumn{3}{c}{SPIRE} \vspace{1mm}\\ \cline{2-6} \cline{7-8}
%%%
\vspace{-2mm} \\
\colhead{} &
\colhead{$f_{\nu}(70\textrm{\um})$} &
\colhead{$f_{\nu}(100\textrm{\um})$} &
\colhead{$f_{\nu}(160\textrm{\um})$} &
\colhead{$I_{\nu}$(\cII\ 158\um)} &
\colhead{$I_{\nu}$(\oI\ 63\um)} &
\colhead{$f_{\nu}(150\textrm{\um})$} &
\colhead{$f_{\nu}(350\textrm{\um})$} &
\colhead{$f_{\nu}(500\textrm{\um})$} \\
%%%
\colhead{Galaxy} &
\colhead{(mJy)} &
\colhead{(mJy)} &
\colhead{(mJy)} &
\colhead{($10^{9}$\ W m$^{-2}$\ sr$^{-1}$)} &
\colhead{($10^{9}$\ W m$^{-2}$\ sr$^{-1}$)} &
\colhead{(mJy)} &
\colhead{(mJy)} &
\colhead{(mJy)} \\
\vspace{-3mm}}
\startdata
0336\_469\_51999&21.41 $\pm$~3.46&36.84 $\pm$~3.93&38.84 $\pm$~3.93&$<$3.34&$\dots$&$\dots$&$\dots$&$\dots$\\
0379\_579\_51789&36.62 $\pm$~6.56&58.26 $\pm$~7.35&66.19 $\pm$~7.35&$<$3.47&$\dots$&$\dots$&$\dots$&$\dots$\\
0413\_238\_51929&124.88 $\pm$~7.71&161.52 $\pm$~7.50&118.26 $\pm$~7.50&$<$5.67&$<$71.56&$\dots$&$\dots$&$\dots$\\
0480\_580\_51989&1115.67 $\pm$~56.25&1368.93 $\pm$~50.50&998.89 $\pm$~50.50&6.86 $\pm$ 1.69&$<$35.06&418.64 $\pm$~32.03&182.59 $\pm$~15.87&52.22 $\pm$~9.26\\
0570\_537\_52266&26.10 $\pm$~4.07&33.06 $\pm$~4.71&48.83 $\pm$~4.71&$<$5.31&81.59 $\pm$ 14.97&22.65 $\pm$~6.31&$<$16.34&$<$15.69\\
0598\_170\_52316&16.81 $\pm$~2.70&18.67 $\pm$~2.69&14.45 $\pm$~2.69&17.67 $\pm$ 1.66&$\dots$&$\dots$&$\dots$&$\dots$\\
0623\_207\_52051&17.09 $\pm$~5.03&66.23 $\pm$~7.15&99.73 $\pm$~7.15&$<$3.45&$\dots$&$\dots$&$\dots$&$\dots$\\
0637\_584\_52174&40.85 $\pm$~4.70&62.64 $\pm$~5.93&79.53 $\pm$~5.93&4.37 $\pm$ 1.20&$\dots$&$\dots$&$\dots$&$\dots$\\
0656\_404\_52148&43.94 $\pm$~6.21&62.11 $\pm$~6.25&36.29 $\pm$~6.25&4.17 $\pm$ 1.18&$\dots$&$\dots$&$\dots$&$\dots$\\
0755\_042\_52235&74.80 $\pm$~5.37&80.83 $\pm$~6.71&108.70 $\pm$~6.71&$<$5.64&$<$47.27&27.62 $\pm$~6.66&$<$16.53&$<$15.87\\
0756\_424\_52577&63.32 $\pm$~4.45&100.79 $\pm$~6.27&107.19 $\pm$~6.27&$<$5.18&$\dots$&$\dots$&$\dots$&$\dots$\\
0815\_586\_52374&27.71 $\pm$~5.96&59.10 $\pm$~7.47&89.67 $\pm$~7.47&6.48 $\pm$ 1.18&$\dots$&$\dots$&$\dots$&$\dots$\\
0870\_208\_52325&7.93 $\pm$~2.23&18.27 $\pm$~2.37&14.35 $\pm$~2.37&$<$3.23&$\dots$&$\dots$&$\dots$&$\dots$\\
0951\_128\_52398&9.98 $\pm$~7.11&14.67 $\pm$~7.34&6.15 $\pm$~7.34&$<$3.60&$\dots$&$\dots$&$\dots$&$\dots$\\
0962\_212\_52620&614.32 $\pm$~30.98&666.23 $\pm$~24.54&483.63 $\pm$~24.54&7.45 $\pm$ 1.87&$<$81.59&173.82 $\pm$~14.80&61.38 $\pm$~7.52&12.70 $\pm$~5.77\\
0986\_468\_52443&231.27 $\pm$~12.21&273.28 $\pm$~9.16&164.03 $\pm$~9.16&19.57 $\pm$ 2.38&88.49 $\pm$ 25.91&79.73 $\pm$~9.72&$<$17.22&$<$16.53\\
1001\_048\_52670&29.70 $\pm$~3.28&38.25 $\pm$~3.69&40.59 $\pm$~3.69&17.25 $\pm$ 1.85&$\dots$&$\dots$&$\dots$&$\dots$\\
1003\_087\_52641&60.69 $\pm$~3.93&60.72 $\pm$~3.23&38.78 $\pm$~3.23&$<$5.15&$<$34.38&22.61 $\pm$~4.61&$<$10.68&$<$10.25\\
1039\_042\_52707&31.44 $\pm$~4.58&40.59 $\pm$~4.52&14.02 $\pm$~4.52&8.04 $\pm$ 1.13&$\dots$&$\dots$&$\dots$&$\dots$\\
1170\_189\_52756&114.85 $\pm$~9.50&142.61 $\pm$~9.40&104.32 $\pm$~9.40&8.79 $\pm$ 2.09&$<$47.81&$\dots$&$\dots$&$\dots$\\
1279\_362\_52736&197.52 $\pm$~11.42&298.26 $\pm$~15.92&295.56 $\pm$~15.92&6.25 $\pm$ 1.94&155.03 $\pm$ 23.60&128.22 $\pm$~12.98&43.96 $\pm$~7.71&10.65 $\pm$~6.68\\
1352\_610\_52819&112.25 $\pm$~7.69&120.71 $\pm$~8.70&136.33 $\pm$~8.70&8.93 $\pm$ 1.94&$<$64.09&49.03 $\pm$~9.19&17.65 $\pm$~6.73&$<$21.74\\
1604\_161\_53078&41.65 $\pm$~4.01&55.24 $\pm$~5.04&72.31 $\pm$~5.04&$<$5.53&$<$51.02&33.85 $\pm$~6.04&12.11 $\pm$~4.35&$<$14.23\\
1616\_071\_53169&15.94 $\pm$~7.21&23.84 $\pm$~7.48&23.49 $\pm$~7.48&4.86 $\pm$ 1.27&$\dots$&$\dots$&$\dots$&$\dots$\\
1853\_070\_53566&40.50 $\pm$~3.79&50.79 $\pm$~3.89&40.17 $\pm$~3.89&$<$3.47&$\dots$&$\dots$&$\dots$&$\dots$\\
1927\_584\_53321&88.27 $\pm$~6.15&93.20 $\pm$~6.25&88.18 $\pm$~6.25&$<$3.42&$\dots$&$\dots$&$\dots$&$\dots$\\
2001\_473\_53493&772.05 $\pm$~38.94&748.79 $\pm$~24.25&473.34 $\pm$~24.25&$<$4.89&$<$32.60&159.44 $\pm$~14.69&62.14 $\pm$~8.32&19.19 $\pm$~7.11\\
2276\_444\_53712&715.96 $\pm$~36.22&952.06 $\pm$~40.64&804.85 $\pm$~40.64&14.79 $\pm$ 2.03&$<$39.42&306.62 $\pm$~24.14&114.22 $\pm$~11.41&29.06 $\pm$~8.18\\
2360\_167\_53728&347.54 $\pm$~17.66&277.00 $\pm$~7.34&132.17 $\pm$~7.34&5.81 $\pm$ 1.67&$<$35.37&30.84 $\pm$~5.48&$<$13.46&$<$12.92\\
2365\_624\_53739&77.00 $\pm$~5.87&86.48 $\pm$~7.53&119.14 $\pm$~7.53&9.77 $\pm$ 1.59&$\dots$&$\dots$&$\dots$&$\dots$\\
2376\_454\_53770&150.73 $\pm$~8.42&197.59 $\pm$~9.58&174.82 $\pm$~9.58&15.49 $\pm$ 1.95&$<$25.09&81.23 $\pm$~9.95&31.37 $\pm$~6.71&$<$15.53\\
2750\_018\_54242&49.26 $\pm$~3.56&32.43 $\pm$~3.01&26.96 $\pm$~3.01&$<$5.74&$\dots$&10.58 $\pm$~4.26&$<$11.29&$<$10.84\\
2777\_258\_54554&975.49 $\pm$~49.44&1098.12 $\pm$~40.04&783.22 $\pm$~40.04&$<$5.33&$<$46.63&296.02 $\pm$~24.80&120.44 $\pm$~13.14&34.43 $\pm$~9.70\\
\enddata
\tablecomments{\textit{Herschel} photometry and \cII\ spectroscopy for the full sample. Upper limits are prefaced with `$<$'  and `$\ldots$' denotes a source unobserved with SPIRE.}
\end{deluxetable*}

\clearpage
\begin{deluxetable*}{rcccc}
\tablecaption{WISE Photometry\label{tab:wise-phot}}
\tablecolumns{5}
\tabletypesize{\small}
\tablehead{%
\colhead{} &
\colhead{$f_{\nu}(3.4\textrm{\um})$} &
\colhead{$f_{\nu}(4.6\textrm{\um})$} &
\colhead{$f_{\nu}(12\textrm{\um})$} &
\colhead{$f_{\nu}(22\textrm{\um})$} \\
%%%
\colhead{Galaxy} &
\colhead{(mJy)} &
\colhead{(mJy)} &
\colhead{(mJy)} &
\colhead{(mJy)} \\
}
\startdata
0336\_469\_51999&1.00 $\pm$~0.02&0.54 $\pm$~0.03&1.22 $\pm$~0.16&$<$4.00\\
0379\_579\_51789&6.53 $\pm$~0.08&3.38 $\pm$~0.06&1.25 $\pm$~0.10&$<$2.45\\
0413\_238\_51929&4.53 $\pm$~0.05&2.80 $\pm$~0.04&6.74 $\pm$~0.17&14.50 $\pm$~0.84\\
0480\_580\_51989&1.41 $\pm$~0.02&0.88 $\pm$~0.02&2.34 $\pm$~0.16&7.72 $\pm$~0.80\\
0570\_537\_52266&3.14 $\pm$~0.04&1.81 $\pm$~0.03&2.30 $\pm$~0.13&4.93 $\pm$~0.66\\
0598\_170\_52316&0.90 $\pm$~0.02&0.60 $\pm$~0.02&1.42 $\pm$~0.10&4.37 $\pm$~0.52\\
0623\_207\_52051&1.90 $\pm$~0.03&1.04 $\pm$~0.03&1.65 $\pm$~0.07&$<$1.75\\
0637\_584\_52174&5.47 $\pm$~0.07&3.11 $\pm$~0.05&2.45 $\pm$~0.13&$<$3.20\\
0656\_404\_52148&12.05 $\pm$~0.13&6.50 $\pm$~0.08&2.62 $\pm$~0.13&$<$3.25\\
0755\_042\_52235&4.46 $\pm$~0.05&3.23 $\pm$~0.04&5.36 $\pm$~0.17&9.43 $\pm$~0.82\\
0756\_424\_52577&2.48 $\pm$~0.03&1.48 $\pm$~0.03&1.64 $\pm$~0.10&$<$2.55\\
0815\_586\_52374&8.21 $\pm$~0.09&4.31 $\pm$~0.06&3.02 $\pm$~0.08&2.97 $\pm$~0.53\\
0870\_208\_52325&1.48 $\pm$~0.02&0.86 $\pm$~0.02&0.80 $\pm$~0.13&$<$3.20\\
0951\_128\_52398&8.12 $\pm$~0.09&4.46 $\pm$~0.06&1.25 $\pm$~0.08&$<$2.35\\
0962\_212\_52620&4.17 $\pm$~0.05&3.32 $\pm$~0.05&13.02 $\pm$~0.25&32.39 $\pm$~1.64\\
0986\_468\_52443&6.29 $\pm$~0.07&3.63 $\pm$~0.05&8.34 $\pm$~0.29&15.85 $\pm$~1.33\\
1001\_048\_52670&4.00 $\pm$~0.05&2.28 $\pm$~0.04&1.37 $\pm$~0.12&$<$2.90\\
1003\_087\_52641&3.19 $\pm$~0.04&2.00 $\pm$~0.04&3.28 $\pm$~0.15&5.16 $\pm$~0.73\\
1039\_042\_52707&6.66 $\pm$~0.07&3.92 $\pm$~0.05&1.35 $\pm$~0.08&$<$2.15\\
1170\_189\_52756&11.65 $\pm$~0.13&6.76 $\pm$~0.08&3.12 $\pm$~0.09&3.56 $\pm$~0.42\\
1279\_362\_52736&9.79 $\pm$~0.11&5.50 $\pm$~0.07&6.54 $\pm$~0.16&4.50 $\pm$~0.76\\
1352\_610\_52819&3.92 $\pm$~0.05&2.31 $\pm$~0.04&5.80 $\pm$~0.18&12.11 $\pm$~0.88\\
1604\_161\_53078&9.32 $\pm$~0.11&5.73 $\pm$~0.08&4.66 $\pm$~0.17&8.95 $\pm$~0.78\\
1616\_071\_53169&7.15 $\pm$~0.09&3.75 $\pm$~0.07&0.83 $\pm$~0.12&\ldots\\
1853\_070\_53566&6.45 $\pm$~0.07&3.64 $\pm$~0.05&2.28 $\pm$~0.20&2.46 $\pm$~0.41\\
1927\_584\_53321&11.09 $\pm$~0.12&6.20 $\pm$~0.08&4.04 $\pm$~0.17&$<$4.20\\
2001\_473\_53493&0.64 $\pm$~0.01&0.42 $\pm$~0.02&1.89 $\pm$~0.12&12.61 $\pm$~0.86\\
2276\_444\_53712&5.02 $\pm$~0.06&3.25 $\pm$~0.07&11.51 $\pm$~0.43&18.67 $\pm$~2.08\\
2360\_167\_53728&0.43 $\pm$~0.01&1.50 $\pm$~0.03&3.34 $\pm$~0.17&13.53 $\pm$~1.06\\
2365\_624\_53739&2.63 $\pm$~0.04&1.72 $\pm$~0.04&3.18 $\pm$~0.14&$<$3.50\\
2376\_454\_53770&1.63 $\pm$~0.02&1.17 $\pm$~0.03&2.94 $\pm$~0.19&7.60 $\pm$~0.92\\
2750\_018\_54242&1.08 $\pm$~0.02&0.72 $\pm$~0.02&2.23 $\pm$~0.10&6.16 $\pm$~0.61\\
2777\_258\_54554&1.53 $\pm$~0.02&0.97 $\pm$~0.03&3.91 $\pm$~0.15&16.60 $\pm$~1.12\\
\enddata
\tablecomments{WISE photometry. 5$\sigma$\ upper limits are prefaced with `$<$'. Non-detections are denoted by `\ldots'.}
\end{deluxetable*}

\begin{figure*}[t]
\centering
\leavevmode
\includegraphics[width={0.95\linewidth}]{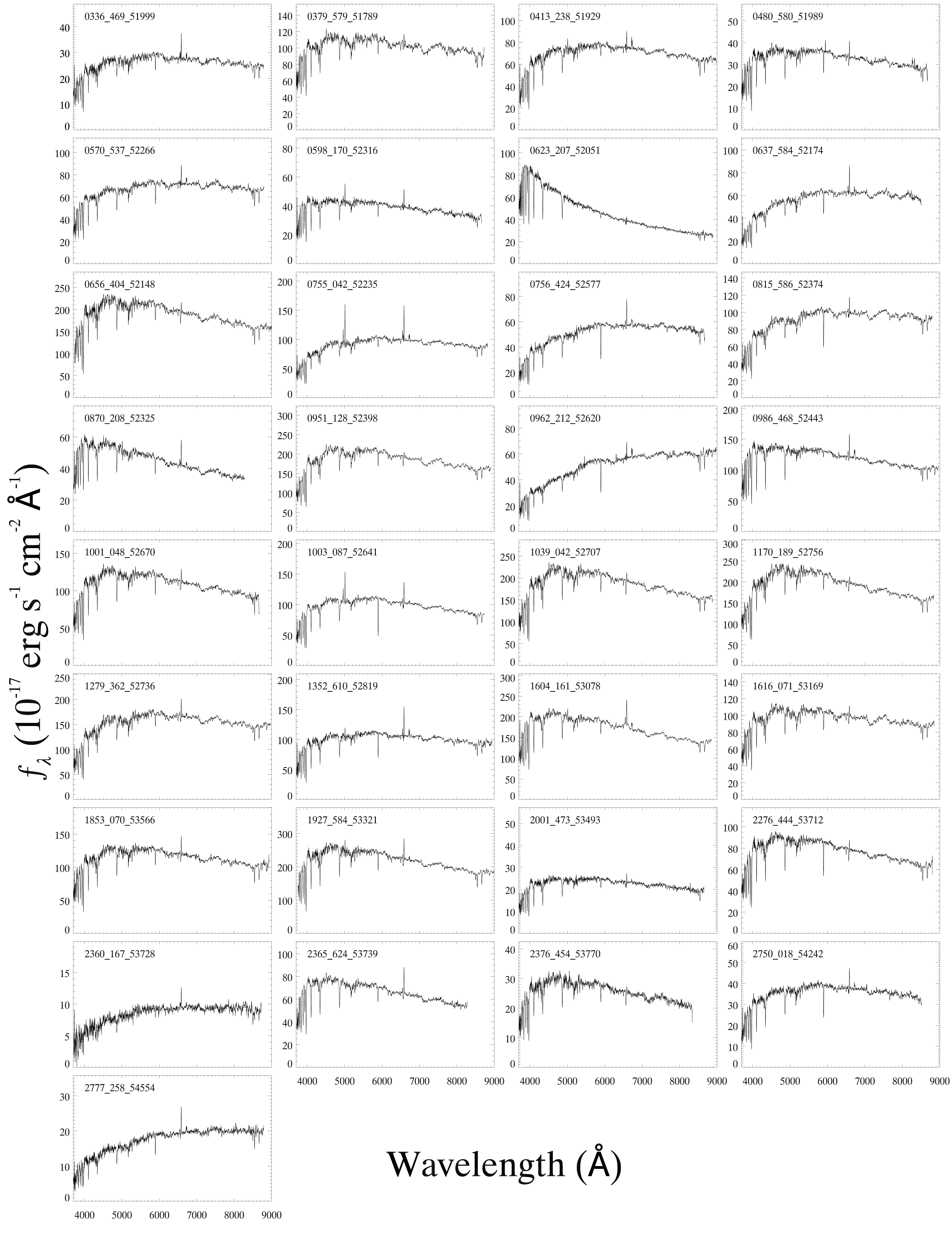}
\caption{SDSS spectra for the sample. Each galaxy's ID is shown in the upper left corner, with spectra sorted by plate ID.}
\label{fig:sdss_spec}
\end{figure*}

\clearpage

\newpage
\section{PAHFIT Decomposition Results}
\begin{figure*}[t]
\centering
\leavevmode
\includegraphics[width={0.95\linewidth}]{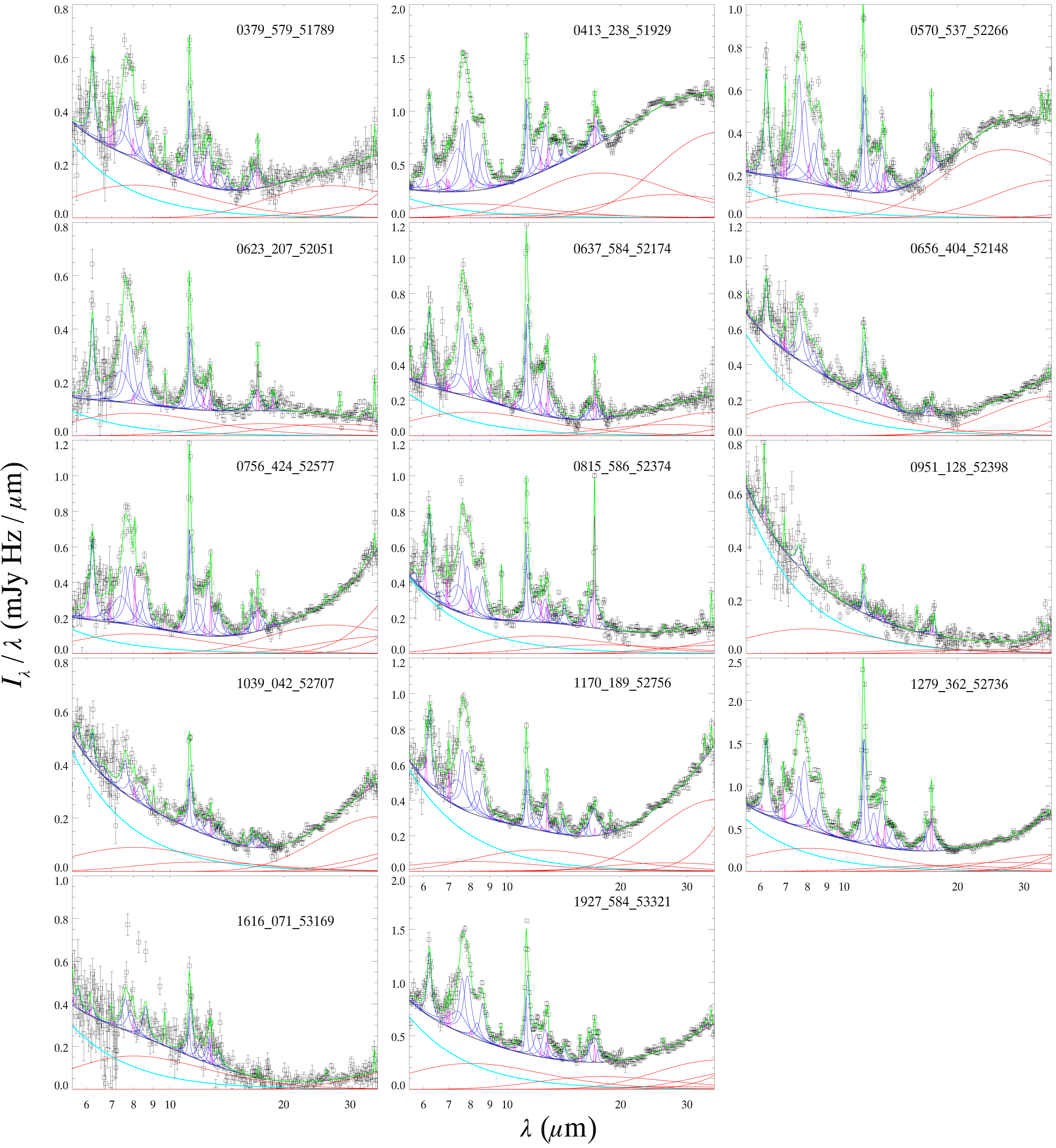}
\caption{PAHFIT decompositions of the IRS spectra for each of the remaining \textit{Spitzer} sources (minus 0962), ordered by plate number. The cyan curve represents the stellar continuum, the red curve the various dust continua, and the gray curve the total continuum. The blue curves indicate PAH emission features, while the magenta curves indicate atomic and molecular emission lines. The green curve shows the total fit. The silicate optical depth has been set to 0 for these sources, for the purpose of a more reliable fit. No silicate extinction can be seen by eye in the spectra, except for 0962\_212\_52620 (see \S~\ref{sec:specdecomp}).}
\label{fig:pahfits}
\end{figure*}

In Figure \ref{fig:pahfits} we present the spectral decompositions of each galaxy's IRS spectrum, using PAHFIT (see SDD07; \S\,\ref{sec:specdecomp}). In Tables \ref{tab:pah} and \ref{tab:lines}, we present the integrated PAH strengths and line fluxes from this decomposition. Note that only 15 of the 33 galaxies in the sample have IRS coverage and, thus, PAHFIT-derived results. 

%\cleardoublepage
\begin{deluxetable*}{lppppppp}
\tablecaption{PAH Feature Strengths\label{tab:pah}}
\tablecolumns{8}
\tabletypesize{\small}
\tablehead{%
\colhead{Galaxy} &
\colhead{6.2um} &
\colhead{7.7\um\ Complex} &
\colhead{8.6\um} &
\colhead{11.3\um\ Complex} &
\colhead{12.6\um\ Complex} &
\colhead{17\um\ Complex} &
\colhead{$\Sigma \PAH$} \\
%%%
\colhead{(1)} & 
\colhead{(2)} & 
\colhead{(3)} & 
\colhead{(4)} & 
\colhead{(5)} & 
\colhead{(6)} & 
\colhead{(7)} & 
\colhead{(8)} \\
}
\startdata
0379\_579\_51789 & $0.44 \pm 0.03$ & $1.45 \pm 0.09$ & $0.28 \pm 0.02$ & $0.59 \pm 0.02$ & $0.32 \pm 0.02$ & $0.29 \pm 0.03$ & $3.96 \pm 0.17$ \\
0413\_238\_51929 & $1.18 \pm 0.04$ & $5.39 \pm 0.15$ & $0.79 \pm 0.03$ & $1.47 \pm 0.02$ & $0.92 \pm 0.03$ & $0.82 \pm 0.03$ & $13.32 \pm 0.24$ \\
0570\_537\_52266 & $0.68 \pm 0.03$ & $2.43 \pm 0.07$ & $0.48 \pm 0.02$ & $0.98 \pm 0.02$ & $0.49 \pm 0.02$ & $0.35 \pm 0.03$ & $6.39 \pm 0.15$ \\
0623\_207\_52051 & $0.43 \pm 0.03$ & $1.82 \pm 0.09$ & $0.39 \pm 0.02$ & $0.57 \pm 0.01$ & $0.21 \pm 0.02$ & $0.27 \pm 0.01$ & $4.19 \pm 0.19$ \\
0637\_584\_52174 & $0.60 \pm 0.04$ & $2.47 \pm 0.09$ & $0.50 \pm 0.03$ & $1.21 \pm 0.02$ & $0.59 \pm 0.03$ & $0.43 \pm 0.04$ & $6.67 \pm 0.20$ \\
0656\_404\_52148 & $0.45 \pm 0.05$ & $1.02 \pm 0.04$ & $0.12 \pm 0.03$ & $0.55 \pm 0.02$ & $0.27 \pm 0.03$ & $0.23 \pm 0.04$ & $3.56 \pm 0.17$ \\
0756\_424\_52577 & $0.65 \pm 0.03$ & $2.41 \pm 0.08$ & $0.44 \pm 0.03$ & $1.18 \pm 0.02$ & $0.60 \pm 0.02$ & $0.55 \pm 0.04$ & $7.27 \pm 0.18$ \\
0815\_586\_52374 & $0.66 \pm 0.03$ & $2.32 \pm 0.10$ & $0.43 \pm 0.03$ & $0.85 \pm 0.02$ & $0.32 \pm 0.02$ & $0.75 \pm 0.02$ & $6.61 \pm 0.22$ \\
0951\_128\_52398 & $<$0.15 & $0.22 \pm 0.04$ & $<$0.06 & $0.17 \pm 0.01$ & $0.05 \pm 0.02$ & $0.10 \pm 0.03$ & $0.86 \pm 0.12$ \\
0962\_212\_52620 & $4.44 \pm 0.04$ & $16.66 \pm 0.21$ & $4.33 \pm 0.06$ & $4.60 \pm 0.04$ & $1.79 \pm 0.03$ & $4.71 \pm 0.07$ & $40.04 \pm 0.36$ \\
1039\_042\_52707 & $0.17 \pm 0.03$ & $0.43 \pm 0.09$ & $0.08 \pm 0.02$ & $0.42 \pm 0.01$ & $0.13 \pm 0.02$ & $0.18 \pm 0.02$ & $2.19 \pm 0.18$ \\
1170\_189\_52756 & $0.60 \pm 0.04$ & $2.44 \pm 0.15$ & $0.37 \pm 0.03$ & $0.60 \pm 0.02$ & $0.31 \pm 0.02$ & $0.56 \pm 0.03$ & $5.85 \pm 0.24$ \\
1279\_362\_52736 & $1.26 \pm 0.04$ & $5.32 \pm 0.10$ & $0.86 \pm 0.02$ & $2.56 \pm 0.02$ & $1.22 \pm 0.02$ & $1.26 \pm 0.02$ & $14.74 \pm 0.21$ \\
1616\_071\_53169 & $<$0.21 & $0.57 \pm 0.06$ & $0.28 \pm 0.05$ & $0.48 \pm 0.04$ & $0.22 \pm 0.04$ & $<$0.06 & $2.39 \pm 0.22$ \\
1927\_584\_53321 & $0.86 \pm 0.04$ & $3.58 \pm 0.12$ & $0.65 \pm 0.03$ & $1.42 \pm 0.02$ & $0.62 \pm 0.02$ & $0.92 \pm 0.05$ & $9.25 \pm 0.24$ \\
\enddata
\tablecomments{(1) Galaxy ID. \\
(2)-(7) Integrated fluxes of the primary PAH emission features returned by PAHFIT, in units of $10^{-16}$\ W m$^{-2}$.\\
(8) Total integrated PAH flux returned by PAHFIT, including minor emission features, in units of $10^{-16}$\ W m$^{-2}$.\\
3$\sigma$\ upper limits are prefaced by $<$.}
\end{deluxetable*}

\floattable
\begin{deluxetable*}{rccccccccc}
\rotate
\tablecaption{PAHFIT Emission Line Fluxes\label{tab:lines}}
\tablecolumns{10}
\tabletypesize{\small}
\tablehead{%
\colhead{} &
\colhead{[Ar\,\textsc{ii}]} &
\colhead{[S\,\textsc{iv}]} &
\colhead{\neII} &
\colhead{\neIII} &
\colhead{[S\,\textsc{iii}]} &
\colhead{[O\,\textsc{iv}]} &
\colhead{[Fe\,\textsc{ii}]} &
\colhead{[S\,\textsc{iii}]} &
\colhead{[Si\,\textsc{ii}]} \\
%%%
\colhead{Galaxy} &
\colhead{7.0\um} &
\colhead{10.5\um} &
\colhead{12.8\um} &
\colhead{15.6\um} &
\colhead{18.7\um} &
\colhead{25.9\um} &
\colhead{26.0\um} &
\colhead{33.5\um} &
\colhead{34.9\um} \\
}
\startdata
0379\_579\_51789 & $<$5.86 & $<$3.34 & $<$2.16 & $<$2.86 & $<$3.45 & \ldots & \ldots & $<$4.69 & $<$5.59 \\
0413\_238\_51929 & $<$6.52 & \ldots & 5.18 $\pm$~0.72 & 2.08 $\pm$~0.40 & \ldots & $<$4.15 & $<$3.09 & $<$2.73 & $<$3.30 \\
0570\_537\_52266 & 8.03 $\pm$~0.96 & $<$2.29 & $<$2.16 & $<$2.55 & \ldots & \ldots & $<$2.25 & 4.06 $\pm$~0.67 & $<$5.08 \\
0623\_207\_52051 & \ldots & \ldots & $<$3.15 & \ldots & $<$1.89 & \ldots & $<$0.71 & $<$1.83 & 5.26 $\pm$~0.74 \\
0637\_584\_52174 & $<$9.29 & $<$3.05 & 5.32 $\pm$~0.58 & \ldots & $<$3.48 & \ldots & $<$2.41 & $<$5.29 & \ldots \\
0656\_404\_52148 & $<$14.61 & $<$3.70 & $<$3.81 & $<$2.65 & \ldots & $<$1.61 & \ldots & $<$2.57 & $<$3.45 \\
0756\_424\_52577 & \ldots & $<$2.64 & 5.77 $\pm$~0.50 & $<$4.27 & $<$3.11 & 1.61 $\pm$~0.29 & \ldots & $<$4.04 & $<$6.97 \\
0815\_586\_52374 & $<$7.91 & $<$2.52 & 2.90 $\pm$~0.48 & 2.69 $\pm$~0.35 & \ldots & $<$1.96 & $<$2.54 & \ldots & 5.24 $\pm$~0.68 \\
0951\_128\_52398 & \ldots & $<$2.14 & $<$1.70 & $<$2.91 & \ldots & $<$1.22 & \ldots & $<$2.41 & $<$3.41 \\
0962\_212\_52620 & 13.29 $\pm$~1.34 & \ldots & 18.15 $\pm$~0.51 & 10.84 $\pm$~0.70 & 8.53 $\pm$~1.05 & \ldots & 4.84 $\pm$~0.43 & \ldots & \ldots \\
1039\_042\_52707 & \ldots & $<$1.94 & $<$1.56 & $<$1.10 & \ldots & $<$1.68 & $<$1.61 & $<$2.62 & $<$3.19 \\
1170\_189\_52756 & $<$7.41 & \ldots & $<$5.86 & $<$2.53 & $<$2.76 & $<$1.89 & \ldots & 4.08 $\pm$~0.57 & 3.58 $\pm$~0.65 \\
1279\_362\_52736 & $<$8.20 & 2.72 $\pm$~0.48 & 7.45 $\pm$~0.47 & $<$2.14 & \ldots & $<$4.66 & $<$3.83 & $<$2.55 & $<$2.93 \\
1616\_071\_53169 & $<$10.96 & \ldots & $<$6.76 & $<$3.55 & \ldots & $<$4.00 & $<$3.69 & \ldots & $<$4.66 \\
1927\_584\_53321 & $<$8.60 & $<$5.12 & 6.55 $\pm$~0.50 & 6.29 $\pm$~0.62 & $<$2.60 & $<$4.37 & $<$5.55 & $<$3.51 & $<$4.21 \\
\enddata
\tablecomments{Integrated fine-structure emission line fluxes from PAHFIT, reported in units of $10^{-18}$\ W m$^{-2}$. $5\sigma$\ upper limits are prefaced with `$<$'. Non-detections are denoted by `$\ldots$'.}
\end{deluxetable*}

\floattable
\begin{deluxetable*}{rcccccccc}
\rotate
\tablecaption{PAHFIT H$_{2}$\ Rotational Line Fluxes\label{tab:h2lines}}
\tablecolumns{9}
\tabletypesize{\small}
\tablehead{%
\colhead{} &
\colhead{H$_{2}$S(0)} &
\colhead{H$_{2}$S(1)} &
\colhead{H$_{2}$S(2)} &
\colhead{H$_{2}$S(3)} &
\colhead{H$_{2}$S(4)} &
\colhead{H$_{2}$S(5)} &
\colhead{H$_{2}$S(6)} &
\colhead{H$_{2}$S(7)} \\
%%%
\colhead{Galaxy} &
\colhead{28.2\um} &
\colhead{17.0\um} &
\colhead{12.3\um} &
\colhead{9.7\um} &
\colhead{8.0\um} &
\colhead{6.9\um} &
\colhead{6.1\um} &
\colhead{5.5\um} \\
}
\startdata
0379\_579\_51789 & \ldots & 0.35 $\pm$~0.011 & 0.18 $\pm$~0.018 & 0.36 $\pm$~0.019 & 0.26 $\pm$~0.035 & 0.38 $\pm$~0.037 & \ldots & \ldots \\
0413\_238\_51929 & 0.08 $\pm$~0.006 & 0.86 $\pm$~0.012 & 0.46 $\pm$~0.022 & 0.49 $\pm$~0.025 & 0.55 $\pm$~0.041 & 0.36 $\pm$~0.051 & 0.42 $\pm$~0.054 & \ldots \\
0570\_537\_52266 & $<$0.04 & 0.81 $\pm$~0.014 & 0.12 $\pm$~0.020 & 0.59 $\pm$~0.029 & 0.36 $\pm$~0.038 & 0.38 $\pm$~0.071 & \ldots & $<$0.18 \\
0623\_207\_52051 & 0.28 $\pm$~0.003 & 0.50 $\pm$~0.005 & 0.14 $\pm$~0.009 & 0.45 $\pm$~0.013 & $<$0.14 & \ldots & \ldots & \ldots \\
0637\_584\_52174 & 0.10 $\pm$~0.011 & 0.68 $\pm$~0.014 & 0.19 $\pm$~0.027 & 0.65 $\pm$~0.028 & 0.38 $\pm$~0.042 & $<$0.25 & 0.28 $\pm$~0.052 & 0.53 $\pm$~0.053 \\
0656\_404\_52148 & \ldots & 0.17 $\pm$~0.019 & \ldots & 0.15 $\pm$~0.029 & \ldots & $<$0.36 & $<$0.25 & $<$0.27 \\
0756\_424\_52577 & 0.14 $\pm$~0.007 & 0.49 $\pm$~0.014 & 0.15 $\pm$~0.026 & 0.35 $\pm$~0.031 & 0.88 $\pm$~0.038 & $<$0.41 & 0.29 $\pm$~0.048 & \ldots \\
0815\_586\_52374 & 0.14 $\pm$~0.005 & 1.75 $\pm$~0.006 & 0.36 $\pm$~0.014 & 1.01 $\pm$~0.019 & 0.68 $\pm$~0.034 & 0.42 $\pm$~0.077 & 0.57 $\pm$~0.045 & \ldots \\
0951\_128\_52398 & \ldots & 0.14 $\pm$~0.009 & $<$0.08 & $<$0.12 & \ldots & 0.25 $\pm$~0.048 & 0.70 $\pm$~0.065 & $<$0.41 \\
0962\_212\_52620 & 1.80 $\pm$~0.010 & 5.67 $\pm$~0.022 & 1.36 $\pm$~0.046 & 12.59 $\pm$~0.207 & 1.03 $\pm$~0.084 & 1.50 $\pm$~0.075 & \ldots & 0.54 $\pm$~0.069 \\
1039\_042\_52707 & \ldots & $<$0.03 & $<$0.09 & 0.23 $\pm$~0.019 & 0.19 $\pm$~0.031 & \ldots & \ldots & \ldots \\
1170\_189\_52756 & 0.04 $\pm$~0.004 & 0.11 $\pm$~0.008 & $<$0.06 & $<$0.09 & \ldots & \ldots & 0.45 $\pm$~0.060 & \ldots \\
1279\_362\_52736 & 0.23 $\pm$~0.004 & 1.38 $\pm$~0.008 & 0.38 $\pm$~0.027 & 0.91 $\pm$~0.025 & 0.42 $\pm$~0.059 & 1.03 $\pm$~0.071 & 0.29 $\pm$~0.039 & \ldots \\
1616\_071\_53169 & \ldots & 0.13 $\pm$~0.013 & \ldots & 0.61 $\pm$~0.025 & 0.19 $\pm$~0.036 & $<$0.22 & $<$0.21 & 0.34 $\pm$~0.038 \\
1927\_584\_53321 & $<$0.03 & 0.47 $\pm$~0.014 & 0.13 $\pm$~0.021 & 0.56 $\pm$~0.030 & $<$0.24 & $<$0.38 & $<$0.25 & \ldots \\
\enddata
\tablecomments{Integrated H$_{2}$\ pure rotational emission line fluxes from PAHFIT, reported in units of $10^{-18}$\ W m$^{-2}$. $5\sigma$\ upper limits are prefaced with `$<$'. Non-detections are denoted by `$\ldots$'.}
\end{deluxetable*}

\end{document}